\documentclass[a4paper,fleqn]{cas-dc}

\usepackage{bm}

\usepackage{booktabs}
\usepackage{tabularx}
\usepackage{lineno}
\usepackage{hyperref}
\usepackage{cleveref}
\usepackage{amsthm}

\usepackage[numbers]{natbib}
\usepackage{algorithmic}
\usepackage{algorithm}
\usepackage{graphicx}
\usepackage{epstopdf}
\usepackage{appendix}

\def\tsc#1{\csdef{#1}{\textsc{\lowercase{#1}}\xspace}}
\tsc{WGM}
\tsc{QE}
\tsc{EP}
\tsc{PMS}
\tsc{BEC}
\tsc{DE}

\begin{document}
\let\WriteBookmarks\relax
\def\floatpagepagefraction{1}
\def\textpagefraction{.001}

\shorttitle{Robust Longitudinal-lateral Look-ahead Pursuit Path-Following Control: Fast Finite-Time Stability Guarantee}

\author[a]{Zimao Sheng}[style=chinese,
						type=editor,
                        auid=000,
                        orcid=0000-0001-6067-3379
                        ]
\ead{hpShengZimao@163.com}

\author[a]{Hong’an Yang}[style=chinese,
                        orcid=0000-0001-5426-0041]
 \cortext[cor1]{Corresponding author}
 \cormark[1]
 
\ead{yhongan@nwpu.edu.cn}

\author[a]{Shuxiang Yang}

\author[b]{Zirui Yu}

\shortauthors{ZM. S and HA. Y, etc.}

\affiliation[a]{
	organization = {School of Mechanical Engineering, Northwestern Polytechnical University},
	city         = {Xi'an, Shaanxi},
	postcode     = {710072},
	country      = {China},
}

\affiliation[b]{
	organization = { School of Artificial Intelligence, Tianjin University of Science and Technology},
	city         = {Tianjin},
	postcode     = {300457},
	country      = {China},
}

\title [mode = title]{Robust Longitudinal-lateral Look-ahead Pursuit Path-Following Control: Fast Finite-Time Stability Guarantee}                      

\begin{abstract}
This paper addresses the challenging problem of robust path-following for fixed-wing unmanned aerial vehicles (UAVs) in complex environments with bounded external disturbances and non-smooth predefined paths. Due to the unique aerodynamic characteristics and flight constraints of fixed-wing UAVs, achieving accurate and fast stable path following remains difficult, especially in low-altitude mountainous terrains, urban landscapes, and under wind disturbances. Most existing path-following guidance laws often struggle to ensure fast stabilization under unknown bounded disturbances while maintaining sufficient robustness, and there is a lack of research on optimizing robustness for non-smooth paths under flight constraints. This paper addresses these issues by proposing a constraints-based robust path-following controller. Firstly, from the perspective of global random attractor, we innovatively introduce robustness metrics that quantify both the exponential convergence rate and the range of the ultimate attractor set. Secondly, building on these metrics, we develop a robust longitudinal-lateral look-ahead pursuit (RLLP) guidance law for fixed-wing UAVs, specifically considering the flight path angle and track angle under external disturbances. Thirdly, we also derive an optimized version (Optimal-RLLP) to enhance the robustness metrics, and elaborate on the sufficient conditions for fast finite-time stability, ensuring the guidance law achieves finite-time stability and robustness with reduced sensitivity to constrained uncertainties. The simulation results validate the proposed guidance law's feasibility, optimality and robustness under atmospheric disturbances using a high-fidelity  simulation platform and provide key principle for practical deployment.
\end{abstract}


\maketitle
\section{Introduction}
\label{sec:introduction}
\subsection{Motivation}
In recent years, fixed-wing unmanned aerial vehicles (UAVs) are widely used in extensive range of applications encompassing environmental monitoring\cite{WOS:000862429800049}, rescue operations\cite{WOS:001096585700001}, military reconnaissance\cite{ren2024research}, etc., primarily due to their attributes of large payload capacity, high speed, and long flight endurance. The autonomous missions execution of many fixed-wing UAVs heavily relies on predefined geometric path following \cite{aguiar2007trajectory}\cite{aguiar2008performance}\cite{sujit2014unmanned} guidance law, which is deployed in the flight control system (FCS) and outputs the acceleration commands according to the predefined path and constraints on the vehicle’s dynamics. However, due to the unique aerodynamic characteristics and flight constraints of fixed-wing UAVs, achieving robust path following in complex low-altitude mountainous terrains, urban landscapes, and under wind disturbances remains a challenging problem. The conventional path-following problem under perturbed conditions tends to impose external disturbances on the three-axis velocities, neglecting the impact on flight-path angle and track angle. Furthermore, most path-following guidance laws struggle to ensure fast stabilization under unknown bounded disturbances while maintaining a certain robustness threshold. Additionally, there is a lack of further research on how to optimally design the robustness for non-smooth paths based on the flight constraints. These issues inspire us to design a robust constraints-based path-following controller that can achieve fast robust stability and fast finite-time stability in the case of unstructured predefined path and constrained disturbances.
\subsection{Related work}
Classical path-following algorithms have predominantly focused on 2D contexts\cite{sujit2014unmanned}\cite{kumar2023robust}\cite{ranjan2024robust}. In order to tackle the challenge of achieving high-robustness online 3D UAVs path-following, the primary methodologies\cite{kumar2024three} have encompassed error regulation based\cite{kaminer2010path}\cite{piprek2022optimal}\cite{yamasaki2013separate}, optimal control\cite{rezk2024predictive}\cite{yang2020optimal}, vector field based\cite{yao2021singularity}\cite{yao2023path}, and virtual target based\cite{Wang}\cite{Kumar}\cite{beard2014fixed} approaches. 

Error regulation based method prioritizes the achievement of asymptotic convergence for positional errors. Piprek\cite{piprek2022optimal} decoupled the three-dimensional error model into time-varying linear models in both lateral and longitudinal planes, achieving error asymptotic stability by configuring the pole positions of the linear time-varying system (LTV) at any given moment. Isaac\cite{kaminer2010path} employed the $L_1$ adaptive augmentation algorithm to design an adaptive feedback controller for the linearized kinematic model of UAVs by solving the Lyapunov equation under perturbed conditions, thereby guaranteeing the tuning effectiveness of the error. Takeshi\cite{yamasaki2013separate} , however, constrained the convergence surface of the tracking error by solving for high-order sliding mode variable structure control. Nevertheless, these methods struggles to guarantee a effective stable time, and model-based control is overly complex and challenging to implement into the autopilot for 3D path-following.

Optimal control based method tends to utilize optimal control techniques to minimize both the quadratic form of tracking deviation and the quadratic form of control input over a future time horizon. Ahmed\cite{rezk2024predictive} employed quasi-linear parameter-varying model predictive control (ql-MPC) with time-varying semilinear characteristics to perform rolling optimization for solving the optimal strategy of a three-dimensional perturbed velocity form kinematic model, achieving robust path following under high disturbance conditions. Yang\cite{yang2020optimal} designed a finite-time-horizon optimal control law for an affine kinematic model under perturbed conditions, enabling exponential convergence of path tracking deviations. However, such optimal control methods have high computational costs, making it difficult to meet the requirements for real-time path following.

Vector field based method involves designing a vector field to compel the  UAV flight trajectory to follow a path formed by the intersection of multiple planes\cite{yao2023path}. However, this method is primarily suited for relatively regular and simple paths\cite{yao2021singularity} such as straight lines and arcs, and it falls short when dealing with complex path following tasks.

Compared to the aforementioned methods, due to its simplicity, the virtual target based method is more commonly used in practice\cite{beard2014fixed}. In this approach, the UAV continuously approximates a virtual target point moving along a predefined path. The virtual target can be set in front of the current position and on the extension of the desired path. Its distance is related to the current tracking error. By real-timely measuring position and angle errors, and correcting the attitude input of the UAV between the current position and the virtual target, the actual trajectory and the predefined path will exhibit a high degree of congruence, thereby realizing efficient path-following. This characteristic makes the algorithm suitable for following a relatively general 3D path. Zian\cite{Wang} employed the $L_1$ guidance law in both the horizontal and vertical planes for a 3D UAV to achieve optimal path point approximation and thus path following after selecting the optimal virtual approach point, which led to the successful completion of real-flight experiments. However, this method struggles to guarantee positional and angular tracking errors, and it may not robustly and stably converge under external disturbances caused by factors such as wind fields and the movement of virtual points. Despite existing research\cite{Kumar} attempts to address the finite-time stability of path-following errors under wind disturbances, these methods still primarily consider relatively simple 2D particle model and the wind disturbances are mainly imposed on the impact on velocity, instead of angular rates. Additionally, they seldom take into account the input command saturation and the effectiveness of following non-smooth path.

The impact of disturbance factors, including wind field disturbances and model simplification, on the path following performance under constrained kinetic limitations, is often substantial. These disturbances directly influence the flight velocity and angular rate of UAVs. Therefore, a relatively straightforward approach is to correct the disturbance term by incorporating a compensation quantity. As a result, Thomas\cite{stastny2018l1} introduces an airspeed reference command increment according to the wind speed ratio, and  the $L_1$ ratio and feasibility parameters are adaptively compensated in an attempt to mitigate run-away phenomena in over-wind scenarios. The authors\cite{yang2020optimal}\cite{liu2013path} seeks to employ disturbance observers to dynamically compensate for the estimated disturbances, thereby bolstering the robustness of the guidance rate in the presence of wind disturbances. Similarly, compensation effects can also be applied to classical feedback controllers to address external disturbances. Souanef\cite{souanef2022mathcal} employs feedback linearized dynamics compensation based on $L_1$ adaptive control for UAVs path-following controller with parametric uncertainties and external disturbances. Wu\cite{WOS:Wu} proposed a robust fuzzy control scheme that employs a proportional-integral-derivative(PID) controller based on feedback linearization to compensate nonlinear deviations caused by system uncertainties and modeling parameters. 

The previously mentioned efforts have yielded satisfactory performance improvements in mitigating the effects of disturbances, yet most research frameworks still rely on controllers that guarantee asymptotic convergence. Such controllers may be hard to ensure rapid convergence time and robustness metric\cite{Kumar}. When virtual target points are densely distributed, these controllers may find it challenging to handle disturbances and uncertainties, leading to deviations in the system's steady state. The system may exhibit oscillations and unpredictable behaviors while abrupt changes in state occur due to external disturbances such as wind fields, measurement noise, and other error sources. Furthermore, the system's path-following error is further amplified by factors such as external disturbances, if the UAV is initially far away from the desired path, the movement of virtual target points, and sharp curvatures, resulting in an inability to provide the expected following performance. Nevertheless, the above-mentioned model-based methodologies pose challenges in guaranteeing error stabilization within a finite timeframe in the presence of external perturbations.

\subsection{Contribution}
In this paper, we further augment the preliminary results obtained in\cite{Wang} with additional in-depth analysis, deriving a robust guidance law in the presence of bounded external disturbances and non-smooth predefined path within finite-time stability guarantee, and summarize our main contributions below. 
\begin{itemize}
	\item Addressing the quantification of robust stability for perturbed nonlinear dynamic systems, we innovatively propose robustness metrics from the perspective of global random attractor—metrics that quantify both the exponential convergence rate and the range of the ultimate attractor set. This work establishes a foundation for subsequent robustness analysis of the robust longitudinal and lateral look-ahead pursuit path-following guidance law;
	\item To quantify and enhance the robustness of traditional look-ahead pursuit guidance laws, we further elaborate on the sufficient conditions for their fast finite-time stability. Building upon the aforementioned robustness metrics, we present a robust longitudinal and lateral look-ahead pursuit (RLLP) guidance law for fixed-wing UAVs, with specific consideration of the flight path angle and track angle under external disturbances. Additionally, we derive an optimized version of the RLLP (Optimal-RLLP) that further improves its robustness metrics. Consequently, the proposed guidance law guarantees finite-time stability, robustness, reduced sensitivity to constrained uncertainties compared to traditional asymptotic convergence controllers. Results validate the guidance law’s robustness under small disturbances and its feasibility for practical deployment;
	\item We verify the feasibility and robustness of the RLLP guidance law based on a UAV path-following platform integrating realistic low-altitude urban scenarios, actual UAV dynamic models, aerodynamic models, and atmospheric disturbances. Through detailed analysis, we provide principles for adjusting algorithm parameters when deploying the RLLP in practical applications.
\end{itemize}

The remainder of this paper is organized as follows. After an overview in Section \ref{sec:introduction}, the numerous theoretical results regarding the robustness of nonlinear perturbed systems intended for subsequent guidance law design is given in Section \ref{sec:Preliminaries}, and the perturbed model and constrained input commands, along with the problem formulation, is presented in Section \ref{sec:Problem Formulation}. The proposed robust guidance law and its modified optimal version is designed in Section \ref{sec:Path Following Controller with optimal robustness indictor}, followed by a matched autopilot design in Section \ref{sec:The Autopilot} and simulation experiments in Section \ref{sec:Simulation}. Finally, we present concluding remarks and directions for future investigation in Section Conclusions and future work.

\section{Preliminaries}
\label{sec:Preliminaries}
In this section, we present several key definitions, lemmas, and theorems as essential tools for subsequent analyses.
\subsection{Definitions}
\newtheorem{definition}{Definition}
\begin{definition}
	(Robust stability) For perturbed nonlinear system $\dot{y} = f(t,y) + g(t,y)$, $y \in \mathbb{R}^n$,$f(t,0)=0$, $f,g \in C[I\times S_H,\mathbb{R}^n]$,$S_H=\{x|||x||\leq H\}$, if $\forall \varepsilon >0$, there exists $\delta_1(\varepsilon) >0$ and $\delta_2(\varepsilon) >0$ to make $||g(t,y)||\leq \delta_1(\varepsilon)$, $||y(0)||\leq \delta_2(\varepsilon)$, and $||y||\leq \varepsilon$, then the trivial solution of $\dot{y} = f(t,y) + g(t,y),f(t,0)=0$ exhibits robust stability.
\end{definition}

\newtheorem{definition2}[definition]{Definition}
\begin{definition2}
	(Global attractor) For a dynamical system: $\dot{x}(t) = f(x(t))$, $x(t)\in \mathbb{R}^n$, where $f:\mathbb{R}^n\rightarrow \mathbb{R}^n$ is a deterministic vector field (typically Lipschitz continuous). The set $A$ for this state $x(t)$ is defined as the global attractor if $x(t)\notin A$, then the distance between $x(t)$ and $A$: $dist(x(t),A) \rightarrow 0$ as $t\rightarrow \infty$; if $x(t)\in A$, then $x(t+\Delta t) \in A$ for any $\Delta t \geq 0$.
\end{definition2}

\newtheorem{definition3}[definition]{Definition}
\begin{definition3}
	(Global random attractor) For a dynamical system governed by a stochastic differential equation (SDE): $\dot{x} = f(x) + d$, $x\in \mathbb{R}^n$ and $d$ is a stochastic perturbation. Its global attractor can be regarded as the global random attractor of the SDE.
\end{definition3}

\newtheorem{definition4}[definition]{Definition}
\begin{definition4}
	(Co-Lipschitz property) For function $f:X\rightarrow \mathbb{R}^m$, $X\subset \mathbb{R}^n$, if there exists $L>0$ for any $x_1,x_2\in X$ such that $||f(x_1) - f(x_2)||_2\geq L||x_1-x_2||_2$ holds, then $f$ is referred to $L$-co-Lipschitz over $X$.
\end{definition4}

\subsection{Key lemmas and theorem}
\newtheorem{lemma0}{Lemma}
\begin{lemma0}
	\label{thm:lem0}
	(Stability of matrix\cite{bib:Schaft}) If $A$ is stable, which means $Res(A)<0$ or eigenvalue $0$ corresponds to the single characteristic factor, there exists only one positive define $P=P^T=P(A,\delta)>0$ for any $\delta >0$ to make:
	\begin{equation}
		P(A,\delta)A + A^TP(A,\delta) + \delta I  = O
	\end{equation}
\end{lemma0}
\newtheorem{lemma2}[lemma0]{Lemma}
\begin{lemma2}(The fast finite-time stability\cite{khoo2014multi})
	\label{lem:lemma2}
	For a nonlinear system $\dot{x}(t) =f(x(t))$ with $f(0)=0$, $x\in \Omega \subset \mathbb{R}^n$, suppose there is a positive definite function $V(x)$ for any nonzero $x$ satisfying
		\begin{equation}
				\dot{V}(x) + \alpha V(x) + \beta V^{\gamma}(x) \leq 0
			\end{equation}
		where $\alpha >0$, $\beta >0$ and $0\leq \gamma < 1$. Then, the origin of system is fast finite-time stable in $\Omega$, and the settling time, depending on the initial state $x(0)=x_0$, is given by
		\begin{equation}
		T(x_0)\leq \frac{1}{\alpha (1-\gamma)}\ln (1+\frac{\alpha V^{1-\gamma}(x_0)}{\beta})
		\end{equation}
\end{lemma2}	
\newtheorem{lemma}[lemma0]{Lemma}
\begin{lemma}(The fast finite-time stability\cite{khoo2014multi})
	\label{thm:lem1}
	For a nonlinear system $\dot{x}(t) =f(x(t))$ with initial state $x(0)=x_0,f(0)=0$, $x\in \mathbb{R}^n$, suppose there exists a positive definite function $V(x)$ for any nonzero $x$ satisfying
	\begin{equation}
		\label{equ:lem1_equ}
		\dot{V}(x) + \alpha V(x) - \beta V^{\frac{1}{2}}(x)  \leq 0
	\end{equation}
	where $\alpha >0$, $\beta >0$. Then for any state $x(t)$ there exists
	\begin{equation}
		\label{equ:lem2_equ}
		V^{\frac{1}{2}}(x(t)) \leq \frac{\beta}{\alpha} + (V^{\frac{1}{2}}(x_0) - \frac{\beta}{\alpha})e^{-\frac{\alpha}{2}t}
	\end{equation}
\end{lemma}
\begin{proof}
	We define \(s(t) = V^{\frac{1}{2}}(x)\), and then Eq.(\ref{equ:lem1_equ}) can be transformed into
	\begin{equation}
		\dot{s}(t) + \frac{\alpha}{2}s(t) \leq \frac{\beta}{2}
	\end{equation}
	The above equation can be further transformed into 
	\begin{equation}
		e^{\frac{\alpha}{2}t}\dot{s}(t) + \frac{\alpha}{2}e^{\frac{\alpha}{2}t}s(t) \leq \frac{\beta}{2}e^{\frac{\alpha}{2}t}
	\end{equation}
	which means
	\begin{equation}
		\frac{d}{dt}(e^{\frac{\alpha}{2}t}s(t)) \leq \frac{\beta}{2} e^{\frac{\alpha}{2}t}
	\end{equation}
	Integrating both sides of the inequality in the above equation from $0$ to $t$ simultaneously, we can obtain
	\begin{equation}
		e^{\frac{\alpha}{2}t}s(t) - s(0) \leq \frac{\beta}{\alpha} (e^{\frac{\alpha}{2}t} - 1)
	\end{equation}
	Therefore, the above equation can be further derived to obtain Eq.(\ref{equ:lem2_equ}). Thus, the proof is completed. 
\end{proof}

\newtheorem{remark}{Remark}
\begin{remark}
\label{rem:remark1}
Eq.(\ref{equ:lem2_equ}) reveals a fact that,  
as illustrated in Figure \ref{fig:0}, when the initial value \(x_0\) satisfies \(V^{\frac{1}{2}}(x_0)\leq \frac{\beta}{\alpha}\),  \(V^{\frac{1}{2}}(x(t))\leq \frac{\beta}{\alpha}\) holds for all \(t\geq t_0\). Conversely, if \(V^{\frac{1}{2}}(x_0)\geq \frac{\beta}{\alpha}\), \(V^{\frac{1}{2}}(x(t))\) is upper - bounded by a function that converges exponentially to \(\frac{\beta}{\alpha}\) with a rate of \(\alpha\). As \(t\) approaches infinity, once \(V^{\frac{1}{2}}(t_r)\leq \frac{\beta}{\alpha}\) is satisfied at a certain time \(t_r\), the trajectory of \(x(t)\) will remain within this region indefinitely. 
Consequently, it can be deduced that for any initial value $x(t_0) = x_0$, \(x(t)\) will ultimately converge to the global attractor described as:
\begin{equation}
	x(t) \in \{x|V^{\frac{1}{2}}(x)\leq \frac{\beta}{\alpha}\}, t\rightarrow \infty
\end{equation}
\begin{figure*}[htbp]
	\label{fig:0}
	\centering
	\begin{tabular}{cc}
		\includegraphics[width=0.9\columnwidth]{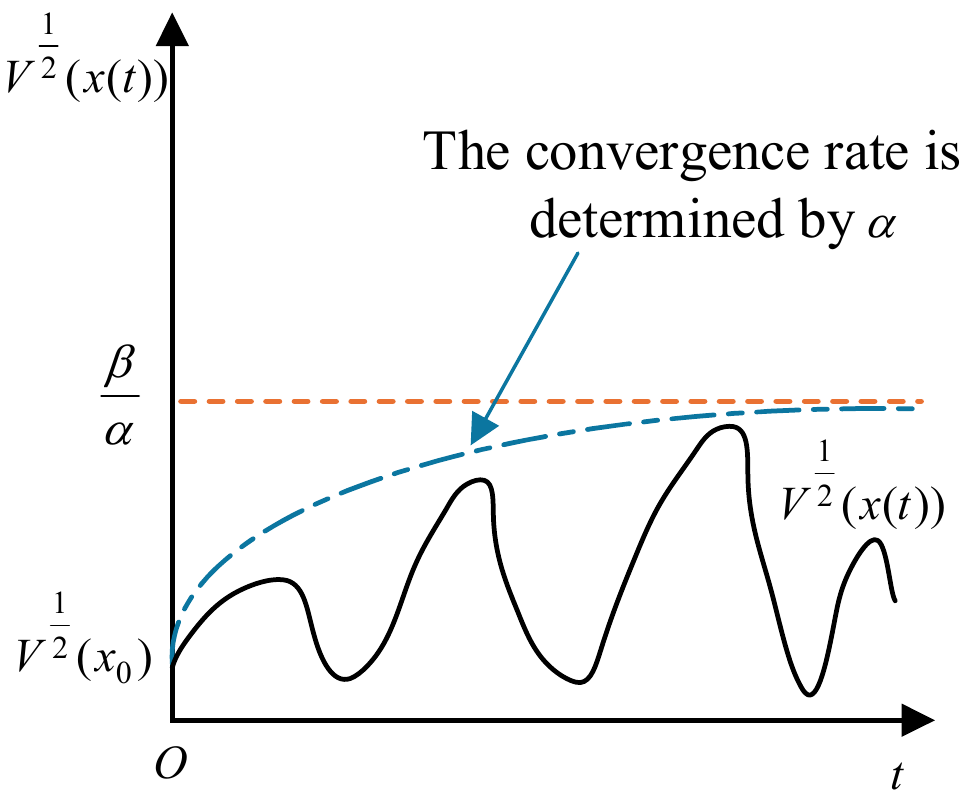} &
		\includegraphics[width=0.9\columnwidth]{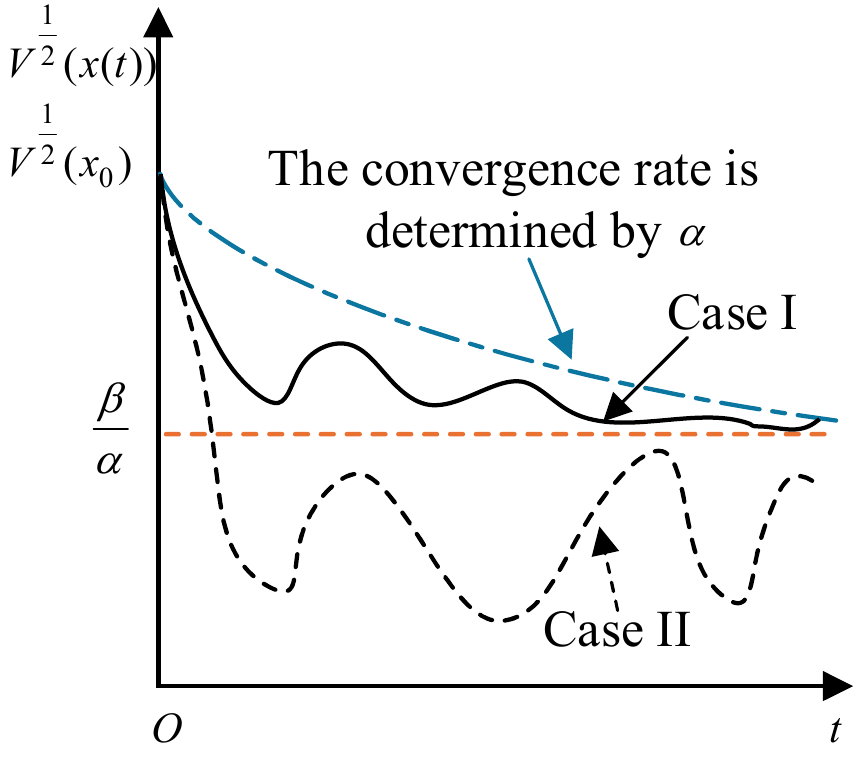}	\\
		(a) & (b)\\
	\end{tabular}
	\caption{The curves depicting the evolution of \(V^{\frac{1}{2}}(x_t)\) are presented under diverse initial values of \(V^{\frac{1}{2}}(x_0)\). (a):$V^{\frac{1}{2}}(x_0)\leq \frac{\beta}{\alpha}$; (b):$V^{\frac{1}{2}}(x_0)> \frac{\beta}{\alpha}$.}
\end{figure*}
\end{remark}

It is worth noting that we generally aspire for the convergence rate \(\alpha\) to be as large as possible. Meanwhile, a larger \(\alpha\) also enables the scope of the random attractor, \(\frac{\beta}{\alpha}\), to be minimized. On the basis of above, we introduce the following theorem as our main result:
\newtheorem{theorem}{Theorem}
\begin{theorem}
	\label{thm:lemma_A1}
	For the perturbed autonomous system \(\dot{x} = f(x)+d\), where $x\in \mathbb{R}^n$,\(x(t_0)=x_0\), \(f(0)=0\), and \(d\in \mathbb{R}^n\) is the perturbation term satisfying \(\|d\|_2\leq L_d\). If there exists a set \(\Omega\) such that when \(x\in\Omega\), the following conditions hold: \\
	1) there exists \(L_f(\Omega)\geq0\) such that \(\left\|\frac{\partial f(x)}{\partial x}\right\|_{2}\leq L_f(\Omega)\);\\ 
	2) there exists a positive-definite matrix \(P = P^T > 0\) and \(\varepsilon(\Omega)>0\) such that \(\left[\frac{\partial f(x)}{\partial x}\right]^T P+P\frac{\partial f(x)}{\partial x}+\varepsilon(\Omega) I\leq O\).\\
	Then the ultimate trajectory of the state \(x(t)\), for any initial value $x_0$, \(x(t)\) will exponentially converge to the following global random attractor $S(\Omega)$ at least the rate of $\varepsilon(\Omega)/\lambda_{\max}(P)$ as $t\rightarrow \infty$:
	\begin{equation}
		\label{eq:lemma_A1_condition}
		x(t)\rightarrow S(\Omega) = \{x: ||f(x)||_2\leq \frac{2L_dL_f(\Omega)\lambda^2_{\max}(P)}{\varepsilon(\Omega) \lambda_{\min}(P)} \}
	\end{equation}
\end{theorem}
\begin{proof}
	Let $V(x)=f(x)^TPf(x)$ for $P=P^T>O$ such that
	\begin{equation}
		\begin{split}
			\dot{V}(x) = & \dot{f}(x)^TPf(x) + f(x)^TP\dot{f}(x) \\
			= & \dot{x}^T[\frac{\partial f(x)}{\partial x}]^TPf(x) + f(x)^TP[\frac{\partial f(x)}{\partial x}]\dot{x} \\
			= & f(x)^T([\frac{\partial f(x)}{\partial x}]^TP + P\frac{\partial f(x)}{\partial x})f(x) \\
			+ & d^T[\frac{\partial f(x)}{\partial x}]^TPf(x) + f(x)^TP\frac{\partial f(x)}{\partial x} d \\
			\leq & -\frac{\varepsilon(\Omega)}{\lambda_{\max}(P)}V(x) + 2d^T[\frac{\partial f(x)}{\partial x}]^TPf(x) \\
			\leq & -\frac{\varepsilon(\Omega)}{\lambda_{\max}(P)}V(x) + 2L_d||\frac{\partial f(x)}{\partial x}||_2||Pf(x)||_2\\
			\leq& -\frac{\varepsilon(\Omega)}{\lambda_{\max}(P)}V(x) + \frac{2L_dL_f(\Omega) \lambda_{\max}(P)}{\sqrt{\lambda_{\min}(P)}}V(x)^{\frac{1}{2}}
		\end{split}
	\end{equation}   
	According to the Lemma \ref{thm:lem1} and Remark \ref{rem:remark1}, as $t\rightarrow \infty$, 
	\begin{equation}
		V^{\frac{1}{2}}(x) \leq \frac{2L_dL_f(\Omega)\lambda^2_{\max}(P)}{\varepsilon(\Omega) \sqrt{\lambda_{\min}(P)}}
	\end{equation}
	Consequently, as $t\rightarrow \infty$,
	\begin{equation}
		||f(x)||_2\leq\frac{V^{\frac{1}{2}}(x)}{\sqrt{\lambda_{\min}(P)}}\leq \frac{2L_dL_f(\Omega)\lambda^2_{\max}(P)}{\varepsilon(\Omega) \lambda_{\min}(P)}
	\end{equation}
	The proof is completed.
\end{proof}
Theorem \ref{thm:lemma_A1} elucidates a pivotal theoretical framework. Let \(\Omega\) be a non-empty subset of the state space \(\mathbb{R}^n\) that encompasses the origin. When the Jacobian of the function \(f(x)\), denoted as \(J_f(x) = \partial f(x)/\partial x\), is uniformly bounded for all \(x\) within the domain \(\Omega\), and \(J_f(x)\) exhibits negative definiteness throughout this region, the following property emerges for the associated dynamical system. For an autonomous dynamical system governed by the differential equation \(\dot{x}(t)=f(x(t))+d\), regardless of the chosen initial condition \(x_0\in\mathbb{R}^n\), the time-derivative of the state trajectory \(f(x(t))\), will converge exponentially to the global random attractor as delineated in Eq.(\ref{eq:lemma_A1_condition}). The spatial extent of this attractor is intricately linked to two critical parameters: the supremum \(L_f(\Omega)\) of the norm of the Jacobian \(J_f(x)\) over the domain \(\Omega\), and the maximum eigenvalue upper bound of \(J_f(x)\) under \(\Omega\). Specifically, a smaller upper bound for the ultimate attractor \(\|f(x)\|_2\) can be achieved through a smaller \(L_f(\Omega)\) and a larger \(\varepsilon(\Omega)\).
\begin{remark}
	\label{rem:rem_lemma_A1}
	According to Lemma \ref{thm:lem0}, there exists a sufficient condition for the validity of condition 2) in Theorem \ref{thm:lemma_A1}, specifically: when \(x \in \Omega\), the Jacobian \( \partial f/\partial x \) is stable. Furthermore, a special case arises when \(\Omega = \{0\}\), in this scenario, determining whether \(\partial f(0)/\partial x\) satisfies conditions 1) and 2) suffices to characterize the global random attractor \(S(0)\) of the perturbed system's trajectory.
\end{remark}
Remark \ref{rem:rem_lemma_A1} establishes the following theoretical principle: if the Jacobian matrix \( \partial f/\partial x \) of \(f(x)\) is continuously bounded and stable over a region \(x \in \Omega\), then the trajectory \(x(t)\) of the perturbed system will exponentially converge to a global random attractor in which the fluctuation amplitude of \(f(x)\) remains bounded. Notably, the extent of this attractor is governed by the eigenvalues of \( \partial f/\partial x \)—larger eigenvalue magnitudes generally correspond to a smaller constrained domain, thereby enabling the system to attain a higher level of robust stability.

To minimize the upper bound of $\|f(x)\|_2$ in Eq.(\ref{eq:lemma_A1_condition}), one must ensure that the ratio $\lambda_{\max}(P)/\lambda_{\min}(P)$ is minimized and the ratio $\varepsilon(\Omega)/\lambda_{\max}(P)$ is maximized. Given that $\lambda_{\max}(P) \geq \lambda_{\min}(P)$, the ratio $\lambda_{\max}(P)/\lambda_{\min}(P)$ reaches its minimum when $\lambda_{\max}(P) = \lambda_{\min}(P) = \varepsilon_P I_n$ (where $I_n$ denotes the identity matrix). Under this condition, when $\Omega = \{0\}$ and $f$ is $L_c$-co-Lipschitz on its neighborhood, the following more specialized conclusion holds.  
\newtheorem{theorem2}[theorem]{Theorem}
\begin{theorem2}
	\label{thm:lemma_A2}
	For the perturbed autonomous system \(\dot{x} = f(x)+d\), where $x\in \mathbb{R}^n$,\(x(t_0)=x_0\), \(f(0)=0\), and \(d\in \mathbb{R}^n\) is the perturbation term satisfying \(\|d\|_2\leq L_d\). 
	If the following conditions hold: \\
	1) there exists \(L_f \geq0\) such that \(\left\|\frac{\partial f(0)}{\partial x}\right\|_{2}\leq L_f\);\\ 
	2) $f$ is $L_c$-co-Lipschitz on the neighbor of origin;\\
	3) there exists \(\varepsilon > 0\) such that \(\left[\frac{\partial f(0)}{\partial x}\right]^T + \frac{\partial f(0)}{\partial x}+ \varepsilon I_n\leq O\).\\
	Then the ultimate trajectory of the state \(x(t)\), for any initial value $x_0$, \(x(t)\) will exponentially converge to the following global random attractor $S(0)$ at least the rate of $\varepsilon$ as $t\rightarrow \infty$:
	\begin{equation}
		\label{eq:lemma_A2_condition}
		x(t)\rightarrow S(0) = \{x: ||x||_2\leq \frac{2L_dL_f}{\varepsilon L_c} \}
	\end{equation}
\end{theorem2}
\begin{proof}
	Let $P=\varepsilon_P I_n>0$, $\varepsilon = \varepsilon(0)/\varepsilon_P$, according to the Theorem \ref{thm:lemma_A1} and Remark \ref{rem:rem_lemma_A1}, 
	\begin{equation}
		x(t)\rightarrow S(0) = \{x: ||f(x)||_2\leq \frac{2L_dL_f}{\varepsilon } \},\  t\rightarrow \infty
	\end{equation}
	Since $f$ is $L_c$-co-Lipschitz on the neighborhood of the origin, i.e., 
	as $t\rightarrow \infty$,
	\begin{equation}
		||x||_2 \leq \frac{||f(x) - f(0)||_2}{L_c} \leq \frac{2L_dL_f}{\varepsilon L_c}
	\end{equation}
	This completes the proof.
\end{proof}
\begin{remark}
	\label{rem:rem_lemma_A3}
	According to Theorem \ref{thm:lemma_A2}, on the one hand, $\varepsilon$ reflects the exponential convergence rate of the trajectory $x(t)$; on the other hand, it also characterizes the size of the global random attractor. This implies that a larger $\varepsilon$ corresponds to a faster convergence rate and a smaller attractor size (ideally, the attractor is the origin), i.e., higher robustness. On the one hand, the robustness of this perturbed dynamical system can be defined by the index $\varepsilon$, which is further described as:
	\begin{equation}
		R(f) = -\lambda_{\max}\left(
		\left[\frac{\partial f(0)}{\partial x}\right]^T + \frac{\partial f(0)}{\partial x} \right) >0
	\end{equation}
	On the other hand, the final attractor size caused by $f$ can be quantified by the following indicator:
	\begin{equation}
		I(f) =\frac{L_f}{L_c} \frac{1}{R(f)}
	\end{equation}
\end{remark}

\section{Problem Formulation}
\label{sec:Problem Formulation}
This paper focuses on developing a robust path-following solution for fixed-wing UAV in the presence of disturbances such as wind fields and model simplifications. Assuming that a UAV intends to adhere to a generic, predefined reference path points denoted as
\begin{equation}
	\mathcal{P} = \{(x_{c,i},y_{c,i},z_{c,i})\}_{i=1}^n
\end{equation} 
The UAV's dynamic equations of motion, while maintaining a given ground speed $V_g$, can be organized as\cite{Wang}
\begin{equation}
	\label{equ:kinematic}
	\begin{cases}
		\begin{split}
			\dot{x}_p &= V_g\cos\gamma\cos\chi \\
			\dot{y}_p &= V_g\cos\gamma\sin\chi \\
			\dot{z}_p &= V_g\sin\gamma \\
		\end{split}
	\end{cases}
\end{equation}
where $(x_p, y_p, z_p)$ denotes the position of the UAV in the north-east-up inertial coordinate system, $\gamma$ is the UAV's flight path angle and $\chi$ is the UAV's course angle.
\subsection{Lateral coordinated turn}
The lateral coordinated turn is a sought-after flight condition in unmanned flight by controlling the aircraft's bank angle $\phi$. During a coordinated turn, the aircraft carves the turn rather than skidding laterally. From an analysis perspective, the assumption of a coordinated turn allows us to develop a simplified expression that relates course rate and bank angle. 
\begin{figure}[htbp]
	\centering
	\includegraphics[width=0.85\columnwidth]{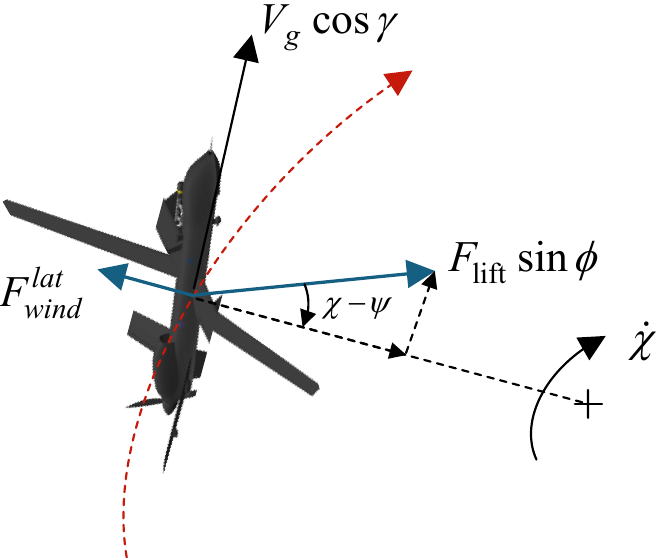}
	\caption{Free-body UAV diagram indicating forces on a UAV in a lateral coordinated turn, where $F_{wind}^{lat}$ is the projection of the wind field force onto the lateral plane.}
	\label{fig:4}
\end{figure}
During a coordinated turn, the bank angle $\phi$ is
set so that there is no net side force acting on the UAV. The centrifugal force acting on the UAV is equal to the summation of horizontal component of the lift force and windy force, as shown in Figure \ref{fig:4}. Summing forces in the horizontal direction gives
\begin{equation}
	\label{eq:lateral dynamic}
	F_{\text{lift}}\sin \phi \cos(\chi - \psi) - F_{wind}^{lat} = m(V_g\cos\gamma)\dot{\chi}
\end{equation}
where $\psi$ is the heading angle, $F_{\text{lift}}$ is the lift force, $F_{wind}^{lat}$ is the windy force projection along the lateral plane. 
\subsection{Longitudinal accelerating climb}
\begin{figure}[htbp]
	\centering
	\includegraphics[width=0.9\columnwidth]{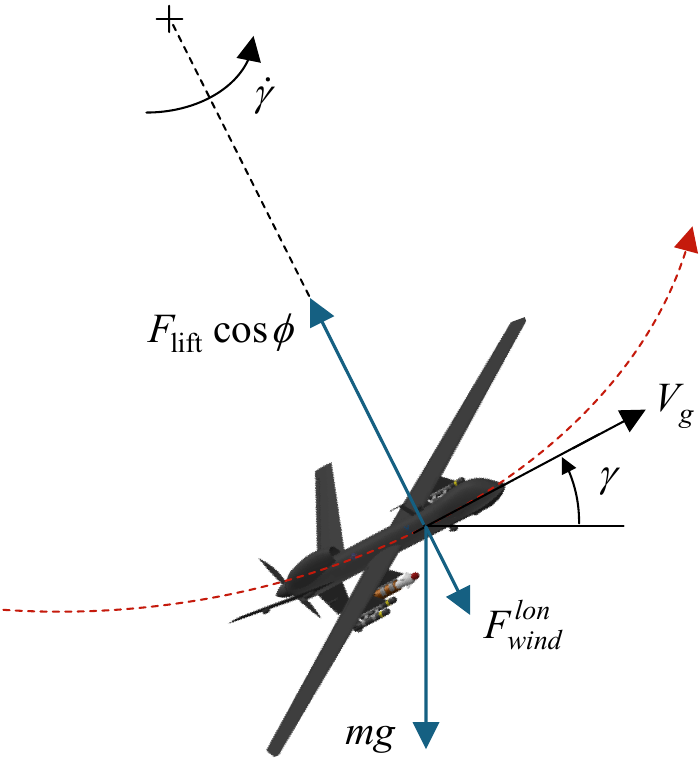}
	\caption{Free-body UAV diagram indicating forces on a UAV in a longitudinal accelerating climb, where $F_{wind}^{lon}$ is the projection of the wind field force onto the longitudinal plane.}
	\label{fig:5}
\end{figure}

To derive the dynamics for the flight-path angle, we will consider a pull-up maneuver in which the aircraft climbs along an arc. Define the two dimensional plane as the plane containing the velocity vector $V_g$ and the vector from the center of mass of the aircraft to the instantaneous center of the circle defined by the pull-up maneuver, as shown in Figure \ref{fig:5}.
Similarly, the vertical component of the lift force has to cancel the summation of the projection of  gravitational force and windy force onto the longitudinal plane to remain at a climb rate, and the
centripetal force due to the pull-up maneuver is $mV_g\dot{\gamma}$. Therefore, summing the forces in the longitudinal plane gives
\begin{equation}
	\label{eq:longitudinal dynamic}
	F_{\text{lift}}\cos \phi - mg\cos\gamma - F_{wind}^{lon} = mV_g\dot{\gamma}
\end{equation}
\subsection{UAV's perturbed dynamic model}
The load factor is defined as the ratio of the lift force acting on an aircraft to its weight:  
\begin{equation}
	n_{\text{lf}} = \frac{F_{\text{lift}}}{mg}
\end{equation}  
From a control engineering perspective, the load factor \(n_{\text{lf}}\) is a critical parameter, as it quantifies the dynamic forces experienced by the aircraft during climbing and turning maneuvers. By treating the load factor as a state variable in control systems, we can enforce command constraints within the aircraft's structural limits, ensuring operational safety and integrity. By reorganizing Eq.(\ref{eq:lateral dynamic}) and Eq.(\ref{eq:longitudinal dynamic}), we derive:
\begin{align}
	\label{eq:UAV's dynamic}
	\begin{cases}
		\dot{\chi} = \frac{g}{V_g}\tan\phi\cos(\chi-\psi) + d_{\chi} \\
		\dot{\gamma} = \frac{g}{V_g}(n_{\text{lf}}\cos\phi - \cos \gamma) + d_{\gamma}
	\end{cases}
\end{align}
where disturbance term $d=(d_{\chi}, d_{\gamma})^T$ can be expressed as
\begin{align}
	\label{eq:UAV's disturbance}
	\begin{cases}
	d_{\chi} = \frac{(mV_g\dot{\gamma} + F_{wind}^{lon})\tan\phi\cos(\chi-\psi)-F_{wind}^{lat}}{mV_g\cos\gamma}\\
	d_{\gamma} = - \frac{F_{wind}^{lon}}{mV_g}
	\end{cases}
\end{align}

\subsection{Flight state constrains}
To enable the UAV to achieve effective rolling and avoid excessive sideslip, the sideslip angle $\beta$ should be ensured to be sufficiently small as
\begin{equation}
	\beta = \psi - \chi \approx 0
\end{equation}
On one hand, as shown in Eq.(\ref{eq:UAV's disturbance}), to avoid excessive nonlinear disturbances, $\phi$ needs to be controlled within an appropriate range:
\begin{equation}
	\phi_{\min} \leq \phi \leq \phi_{\max}
\end{equation}
On the other hand, the lift force is constrained within a specific range, which is manifested as the boundedness of its load factor, i.e.
\begin{equation}
	n_{\text{lf},\min} \leq n_{\text{lf}} \leq n_{\text{lf},\max} 
\end{equation}
Furthermore, the symbols $d_{\chi}$, $d_{\gamma}$ respectively represent the external disturbance imposed on the $\chi$ and $\gamma$-axes, confined within constrained bound $L_d$ as 
\begin{equation}
	\sqrt{d_{\chi}^2 + d_{\gamma}^2} \leq L_d
\end{equation}

\subsection{Look-ahead pursuit guidance law}
This section details the approach to the longitudinal and lateral look-ahead pursuit path following guidance law in the presence of constrained disturbance. 
During the process of the UAV tracking along the current waypoint $(x_c, y_c, z_c)$, the position errors along the $x,y,z$-axis directions are defined as 
\begin{align}
	e_x=x_c-x_p,\quad e_y=y_c-y_p,\quad e_z=z_c-z_p
\end{align}
The longitudinal flight path angle $\gamma_c$ and the lateral flight path angle $\chi_c$ of the Line of Sight (LOS) are defined as
\begin{equation}
	\label{eq:ref_chi_gamma}
	\chi_c= \tan^{-1}\frac{y_c - y_p}{x_c - x_p},
	\gamma_c = \tan^{-1}\frac{z_c - z_p}{\sqrt{(x_c - x_p)^2 + (y_c - y_p)^2}}
\end{equation}

\begin{figure}[hbt!]
	\centering
	\includegraphics[width=\columnwidth]{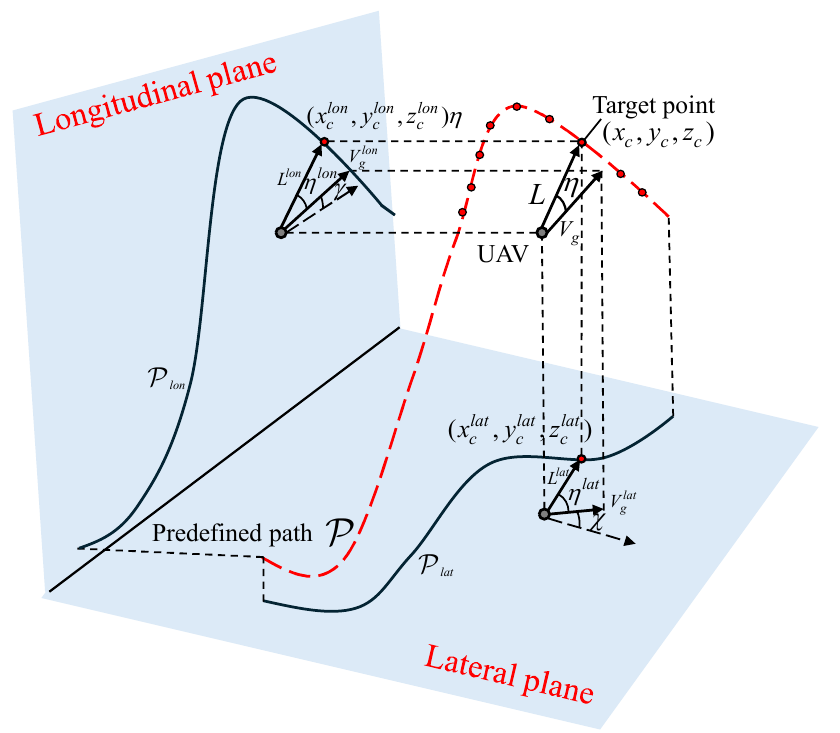}
	\caption{Schematic diagram of look-ahead pursuit guidance law decoupled along the longitudinal and lateral plane projections for a predefined path.}
	\label{fig:1}
\end{figure}

The look-ahead pursuit control of UAV considers the problem of following a predetermined path $\mathcal{P}$ as the tracking problem for moving virtual path points. An imaginary target point $(x_c,y_c,z_c)$ traverses along the reference path $\mathcal{P}$, with LOS vector extending from the UAV to this point, defined as the look-ahead vector $L$ comprised of the lateral component $L^{lat}$ and the longitudinal component $L^{lon}$, as shown in Figure \ref{fig:1},
and the virtual target point $P_c = (x_c,y_c,z_c)$ is defined as the nearest point located ahead along the predefined reference path in the direction of velocity as\cite{Wang}
\begin{equation}
	\label{equ:target points}
	P_c = \underset{(x,y,z) \in \mathcal{P}}{\text{argmin}}| q_LV_g - \sqrt{(x_p-x)^2 + (y_p-y)^2 +(z_p-z)^2} |
\end{equation}
where $q_L$ represents the coefficient that adaptively adjusts the look-ahead length in relation to the UAV's ground velocity.

During the process of tracking a virtual target point, its first-order derivative $\dot{x}_c,\dot{y}_c,\dot{z}_c$ can be considered as zeros under the condition of not considering abrupt target switching, and its lateral look-ahead angle $\eta^{lat}$ and longitudinal look-ahead angle $\eta^{lon}$ can be defined as
\begin{equation}
	\label{eq:eta}
	\eta^{lat} = \chi_c - \chi,\eta^{lon} = \gamma_c - \gamma
\end{equation}
and $\eta^{lat}$ is limited to $\eta^{lat}_{\min} \leq \eta^{lat} \leq \eta^{lat}_{\max}$, $\eta^{lon}$ is limited to $\eta^{lon}_{\min} \leq \eta^{lon} \leq \eta^{lon}_{\max}$. 

Specifically, it is precisely because the velocity direction of the UAV can be ensured to be coincide with the LOS, stablizing the positional errors to zeros in a finite time. This is the foundational principle of look-ahead pursuit control, which will be further elucidated by the following theorem.
\newtheorem{thm}[theorem]{Theorem}
\begin{thm}
	\label{thm:theorem_1}
	(The fast finite-time stability of look-ahead pursuit guidance law) The $e_x$, $e_y$, and $e_z$ will be stablized to 0 within a finite time 
	\begin{equation}
		T(e_0)\leq \frac{\sqrt{e^2_x(0)+e^2_y(0)+e^2_z(0)}}{V_g \cos \delta^{lon}\cos \delta^{lat}}
	\end{equation}
	under
	\begin{align}
		\label{eq:gamma_condition}
		\gamma (z_p - z_c) \leq 0
		\iff
		\begin{cases}
			\eta^{lon} \leq \gamma_c,\ z_p \leq z_c\\
			\eta^{lon} \geq \gamma_c,\ z_p > z_c
		\end{cases}
	\end{align}
	and there exists $\delta^{lon},\delta^{lat}<\frac{\pi}{2}$ such that
	\begin{equation}
		|\eta^{lon}|\leq \delta^{lon}$, $|\eta^{lat}|\leq \delta^{lat}
	\end{equation}
	where initial position error is $e_0 = (e_x(0),e_y(0),e_z(0))^T$.   
\end{thm}
\begin{proof}
	There exists
	\begin{align}
			&\sin \chi = \sin(\chi_c - \eta^{lat})=\frac{e_y\cos \eta^{lat} - e_x \sin \eta^{lat}}{\sqrt{e_x^2 + e_y^2}}\\
			&\cos \chi = \cos(\chi_c - \eta^{lat}) = \frac{e_y\sin\eta^{lat}+e_x\cos\eta^{lat}}{\sqrt{e_x^2 + e_y^2}}\\
			&\sin \gamma = \sin(\gamma_c - \eta^{lon})= \frac{e_z\cos \eta^{lon} - \sqrt{e_x^2+e_y^2} \sin \eta^{lon}}{\sqrt{e_x^2 + e_y^2 + e_z^2}}\\
			& \cos \gamma = \cos(\gamma_c - \eta^{lon}) = \frac{e_z\sin \eta^{lon} + \sqrt{e_x^2+e_y^2}\cos \eta^{lon}}{\sqrt{e_x^2 + e_y^2 + e_z^2}}
	\end{align}
	and $\dot{x}_p=-\dot{e}_x$, $\dot{y}_p=-\dot{e}_y$, $\dot{z}_p=-\dot{e}_z$. Substitute it into Eq.(\ref{equ:kinematic}) to obtain the dynamic of $e_x,e_y,e_z$:
	\begin{equation*}
		\footnotesize
		\begin{cases}
			\begin{split}
				&\dot{e}_x=-V_g \frac{e_z\sin \eta^{lon} + \sqrt{e_x^2+e_y^2}\cos \eta^{lon}}{\sqrt{e_x^2 + e_y^2 + e_z^2}}\frac{e_y\sin\eta^{lat}+e_x\cos\eta^{lat}}{\sqrt{e_x^2 + e_y^2}}\\
				&\dot{e}_y=-V_g\frac{e_z\sin \eta^{lon} + \sqrt{e_x^2+e_y^2}\cos \eta^{lon}}{\sqrt{e_x^2 + e_y^2 + e_z^2}}\frac{e_y\cos \eta^{lat} - e_x \sin \eta^{lat}}{\sqrt{e_x^2 + e_y^2}}\\
				&\dot{e}_z=-V_g\frac{e_z\cos \eta^{lon} - \sqrt{e_x^2+e_y^2} \sin \eta^{lon}}{\sqrt{e_x^2 + e_y^2 + e_z^2}}\\
			\end{split}
		\end{cases}
	\end{equation*}
	we can construct Lyapunov function
	\begin{align}
		V(e)=\frac{1}{2}(e_x^2 + e_y^2 +e_z^2)
	\end{align}
	 and there exists
	 \begin{align}
	 	\label{eq:proof}
	 	\footnotesize
	 	\begin{split}
	 	\dot{V}(e) =& \frac{-V_g}{\sqrt{e_x^2 + e_y^2 + e_z^2}}\times 
	 	[
	 		(e_x^2 + e_y^2+e_z^2)\cos \eta^{lon}\cos \eta^{lat}\\
	 		&+e_z(1-\cos\eta^{lat})\left(
	 		e_z\cos\eta^{lon} -\sqrt{e_x^2+e_y^2}\sin\eta^{lon}
	 		\right)
	 	] \\
	 	=&-V_g[
	 	\sqrt{e_x^2 + e_y^2 + e_z^2}\cos \eta^{lon}\cos \eta^{lat} \\
	 	&+
	 	e_z(1-\cos\eta^{lat})\sin(\gamma_c - \eta^{lon}) ]\\
	 	\leq & -\sqrt{2}V_gV(e)^{\frac{1}{2}} \cos \delta^{lon}\cos \delta^{lat}\\
	 	&-V_g e_z(1-\cos\eta^{lat})\sin(\gamma_c - \eta^{lon})
	 	\end{split}
	 \end{align}
	 Due to the Eq.(\ref{eq:gamma_condition}), here $e_z\gamma \geq 0$, i.e., 
	 \begin{align}
	 	e_z\sin(\gamma_c - \eta^{lon}) = e_z\sin \gamma \geq 0
	 \end{align}
	Substitute above equation into Eq.(\ref{eq:proof}) that
	\begin{align}
		\begin{split}
			\dot{V}(e) \leq& -\sqrt{2}V_gV(e)^{\frac{1}{2}} \cos \delta^{lon}\cos \delta^{lat} \\
			&-V_g e_z(1-\cos\eta^{lat})\sin(\gamma_c - \eta^{lon}) \\
			 \leq& -\sqrt{2}V_gV(e)^{\frac{1}{2}} \cos \delta^{lon}\cos \delta^{lat}
		\end{split}
	\end{align}
	Hence, according to the Lemma \ref{lem:lemma2}, the $e_x$, $e_y$, and $e_z$ will be stablized to 0 within a finite time, and the settling time is:
	\begin{equation*}
		T(e_0)\leq \frac{\sqrt{2}V(e_0)^{\frac{1}{2}}}{V_g\cos \delta^{lon}\cos\delta^{lat}}=\frac{\sqrt{e^2_x(0)+e^2_y(0)+e^2_z(0)}}{V_g\cos \delta^{lon}\cos\delta^{lat}}
	\end{equation*}
	The proof is completed.
\end{proof}
The aforementioned conclusion indicates that look-ahead pursuit guidance guidance law designed for the path angle can guarantee finite-time stabilization of the look-ahead angle errors, thereby ensuring stabilization of the positional errors.

The objective of this paper is to design the desired control commands $\phi$, $n_{\text{lf}}$ to ensure that the path following errors can be robust stablized within a minmum input energy even when subject to bounded path angles disturbances, which is accomplished through the implementation of improved adaptive longitudinal and lateral look-ahead pursuit control strategy\cite{Wang}.

\section{Path Following Controller with Optimal Robustness Indictor}
\label{sec:Path Following Controller with optimal robustness indictor}
\begin{figure*}[hbt!]
	\centering
	\includegraphics[width=0.9\textwidth]{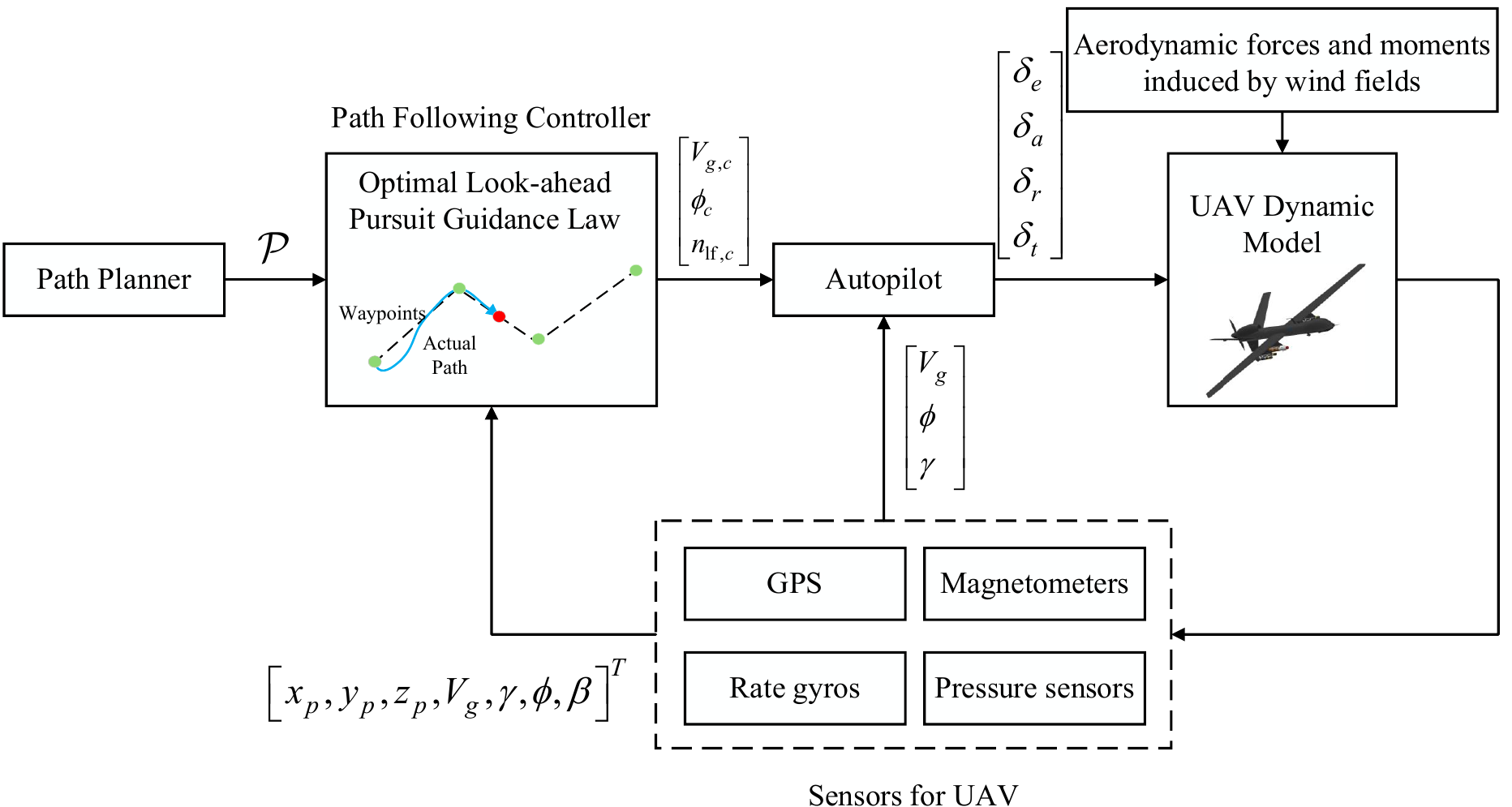}
	\caption{Schematic diagram of path following control via optimal look-ahead pursuit guidance law.}
	\label{fig:3}
\end{figure*}

As the core of this study, an optimal path-following controller is designed in this section, with its role in the UAV path-following process illustrated in Figure \ref{fig:3}. First, the sufficient conditions for ensuring the robust stability of $\eta^{\text{lon}}$ and $\eta^{\text{lat}}$ are derived. Based on these conditions, a guidance law design method is proposed to minimize the UAV's acceleration energy while guaranteeing the predefined robustness indicator. This method can be formulated as a classic optimization model and efficiently solved within a short time.
\subsection{Robustness indictor}
We now formally present the specific formulation of the RLLP guidance law. Define the following reference input commands for the autopilot module.
\begin{align}
	\label{eq:guidance law1}
	\begin{cases}
		\phi_c = \sin^{-1}(a_{yc}/g) \in [\phi_{\min},\phi_{\max}]\\
		n_{\text{lf},c} = a_{zc}/g\cos \phi_c \in [n_{\text{lf},\min},n_{\text{lf},\max}]
	\end{cases}
\end{align}
where the acceleration 
\begin{equation}
	\label{eq:acceleration}
	a=(a_{yc},a_{zc})^T
\end{equation}
is expressed as
\begin{align}
	\label{eq:guidance law2}
	\begin{cases}
		a_{yc} = - V_gf_{\chi}(\eta^{lat},\eta^{lon}) \cos\phi/\cos \beta \in [-g,g] \\
		a_{zc} = -V_g f_{\gamma}(\eta^{lat},\eta^{lon}) + g\cos\gamma\\
	\end{cases}
\end{align}
Here, \( f_{\chi} \) and \( f_{\gamma} \) determine the specific formulation of the guidance law, with their values directly governing the robustness level of the RLLP guidance law. Building upon the preliminaries, as one of the core theorem of this paper, we characterize the relationships between the functions \( f_{\chi} \), \( f_{\gamma} \) and the ultimate convergence set of \( \eta(t) \), and the convergence rate as follows:  
\newtheorem{theorem3}[theorem]{Theorem}
\begin{theorem3}
	\label{thm:lemma_A3}
	(Robust stability conditions for $\eta^{lon}$ and $\eta^{lat}$) For the UAV's dynamic model Eq.(\ref{eq:UAV's dynamic}) using the guidance law Eq.(\ref{eq:guidance law1}) and Eq.(\ref{eq:guidance law2}), let 
	\begin{align}
	f(\eta) = \begin{pmatrix}
		f_{\chi}(\eta^{lat},\eta^{lon}) \\ 
		f_{\gamma}(\eta^{lat},\eta^{lon})
	\end{pmatrix},
	\quad 
	\eta = \begin{pmatrix}
		\eta^{lat} \\ \eta^{lon}
	\end{pmatrix}
	\end{align}
	where $f(0)=(0,0)^T$, the perturbation term is
	\begin{align}
		d=(d_{\chi},d_{\gamma})^T, \quad ||d||_2\leq L_d
	\end{align}
    If $f$ is $L_c$-co-Lipschitz on the neighbor of origin, 
    \begin{align}
    	\left\|\frac{\partial f(0)}{\partial \eta}\right\|_{2}\leq L_f
    \end{align}
    and
    \begin{align}
    	R(f) = -\lambda_{\max}\left(
    	\left[\frac{\partial f(0)}{\partial \eta}\right]^T + \frac{\partial f(0)}{\partial \eta} \right) > 0 
    \end{align}
	Then the ultimate trajectory of the state \(\eta(t)\), for any initial value $\eta(0)$, \(\eta(t)\) will exponentially converge to the following global random attractor $S(0)$ at least the rate of $R(f)$ as $t\rightarrow \infty$:
	\begin{align}
		\label{eq:lemma_A3_condition}
		\eta(t)\rightarrow S(0) = \left\{
		\eta: ||\eta||_2\leq 2L_d I(f)
		\right\}
	\end{align}
	where
	\begin{align}
		I(f) = \frac{L_f}{L_c} \frac{1}{ R(f)}
	\end{align}
\end{theorem3}
\begin{proof}
	By substituting Eq.(\ref{eq:guidance law1}) and Eq.(\ref{eq:guidance law2}) into Eq.(\ref{eq:UAV's dynamic}), the following result can be derived:
	\begin{equation}
		\begin{cases}
			\dot{\chi} = -f_{\chi}(\eta^{lat},\eta^{lon}) + d_{\chi} \\
			\dot{\gamma} = -f_{\gamma}(\eta^{lat},\eta^{lon}) + d_{\gamma}
		\end{cases}
	\end{equation}
	When tracking waypoints within each flight segment, since the every waypoint being tracked is fixed, i.e., $\dot{\gamma}_c = \dot{\chi}_c = 0$, which is equivalent to $\dot{\chi} = -\dot{\eta}_{lat}$, $\dot{\gamma} = -\dot{\eta}_{lon}$. Substituting it into the above equation yields: 
	\begin{equation}
		\begin{cases}
			\dot{\eta}^{lat} = f_{\chi}(\eta^{lat},\eta^{lon}) - d_{\chi} \\
			\dot{\eta}^{lon} = f_{\gamma}(\eta^{lat},\eta^{lon}) - d_{\gamma}
		\end{cases}
	\end{equation}
	By Theorem \ref{thm:lemma_A2} and Remark \ref{rem:rem_lemma_A3}, the conclusion follows, hence the proof is completed.
\end{proof}
\begin{remark}
	\label{rem:remark4}
	If there exists the robust stability constraint for the predefined lowerbound value $R^*>0$ that $	R\left(f \right) \geq R^*$, at this point, the exponential convergence rate is at least $R^*$, and the $\eta(t) = (\eta^{lat}(t),\eta^{lon}(t))^T$ of the final trajectory is converged to the interior of the following region:
	\begin{equation}
		\sqrt{\eta^{lat}(t)^2+\eta^{lon}(t)^2} \leq \frac{2L_dL_f}{R^* L_c},\ t\rightarrow \infty
	\end{equation}
	here $f$ is $L_c$-co-Lipschitz over
	\begin{equation}
		X = \left \{(\eta^{lat},\eta^{lon}): |\eta^{lat}|<\frac{\pi}{2}, |\eta^{lon}|<\frac{\pi}{2} \right \}
	\end{equation}
\end{remark}
We will introduce several classic forms of $f(\eta)$, and based on these, analyze their $R(f)$ and $I(f)$ as tabulated in Table \ref{tab:hyperparameters}.
\begin{table*}
	\begin{center}
		\caption{The $L_f$, $L_c$, $R(f)$, $I(f)$ for some classic types of $f$.}
		\label{tab:hyperparameters}
		\footnotesize
		\begin{tabular}{c|c|c|c|c|c|c}
			\hline
			Type & $f_{\chi}(\eta^{lat},\eta^{lon})$ & $f_{\gamma}(\eta^{lat},\eta^{lon})$ & $L_f$ & $L_c$ & $R(f)$ & $I(f)$\\
			\hline
			$x$ & $-k_{\chi}\eta^{lat},k_{\chi}>0$ & $-k_{\gamma}\eta^{lon},k_{\gamma}>0$ & $\max\{k_{\chi},k_{\gamma}\}$ & 
			$\min\{k_{\chi},k_{\gamma}\}$
			&  $2\min\{k_{\chi},k_{\gamma}\}$ & $
		 \max\{k_{\chi},k_{\gamma}\}/2\min\{k_{\chi},k_{\gamma}\}^2 $\\
			$\tan(x)$ & $-k_{\chi}\tan\eta^{lat},k_{\chi}>0$ & $-k_{\gamma}\tan\eta^{lon},k_{\gamma}>0$ & $\max\{k_{\chi},k_{\gamma}\}$ & 
			$\min\{k_{\chi},k_{\gamma}\}$
			&  $2\min\{k_{\chi},k_{\gamma}\}$ & $ \max\{k_{\chi},k_{\gamma}\}/2\min\{k_{\chi},k_{\gamma}\}^2 $\\
			$\exp(x)$ & $1 - e^{k_{\chi} \eta^{lat}},k_{\chi}>0$ & $1 - e^{k_{\gamma} \eta^{lon}},k_{\gamma}>0$ & $\max\{k_{\chi},k_{\gamma}\}$ & 
			$\min\{k_{\chi}e^{-\frac{\pi}{2}k_{\chi}}, k_{\gamma}e^{-\frac{\pi}{2}k_{\gamma}} \}$
			&  $2\min\{k_{\chi},k_{\gamma}\}$ & $ \max\{k_{\chi},k_{\gamma}\}/2\min\{k_{\chi},k_{\gamma}\}L_c$\\
			$\sin(x)$ & $-k_{\chi}\sin \eta^{lat},k_{\chi}>0$ & $-k_{\gamma}\sin \eta^{lon},k_{\gamma}>0$ &
			$\max\{k_{\chi},k_{\gamma}\}$ &
			$2\min\{k_{\chi},k_{\gamma}\}/\pi$&
			$2\min\{k_{\chi},k_{\gamma}\}$ &
			$\pi\max\{k_{\chi},k_{\gamma}\}/4\min\{k_{\chi},k_{\gamma}\}^2$\\
			linear & $k_{\chi}^{lat} \eta^{lat} + k_{\chi}^{lon}\eta^{lon}$ & 
			$k_{\gamma}^{lat} \eta^{lat} + k_{\gamma}^{lon}\eta^{lon}$ & 
			$||K||_2$ & $||K^{-1}||_2^{-1}$ & $-\lambda_{\max} (K^T + K)>0$ & $-||K||_2||K^{-1}||_2^{-1}/\lambda_{\max} (K^T + K)$\\
			\hline
		\end{tabular}
	\end{center}
\end{table*}

\subsection{The optimization model for $I(f)$}
According to Theorem \ref{thm:theorem_1}, we can minimize the finite convergence time \( T(e_0) \) by reducing the absolute upper bounds \( \delta^{lon} \) and \( \delta^{lat} \) of \( \eta^{lon} \) and \( \eta^{lat} \) as much as possible, which can be achieved by minimizing the attractor size \( I(f) \). In this section, an optimal $f = (f_{\chi}, f_{\gamma})^T$ is selected to minimize the attractor size $I(f)$ under constrained acceleration limits and exponential convergence rate lowerbound, expressed as
\begin{align}
	\label{eq:model_before}
	\begin{split}
		&\min_{f} I(f) \\
		& s.t. 
		\begin{cases}
			R(f) \geq R^* \\
			a_{\min} \leq a\leq a_{\max} 
		\end{cases}
	\end{split}
\end{align}
The optimization model is formulated to minimize the attractor size subject to two constraints: acceleration magnitude limits and a minimum exponential convergence rate to the attractor (predefined as $R^*>0$).

To simplify the presentation, we use a linear model to characterize the form of $f(\eta)$, i.e.,
\begin{align}
	\label{eq:f_eta}
	f(\eta) = K\eta = Fk
\end{align}
where
\begin{align}
	& K = \begin{pmatrix}
		k_{\chi}^{lat} & k_{\chi}^{lon} \\
		k_{\gamma}^{lat} & k_{\gamma}^{lon} \\
	\end{pmatrix},
	\eta = \begin{pmatrix}
		\eta^{lat} \\ \eta^{lon} 
	\end{pmatrix} \\
	& F = \begin{pmatrix}
		\eta^{lat} & 0 & \eta^{lon} & 0\\
		0& \eta^{lat} & 0 & \eta^{lon}
	\end{pmatrix} \\
	& k = \text{vec}(K) = \left( k_{\chi}^{lat},
	k_{\gamma}^{lat},
	k_{\chi}^{lon},
	k_{\gamma}^{lon}\right)^T
\end{align}
The acceleration $a$ can be described as:
\begin{align}
	a = b + \Lambda f(\eta) = b + \Lambda Fk
\end{align}
where
\begin{align}
	b = \begin{pmatrix}
		0 \\ g\cos \gamma
	\end{pmatrix},
	\Lambda = \begin{pmatrix}
		-V_g\frac{\cos\phi}{\cos \beta} & \\
		& - V_g
	\end{pmatrix}
\end{align}
Furthermore, we can estimate the $L_f$ of $f(\eta)$ at the origin by its Jacobian:
\begin{align}
	L_f = \left\|\frac{\partial f(0)}{\partial \eta}\right\|_{2} = \left\|K\right\|_{2}
\end{align}
Under the invertibility of $K$, the co-Lipschitz coefficient $L_c$ of $f(\eta)$ can be estimated by
\begin{align}
	L_c = \frac{1}{||K^{-1}||_2}
\end{align}
which is derived from the following process, $\forall \eta_1,\eta_2$,
\begin{align}
	\begin{split}
		 \left\| \eta_1 - \eta_2  \right\|_{2} &= ||K^{-1}(K \eta_1 - K \eta_2 ) ||_2 \\
		&\leq ||K^{-1}||_2 \left\| K \eta_1 - K \eta_2  \right\|_{2}
	\end{split} 
\end{align}
i.e.,
\begin{align}
	||f(\eta_1) - f(\eta_2)|| \geq \frac{\left\| \eta_1 - \eta_2  \right\|_{2}}{||K^{-1}||_2} 
\end{align}
Here, the exponential convergence rate $R(f)=R(k)$ can be characterized by:
\begin{align}
	\begin{split}
			R(k) &= -\lambda_{\max}(K^T+K)  \\
			     &= -\lambda_{\max} 
				\begin{pmatrix}
						2k_{\chi}^{lat} & k_{\chi}^{lon} + k_{\gamma}^{lat} \\
						k_{\chi}^{lon} + k_{\gamma}^{lat} & 2k_{\gamma}^{lon}
					\end{pmatrix}\\
				 &= -\left( k_{\chi}^{lat} + k_{\gamma}^{lon} + 
				 \sqrt{(k_{\chi}^{lat} - k_{\gamma}^{lon})^2 + (k_{\chi}^{lon} + k_{\gamma}^{lat})^2} \right)
		\end{split} 
\end{align}

To summarize, the final optimization model is transformed into a nonlinear model for parameter $k$-optimization as follows
\begin{align}
	\label{eq:model_after}
	\begin{split}
		    &\min_{k} I(k) =- \frac{\|K\|_{2}\|K^{-1}\|_2}{\lambda_{\max}(K^T+K)} \\
		s.t.&
		\begin{cases}
			R(k) \geq R^* > 0
			\\
			a_{\min} - b  \leq \Lambda Fk \leq a_{\max} - b
		\end{cases}
	\end{split}
\end{align}
The pseudocode of the path following controller with optimal robustness indictor algorithm is as Alg.\ref{alg:optimal_pathfolllowing}.
\begin{algorithm}[htbp]
	\caption{The look-ahead pursuit guidance law with optimal robustness indictor}
	\label{alg:optimal_pathfolllowing}
	\begin{algorithmic}[1]
		\REQUIRE The reference waypoints set $\mathcal{P}$, current UAV's state measured by $(x_p,y_p,z_p,V_g,\gamma,\phi,\beta)$, predefined robustness indictor $R^*$ 
		\ENSURE The optimal guidance law $\phi_c^*$, $n_{\text{lf},c}^*$
		\STATE Calculate the virtual target point $P_c$ by Eq.(\ref{equ:target points}).
		\STATE Calculate the reference angles $\chi_c$, $\gamma_c$ by Eq.(\ref{eq:ref_chi_gamma}).
		\STATE Calculate the look-ahead angles $\eta^{lon}$, $\eta^{lat}$ by Eq.(\ref{eq:eta}).
		\IF{the current waypoint \( P_c \) tracked by the UAV undergoes a change}
		\STATE Solve the optimization model Eq.(\ref{eq:model_after}) to obtain the optimal coefficient \( k^* \). 
		\ELSE
		\STATE Keep \( k^* \) unchanged.
		\ENDIF
		\STATE Calculate optimal acceleration \( a^* \) from \( k^* \) by Eq.(\ref{eq:guidance law2}) and Eq.(\ref{eq:f_eta}).
		\STATE Obtain the $\phi_c^*$, $n_{\text{lf},c}^*$ by Eq.(\ref{eq:guidance law1}).
	\end{algorithmic}
\end{algorithm}
As the core of the algorithm, lines 4-8, the optimization model is resolved upon each switch of reference waypoints. Namely, a fixed optimal coefficient \( k^* \) is employed to derive the optimal guidance law for path following during tracking of each waypoint, whereas re-optimization is necessitated to solve for the optimal \( k^* \) during waypoint transitions. 

\section{The Autopilot}
\label{sec:The Autopilot}
In aerospace applications, autopilot serves as the primary guidance systems for UAVs. Specifically, the autopilot maintains full flight envelope control across all operational phases, including takeoff, waypoint navigation, and landing. The primary function of the autopilot is to accurately track the reference commands $\phi_c$, $n_{\text{lf},c}$ and $V_{g,c}$ calculated by the path following controller. In this section, suitable controller designs for roll angle, load factor, and groundspeed will be presented, which are based on Proportional-Integral (PI) control algorithms tailored to the available sensors and computational resources.

\subsection{Roll angle controller}
For the roll angle controller, the feedback is generated by computing the error between the commanded roll angle $\phi_c$ and the actual roll angle $\phi$ measured by rate gyros. A PI controller is used to calculate the aileron deflection command $\delta_a$ as
\begin{equation}
	\delta_a = k_p^{\phi}(\phi_c - \phi) + k_i^{\phi}\int (\phi_c - \phi) dt - k_d^{\phi} p
\end{equation}
where $k_p^{\phi}$ and $k_i^{\phi}$ are the corresponding PI controller coefficients, $p$ is the roll angle rate measured by rate gyros, and $k_d^{\phi}$ is its corresponding coefficient. The purpose of adding the compensation term for $p$ is to ensure the low-inertia requirement when tracking $\phi_c$. The integral anti-windup mechanism prevents integral term saturation due to prolonged error accumulation, while amplitude limiting is applied to $\delta_a$ to ensure it stays within the actuator's operational range of $[-\pi/4,\pi/4]$.
\begin{figure}[htbp]
	\centering
	\includegraphics[width=\columnwidth]{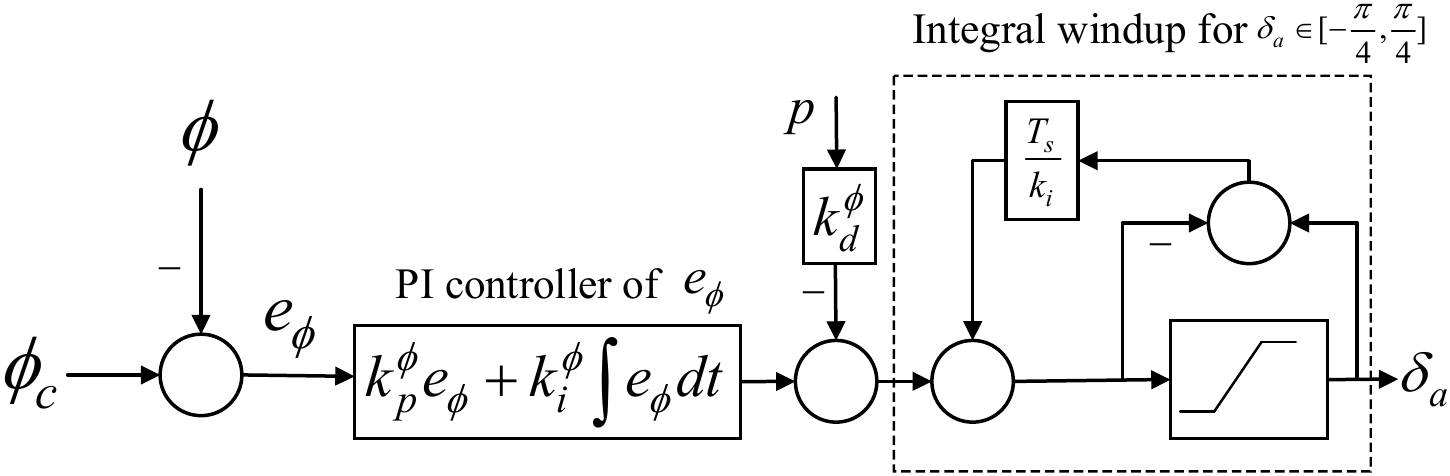}
	\caption{Roll angle control is achieved using a PI controller with integral windup limitation.}
	\label{fig:6}
\end{figure}
The overall control system block diagram is shown in Figure.\ref{fig:6}.
\subsection{Load factor controller}
\begin{figure}[htbp]
	\centering
	\includegraphics[width=\columnwidth]{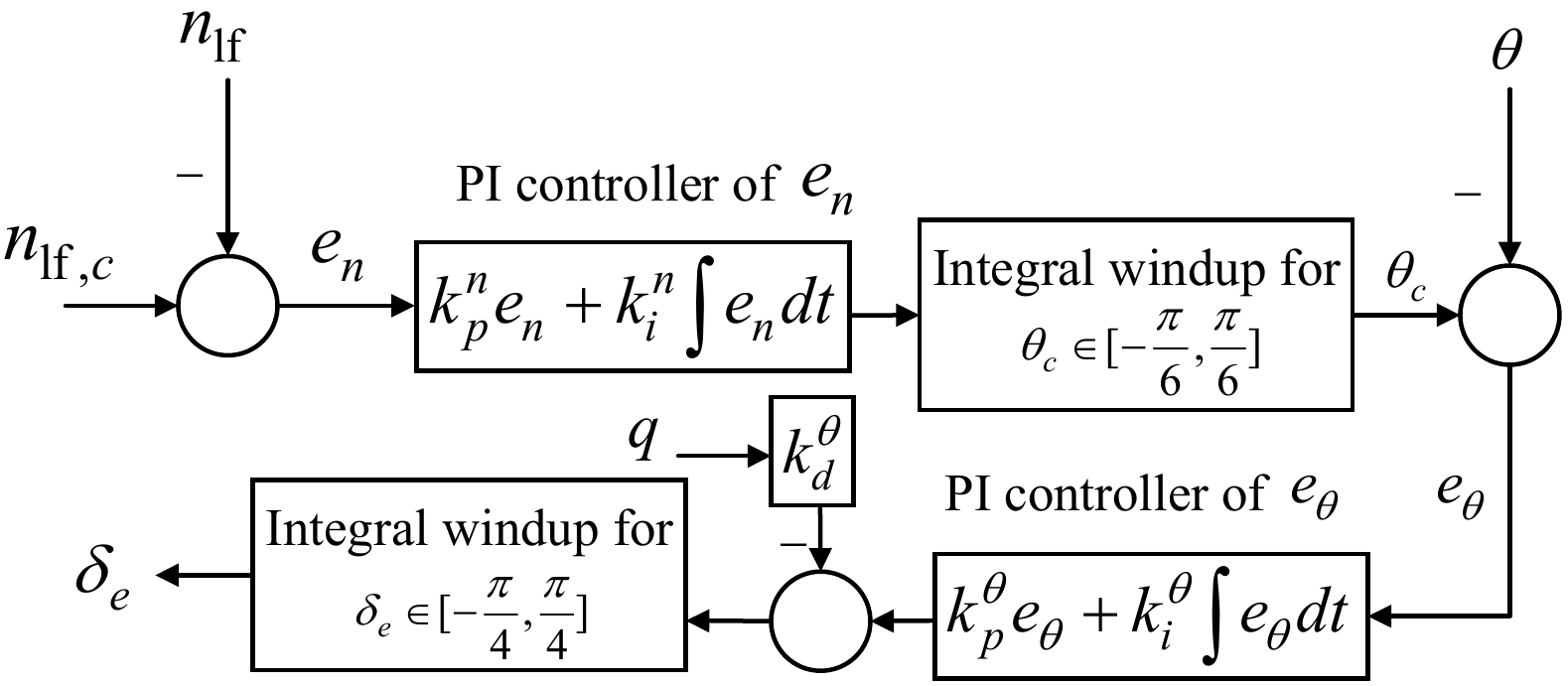}
	\caption{Load factor control is achieved using successive loop closure(SLC) PI controllers with integral windup limitation.}
	\label{fig:7}
\end{figure}

Similarly, the load factor controller also utilizes the feedback error between the reference load factor $n_{\text{lf},c}$ and the measured load factor $n_{\text{lf}}$, where $n_{\text{lf}}$ can be estimated by the following equation:
\begin{equation}
	n_{\text{lf}} = \frac{V_{g}\dot{\gamma}}{g\cos \phi} + \frac{\cos \gamma}{\cos \phi}
\end{equation}
where the flight-path angle $\gamma$ of the UAV and its derivative $\dot{\gamma}$ are measured by GPS and magnetometers, while the roll angle $\phi$ is obtained by integrating the value measured by rate gyros. 
For the tracking control of $n_{\text{lf},c}$, successive loop closure (SLC) PI controllers are implemented. Specifically, the outer loop employs a PI controller with anti-windup to regulate the reference pitch angle $\theta_c$ as follows
\begin{align}
	\hat{\theta}_c &= k_p^n(n_{\text{lf},c} - n_{\text{lf}}) + k_i^n \int (n_{\text{lf},c} - n_{\text{lf}}) dt \\
	\theta_c &= \text{Integral-windup}(\hat{\theta}_c), \theta_c \in [-\frac{\pi}{6},\frac{\pi}{6}]
\end{align}
This design leverages the monotonically increasing relationship between the pitch angle and UAV lift force within a defined operational range, enabling precise tracking of $n_{\text{lf},c}$ via $\theta_c$ control. Concurrently, mirroring the roll angle control architecture, the inner-loop pitch angle controller adopts an identical anti-windup PI control scheme for the elevator $\delta_e$ as follows
\begin{align}
	\hat{\delta}_e &= k_p^{\theta}(\theta_c - \theta) + k_i^{\theta} \int (\theta_c - \theta)dt - k_d^{\theta}q  \\
	\delta_e &= \text{Integral-windup}(\hat{\delta}_e), \delta_e \in [-\frac{\pi}{4},\frac{\pi}{4}]
\end{align}
where $k_p^{\theta}$ and $k_i^{\theta}$ are the corresponding PI controller coefficients, $q$ is the pitch angle rate measured by rate gyros, and $k_d^{\theta}$ is its corresponding coefficient.
The SLC control system configuration is depicted in Figure \ref{fig:7}.
\subsection{Groundspeed controller}
\begin{figure}[htbp]
	\centering
	\includegraphics[width=\columnwidth]{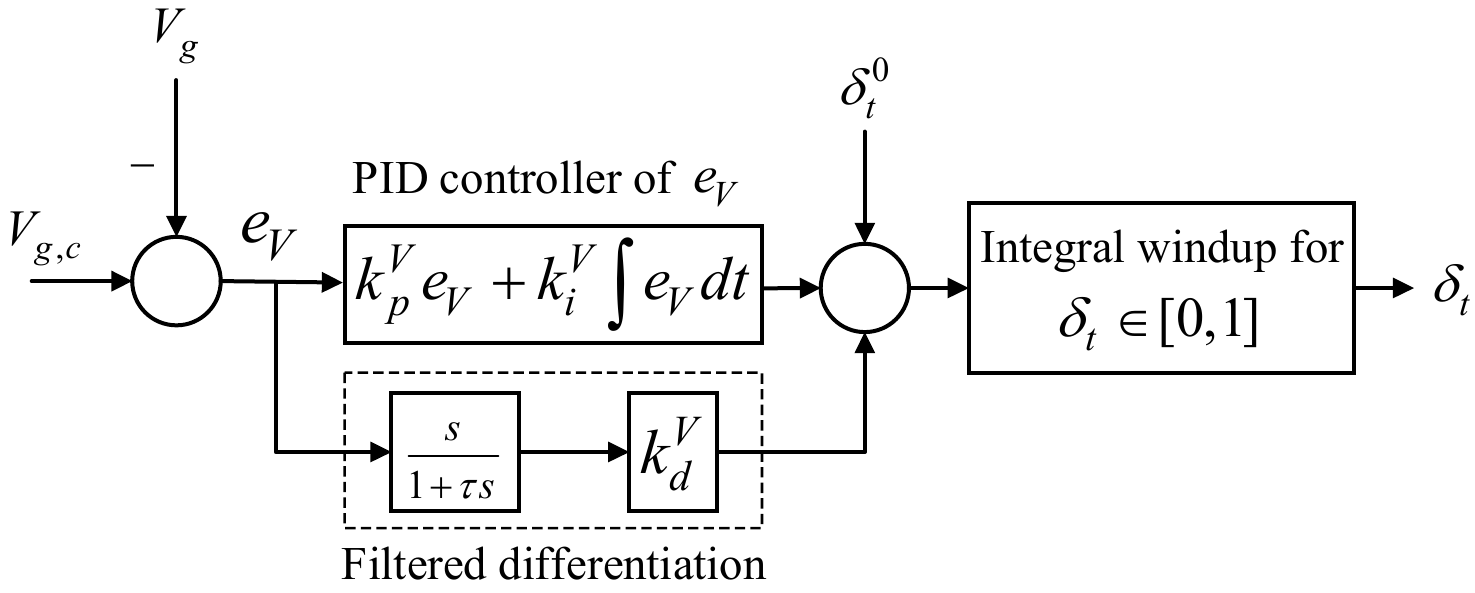}
	\caption{Groundspeed control is achieved using a PI controller with filtered differentiation and integral windup limitation.}
	\label{fig:8}
\end{figure}

For the groundspeed controller, under the constraint of small pitch angles, constant groundspeed control is realized by directly regulating the throttle opening $\delta_t$. As shown in Figure \ref{fig:8}, the control scheme integrates a PI controller with anti-windup, where the velocity error $e_V = V_{g,c} - V_g$ is processed through low-pass filtered differentiation. 
This filtering mechanism serves dual objectives: (1) suppressing high-frequency noise to eliminate controller oscillations triggered by measurement noise, and (2) alleviating abrupt derivative-term surges caused by setpoint transients. The control algorithm is described as follows:
\begin{align}
	\begin{split}
		\hat{\delta}_t &= \delta_t^0 + k_p^V(V_{g,c}-V_g) + k_i^V\int (V_{g,c}-V_g) dt\\
		&+ k_d^V\frac{\dot{V}_{g,c}-\dot{V}_g}{1+\tau s}\\
	\end{split}\\
	\delta_t &= \text{Integral-windup}(\hat{\delta}_t), \delta_t \in [0,1]
\end{align}
where $k_p^V$, $k_i^V$, $k_d^V$ are the corresponding controller coefficients, and $\tau$ is its time coefficient, $\delta_t^0$ is the initial throttle openin.

\section{Simulation}
\label{sec:Simulation}
\begin{figure*}[htbp]
	\centering
	\begin{tabular}{cc}
		\includegraphics[width=0.48\textwidth]{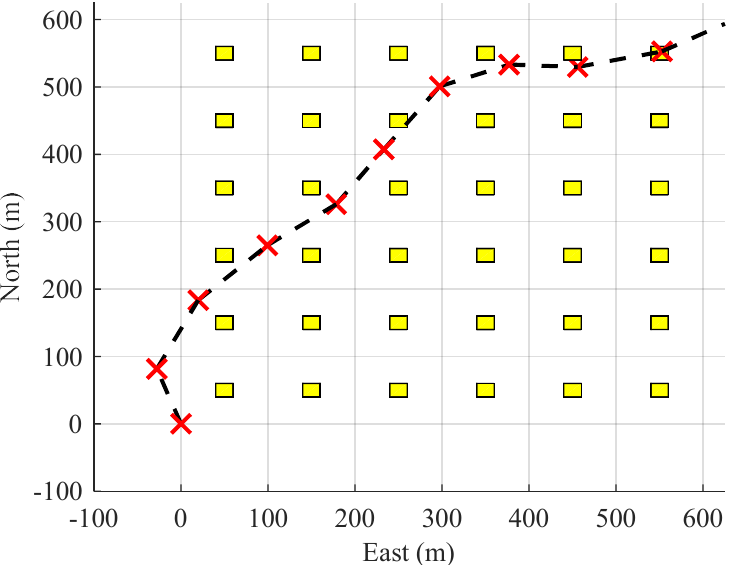} & 
		\includegraphics[width=0.48\textwidth]{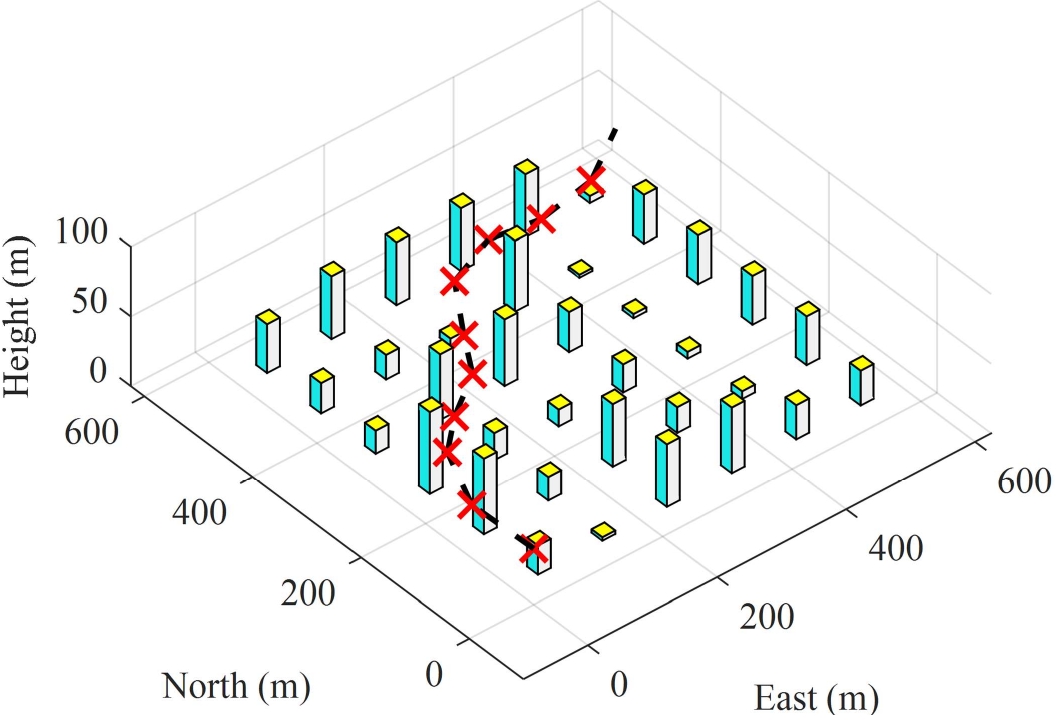} \\
		(a) & (b) \\ 
	\end{tabular}
	\caption{Top view and side view of a sophisticated non-smooth low-altitude urban path used for path-following experiments, where the red crosses symbolize waypoints. (a): Top view; (b): Side view.}
	\label{fig:9}
\end{figure*}

In this section, we first validate the proposed robustness indicator theory in a complex urban terrain environment. Second, a comprehensive experimental validation is performed on the path-following performance of the Robust Longitudinal and Lateral Look-ahead Pursuit (RLLP) guidance law. Additionally, by comparing RLLP with its optimal robustness indicator counterpart (Optimal-RLLP), we demonstrate significant enhancements in multiple critical performance metrics, thus verifying the rationality and superiority of Optimal-RLLP. Finally, the path-following performance across different key parameters is systematically investigated.
\begin{table*}[htbp]
	\centering
	\caption{Definition of some parameters.}
	\footnotesize
	\begin{tabularx}{\textwidth}{lll|lll}
		\toprule
		Param & Value & Implication &
		Param & Value & Implication \\
		\midrule
		$g$ & 9.81$\text{m}/\text{s}^2$ & Gravitational coefficient & 
		$\tau$ & 5$\text{s}$ & Time coefficient for groundspeed controller \\
		
		$x_p(0)$ & 0 $\text{m}$ &  UAV's $x_p$ along the north at initial time & 
		$k_{\chi}$ & 0.5 & Hyperparameter for $f$ in the RLLP \\
		
		$y_p(0)$ & 0 $\text{m}$ &  UAV's $y_p$ along the east at initial time & 
		$k_{\gamma}$ & 0.5 & Hyperparameter for $f$ in the RLLP \\
		
		$z_p(0)$ & 40 $\text{m}$ &  UAV's $z_p$ along the height at initial time &
		$\delta_t^0$ & 0.06 & Initial throttle openings \\
	
		$\phi(0)$ & 0 $\text{deg}$ &  UAV's roll angle at initial time &
		$k_p^{V}$ & 7.06 & Proportional gain for groundspeed controller\\
		
		$\theta(0)$ & 4.09 $\text{deg}$ &  UAV's pitch angle at initial time & 
		$k_i^{V}$ & 0.88 & Integral gain for groundspeed controller\\
		
		$\psi(0)$ & 0 $\text{deg}$ &  UAV's yaw angle at initial time &
		$k_d^{V}$ & 0 & Derivative gain for groundspeed controller\\

		$\gamma(0)$ & 0 $\text{deg}$ &  UAV's flight-path angle at initial time & 
		$k_p^{\phi}$ & 2.14 & Proportional gain for roll angle controller\\
		
		$\chi(0)$ & 0 $\text{deg}$ &  UAV's course angle at initial time&
		$k_i^{\phi}$ & 0.10 & Integral gain for roll angle controller\\

		$q_L$ & 0.25 & Adaptive coefficient for look-ahead length &
		$k_d^{\phi}$ & 1.23 & Derivative gain for roll angle controller\\
		
		$V_{g,c}$ & 13$\text{m}/\text{s}$ & Reference groundspeed & 
		$k_p^{n}$ & 1.50 & Proportional gain for load factor controller\\
		
		$\phi_{\min}$ & -45$\text{°}$ & The minimum value of roll angle & $k_i^{n}$ & 0.50 & Integral gain for load factor controller\\
		
		$\phi_{\max}$ & 45$\text{°}$ & The maximum value of roll angle & $k_p^{\theta}$ & 0.55 & Proportional gain for pitch angle controller\\
		$n_{\text{lf},\min}$ & 0 & The minimum value of load factor &
		$k_i^{\theta}$ & 0 & Integral gain for pitch angle controller\\
		
		$n_{\text{lf},\max}$ & 2.1 & The maximum value of load factor & 	$k_d^{\theta}$ & 0.14 & Derivative gain for pitch angle controller\\
		
		$a_{yc,\min}$ & -9.81$\text{m}/\text{s}^2$ & The minimum value of $a_{yc}$ &
		$a_{yc,\max}$ & 9.81$\text{m}/\text{s}^2$  & The maximum value of $a_{yc}$ \\
		\bottomrule
	\end{tabularx}
	\label{tab:0}
\end{table*}
\begin{figure}[htbp]
	\centering
	\begin{tabular}{c}
		\includegraphics[width=0.95\columnwidth]{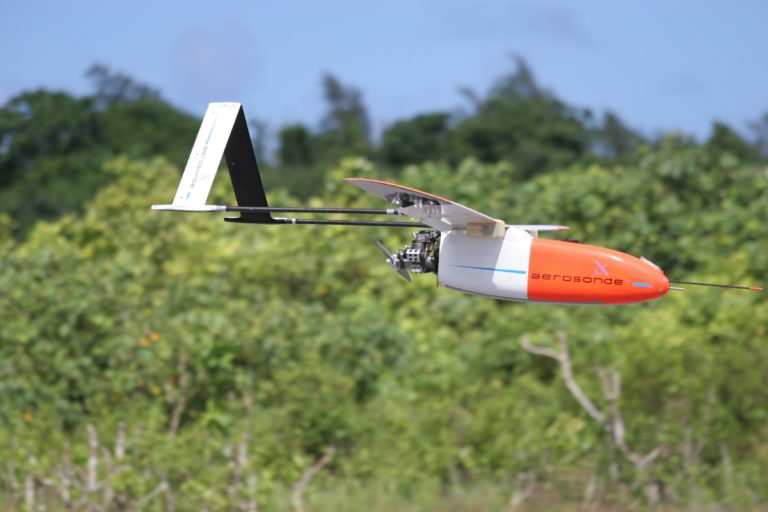}
	\end{tabular}
	\caption{The physical demonstration of Aerosonde aircraft.}
	\label{fig:21}
\end{figure}
The experimental environment is configured as a low-altitude path-following experiment in a dense urbanization environment with 36 high-rise buildings. The waypoints of the path, designed to mimic real-world urban navigation tasks, are shown in Figure \ref{fig:9}, featuring sharp turns and altitude changes. Meanwhile, the key hyperparameter configurations in this experiment, including lookahead distance, control gain, and UAV's initial state, etc., are presented in Table \ref{tab:0}, which are carefully tuned based on prior robustness analysis. 
The UAV model required for simulations is considered as a type of small and Miniature Air Vehicle (MAV), referred to as Aerosonde, specifically a fixed-wing configuration with a 1.2m wingspan, and its physical and aerodynamic parameters specified below are reported in Ref.\cite{burston2014reverse}. The Aerosonde is powered by a small four-stroke engine. Originally designed for scientific missions, the Aerosonde gained notoriety in 1998 as the first UAV to cross the Atlantic Ocean. Denoted by the acronym MAV, its six-degree-of-freedom dynamic model, aerodynamic coefficient tables, autopilot control architecture, and extended Kalman filter-based state estimation are all derived from the benchmark Ref.\cite{beard2012small}, ensuring the simulation fidelity in urban low-altitude scenarios.

\subsection{Validation of the theoretical framework}
\begin{figure*}[htbp]
	\centering
	\begin{tabular}{c}
		\includegraphics[width=0.92\textwidth]{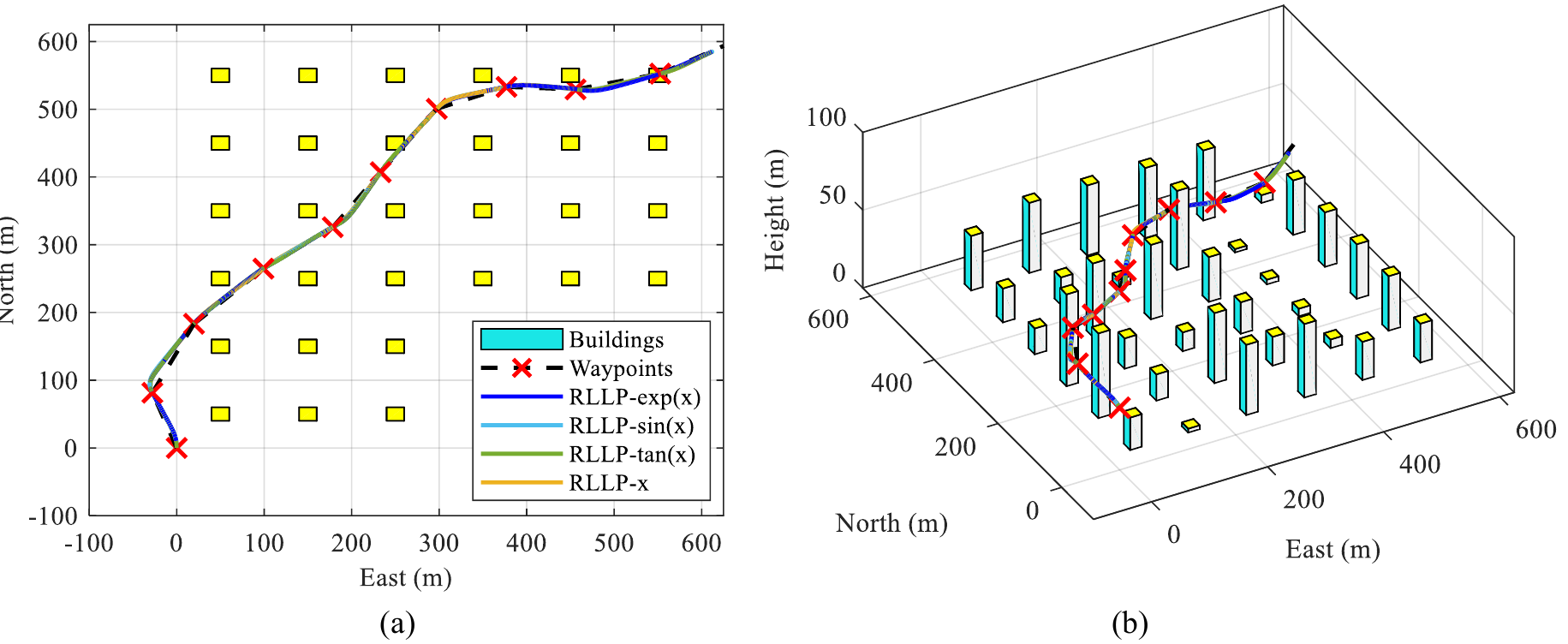} 
	\end{tabular}
	\caption{Simulation comparsion of UAV tracking trajectory for reference waypoints based on different $f$. (a): Top view; (b): Side view.}
	\label{fig:18}
\end{figure*}

To validate the correctness of the proposed theory involving \(I(f)\), including Theorem \ref{thm:theorem_1}, Theorem \ref{thm:lemma_A3}, and Remark \ref{rem:remark4}, we conducted comparative experiments on the RLLP guidance law with four distinct function forms \(f\) (including the $x$, $\tan(x)$, $\exp(x)$, $\sin(x)$, as listed in Table \ref{tab:hyperparameters}). The experiments were executed under \(k_{\chi} = 0.5\) and \(k_{\gamma} = 0.5\), and the corresponding path-following trajectories are visualized in Figure \ref{fig:18}.

\begin{figure*}[htbp]
	\centering
	\begin{tabular}{cc}
		\includegraphics[width=0.48\textwidth]{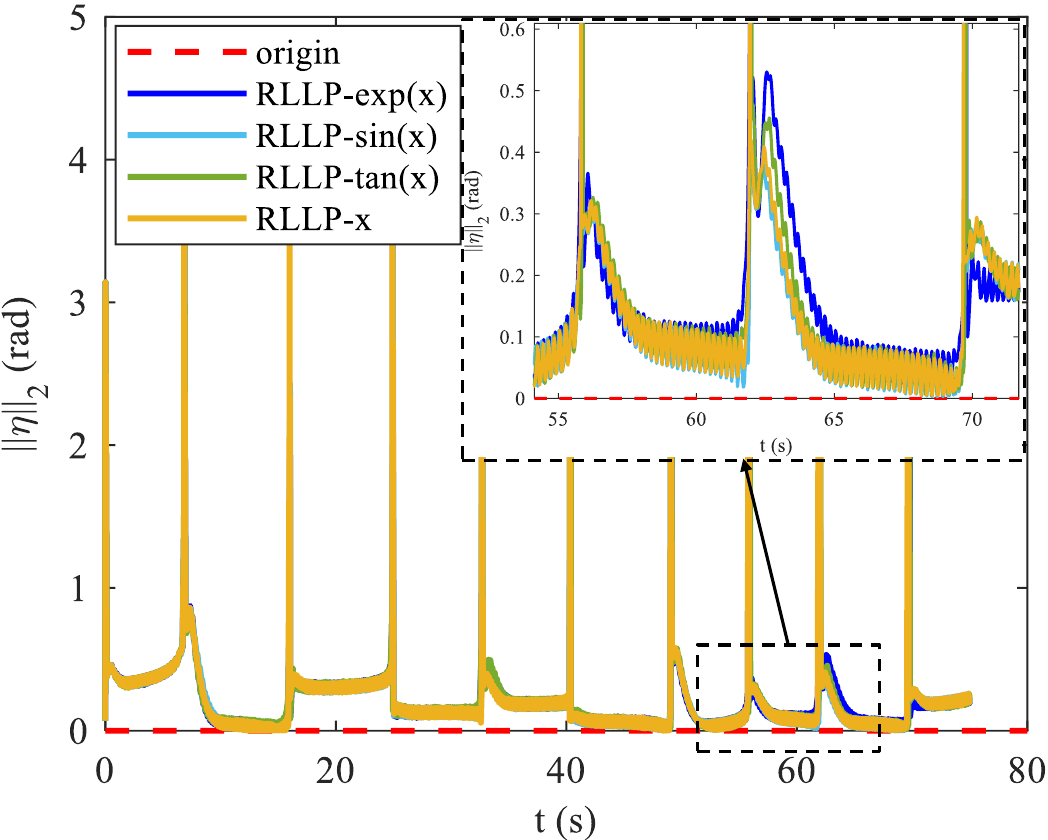} &
		\includegraphics[width=0.48\textwidth]{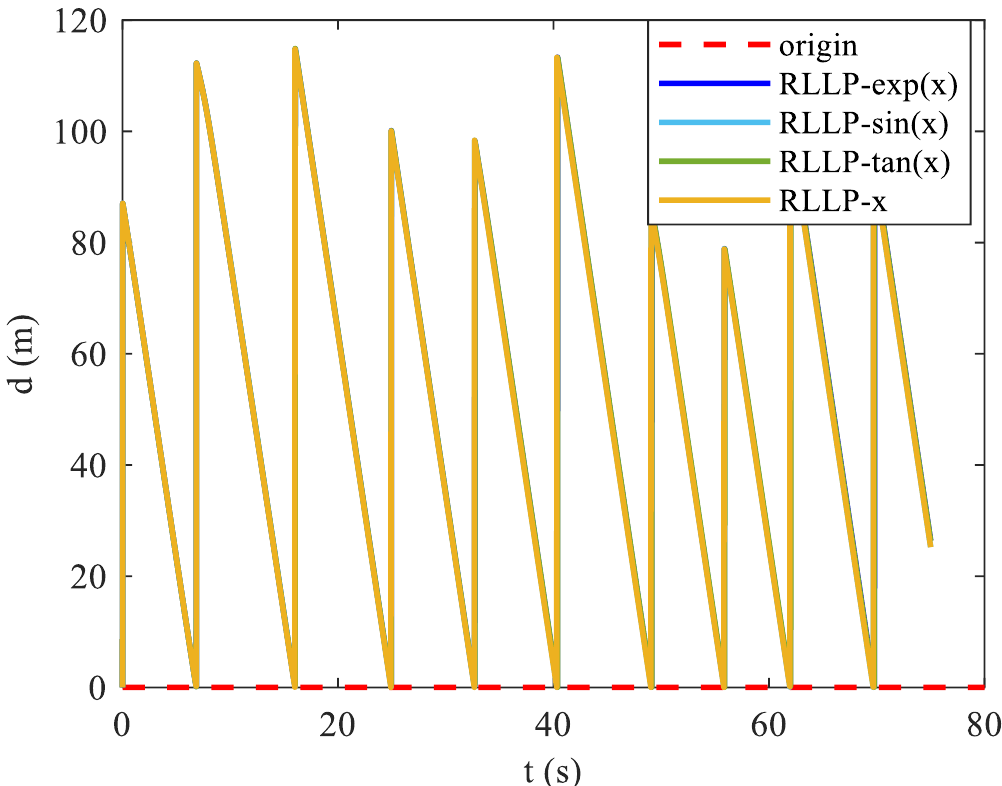} \\
		(a) & (b)
	\end{tabular}
	\caption{Comparison experiments of four algorithms including RLLP-$\exp(x)$, RLLP-$\sin(x)$, RLLP-$\tan(x)$, RLLP-$x$ on convergence indicators $||\eta||_2$ and $d$. (a) Comparison experiment on $||\eta||_2$; (b) Comparison experiment on $d$.}
	\label{fig:19}
\end{figure*}
As demonstrated in Figure \ref{fig:18}, all RLLP guidance laws corresponding to different \(f\) achieve precise waypoint tracking with near-straight trajectories in each segment. To quantify the performance discrepancies, we analyze two key metrics during path following:
\begin{align}
	||\eta||_2 &= \sqrt{(\eta^{lon})^2 + (\eta^{lat})^2} \\
	d &= \sqrt{(x_c - x_p)^2 + (y_c - y_p)^2 + (z_c - z_p)^2}
\end{align} 
as presented in Figure \ref{fig:19}.  
First, subplot (a) shows that for any \(f\), \(||\eta||_2\) converges to a minimal neighborhood (not necessarily zero) in each segment. The final convergence invariant sets of all \(f\) nearly coincide, except for RLLP-\(\exp(x)\), which exhibits a slightly higher invariant set—this is attributed to its higher \(I(f)\), validating Theorem \ref{thm:lemma_A3}.  
Second, subplot (b) reveals that despite \(||\eta||_2\) not converging to the origin, \(d\) for all RLLP variants converges to zero within finite time with linear convergence profiles. The convergence slopes across segments are consistent, corresponding to a UAV groundspeed of ~13 m/s, confirming the validity of Theorem \ref{thm:theorem_1}.
\begin{figure*}[htbp]
	\centering
	\begin{tabular}{ccc}
		\includegraphics[width=0.32\textwidth]{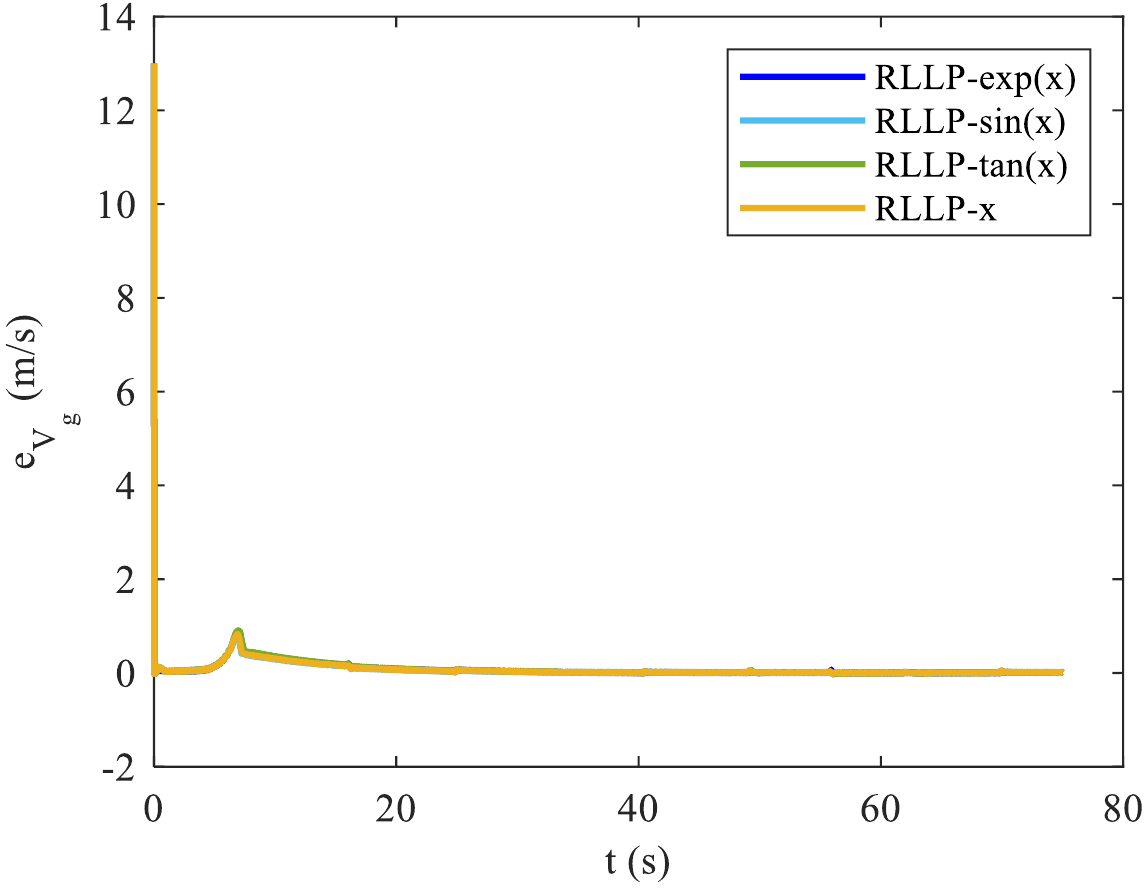} 
		&
		\includegraphics[width=0.32\textwidth]{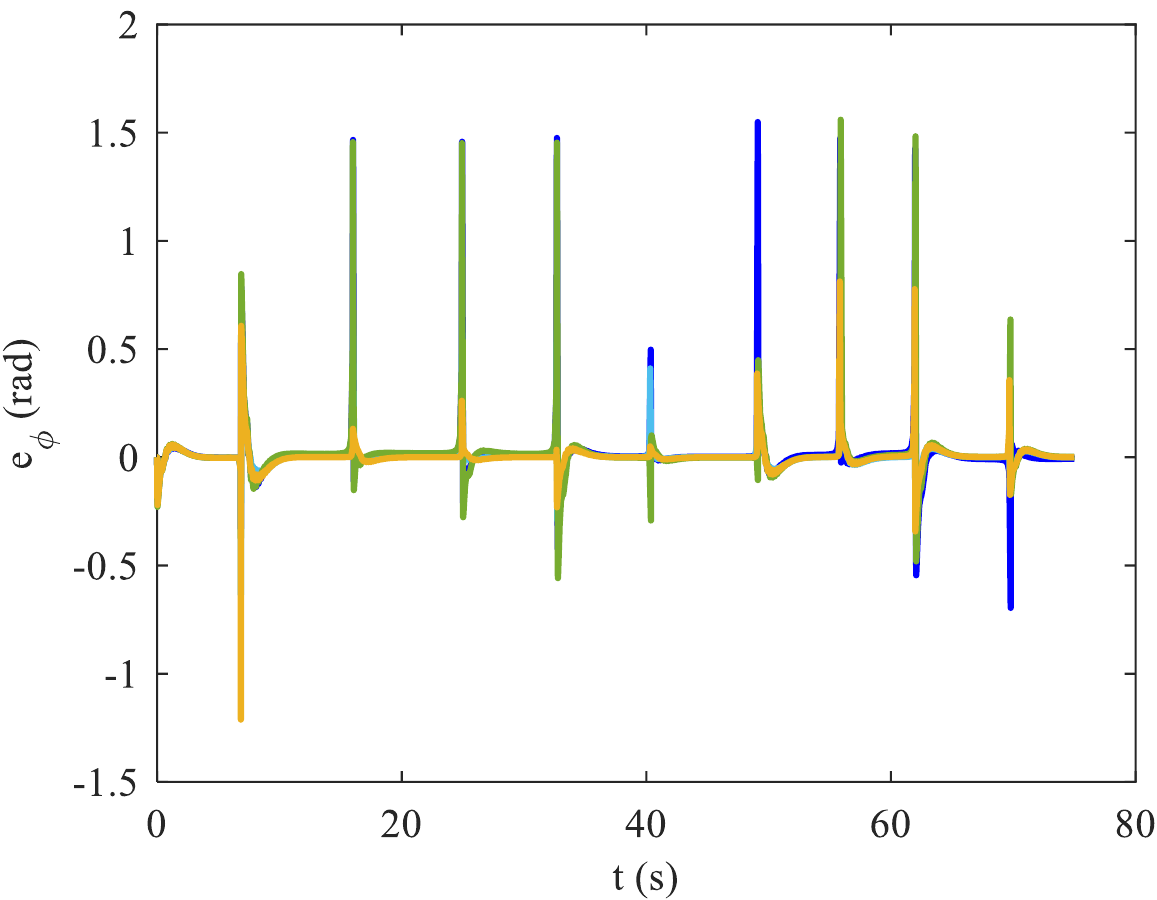}
		&
		\includegraphics[width=0.32\textwidth]{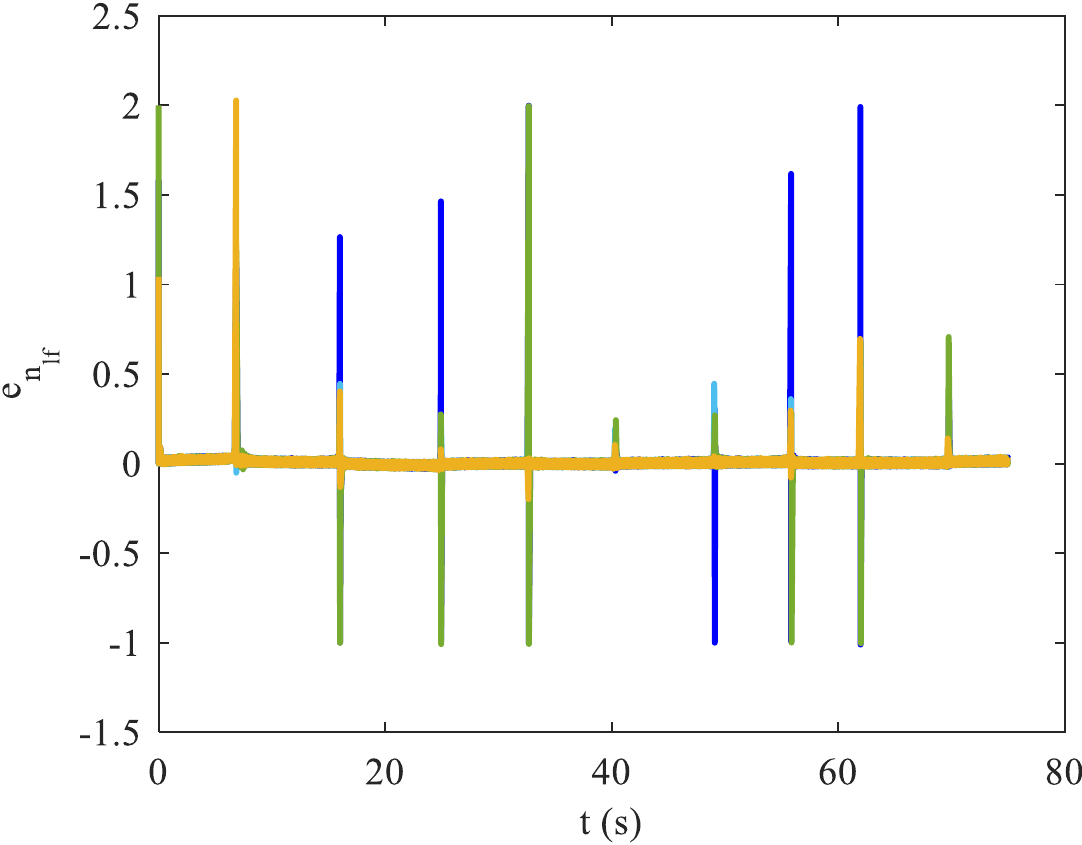}\\
		(a) 
		&
		(b) 
		&
		(c)
	\end{tabular}
	\caption{The tracking error curves of the autopilot for the reference signals \(V_{g,c}\), \(\phi_c\), and \(n_{\text{lf},c}\). (a): Tracking error for \(V_{g,c}\); (b): Tracking error for \(\phi_c\); (c): Tracking error for \(n_{\text{lf},c}\).}
	\label{fig:20}
\end{figure*}
Finally, to quantitatively evaluate the trackability of the autopilot to the proposed RLLP guidance law commands with different \(f\), we analyzed the tracking errors of \(V_g\), \(\phi\), and \(n_{\text{lf}}\) against their reference values \(V_{g,c}\), \(\phi_c\), \(n_{\text{lf},c}\) during path following, as depicted in Figure \ref{fig:20}. The results show that across all trajectory segments and waypoint tracking tasks, RLLP guidance laws with different \(f\) consistently ensure rapid stabilization of tracking errors, demonstrating the autopilot's robust control performance for the proposed RLLP framework.

\subsection{Path-following experiment of RLLP}
In this section, we will study the path-following performance of the Optimal-RLLP guidance law and conduct a comparative analysis with classical algorithms.
\subsubsection{The performance of Optimal-RLLP}
\begin{figure*}[htbp]
	\centering
	\begin{tabular}{cc}
		\includegraphics[width=0.45\textwidth]{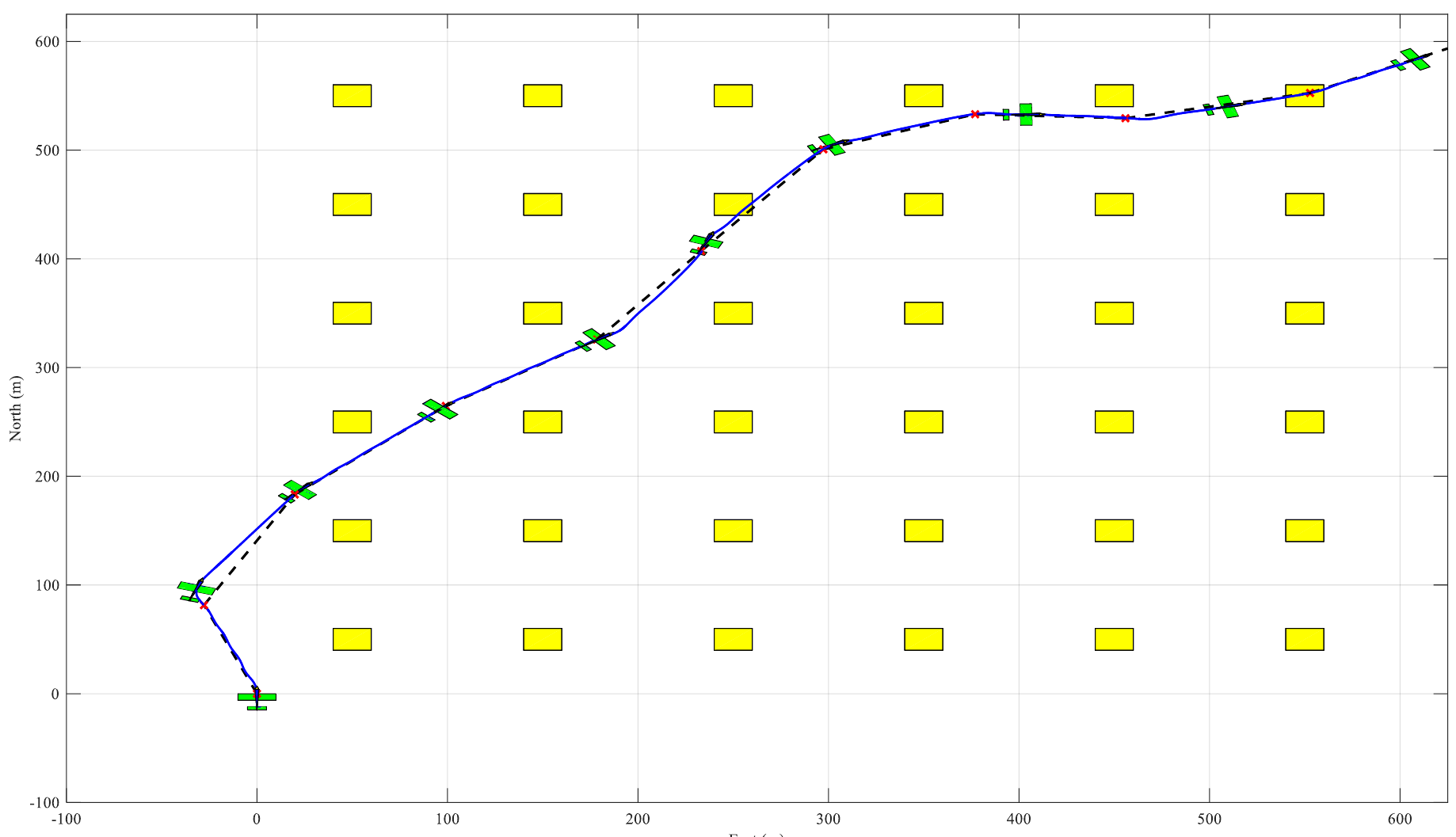} 
		&
		\includegraphics[width=0.48\textwidth]{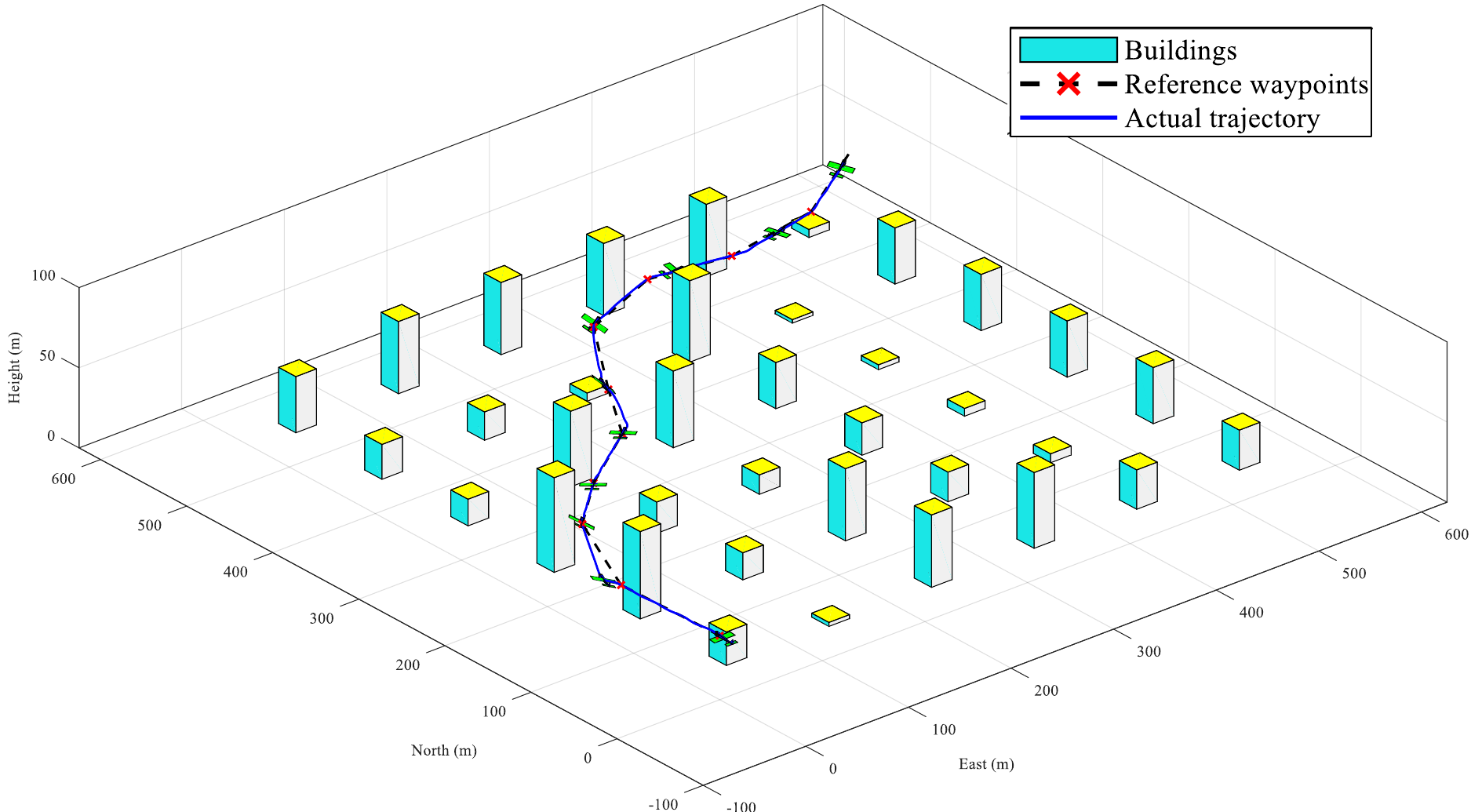} \\
		(a)
		&
		(b)
	\end{tabular}
	\caption{Simulation of UAV tracking trajectory for reference waypoints based on Optimal RLLP. (a): Top view; (b): Side view.}
	\label{fig:10}
\end{figure*}

\begin{figure*}[htbp]
	\centering
	\begin{tabular}{ccc}
		\includegraphics[width=0.32\textwidth]{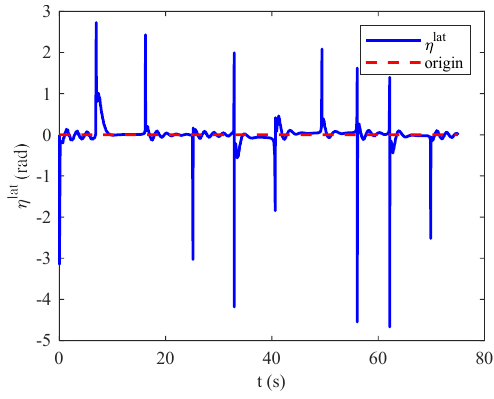} 
		&
		\includegraphics[width=0.32\textwidth]{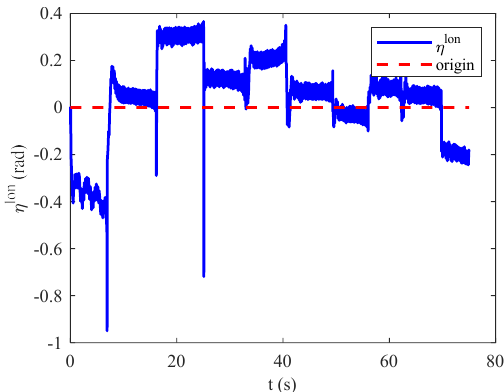}
		&
		\includegraphics[width=0.32\textwidth]{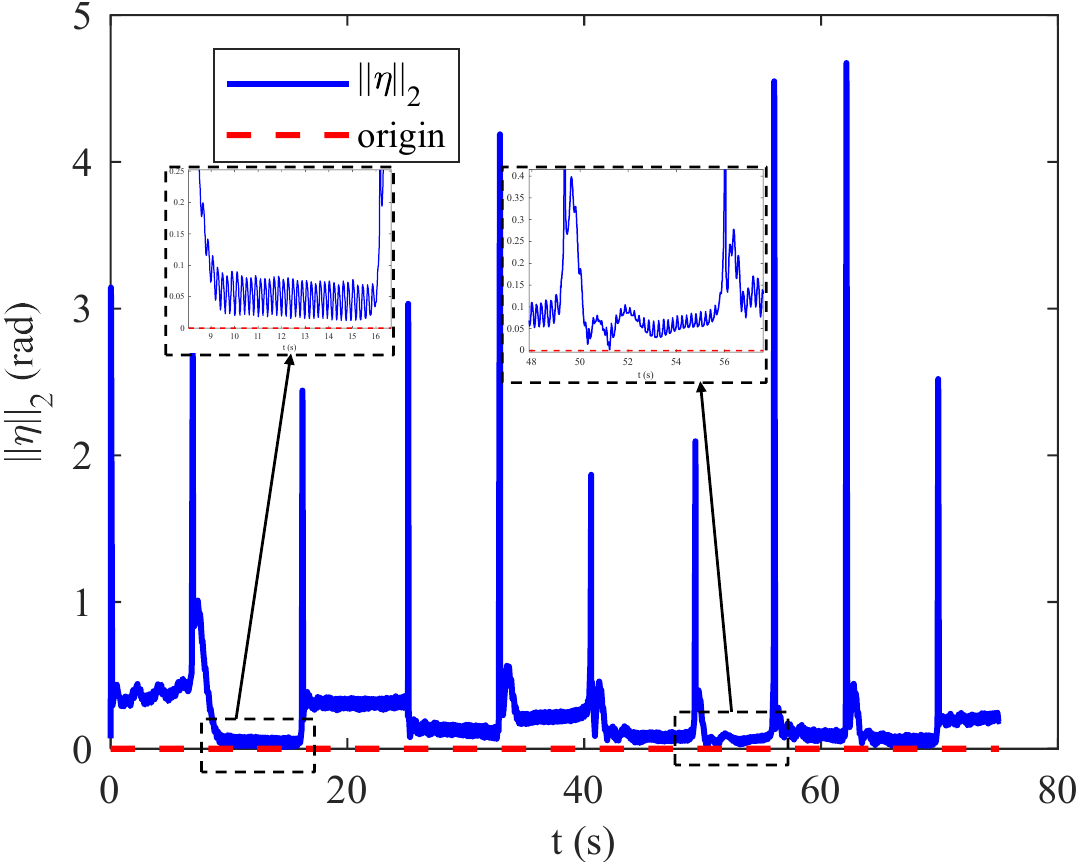}
		\\
		(a) 
		&
		(b)
		&
		(c)
	\end{tabular}
	\caption{Time-series curves of \(\eta^{\text{lat}}\), \(\eta^{\text{lon}}\), and \(||\eta||_2\) during path-following process. Notably, \(||\eta||_2\) converges nearly to the origin in each tracking instance. (a): Time-series curve of \(\eta^{\text{lat}}\); (b): Time-series curve of \(\eta^{\text{lon}}\); (c): Time-series curve of \(||\eta||_2\).}
	\label{fig:12}
\end{figure*}

\begin{figure*}[htbp]
	\centering
	\begin{tabular}{c}
		\includegraphics[width=0.98\textwidth]{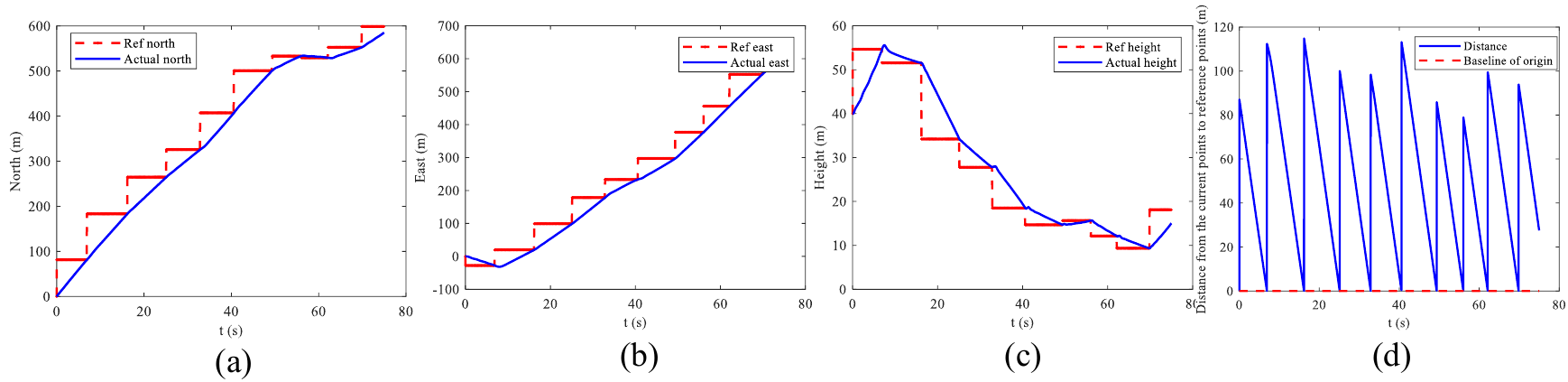}
	\end{tabular}
	\caption{Time-series curves of \(x_p\), \(y_p\), \(z_p\) and the relative distance \(d = \sqrt{(x_c - x_p)^2 + (y_c - y_p)^2 + (z_c - z_p)^2}\) during path-following process when UAV follows the reference waypoints. Notably, \(d\) converges nearly to the origin in each tracking instance. (a): Time-series curve of \(x_p\); (b): Time-series curve of \(y_p\); (c): Time-series curve of \(z_p\); (d): Time-series curve of $d$.}
	\label{fig:13}
\end{figure*}

Figure \ref{fig:10} demonstrates the path-following process of the Optimal-RLLP guidance law, presenting the UAV's actual trajectory and attitude. As clearly shown in the figure, the actual trajectory closely adheres to the straight-line segments of reference waypoints with high precision during path following. The autopilot effectively tracks the guidance law's reference commands \( V_{g,c} \), \( \phi_c \), and \( n_{\text{lf},c} \). Since the optimization objective prioritizes minimizing \( I(f) \), the reference commands tend to exhibit pronounced oscillations, a characteristic indicating that Optimal-RLLP possesses superior sensitivity.
\begin{figure*}[htbp]
	\centering
	\begin{tabular}{ccc}
		\includegraphics[width=0.32\textwidth]{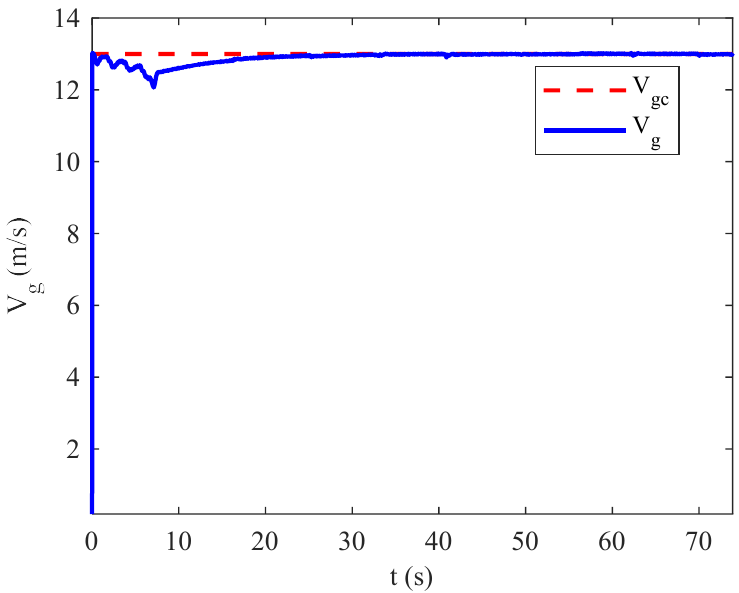} 
		&
		\includegraphics[width=0.32\textwidth]{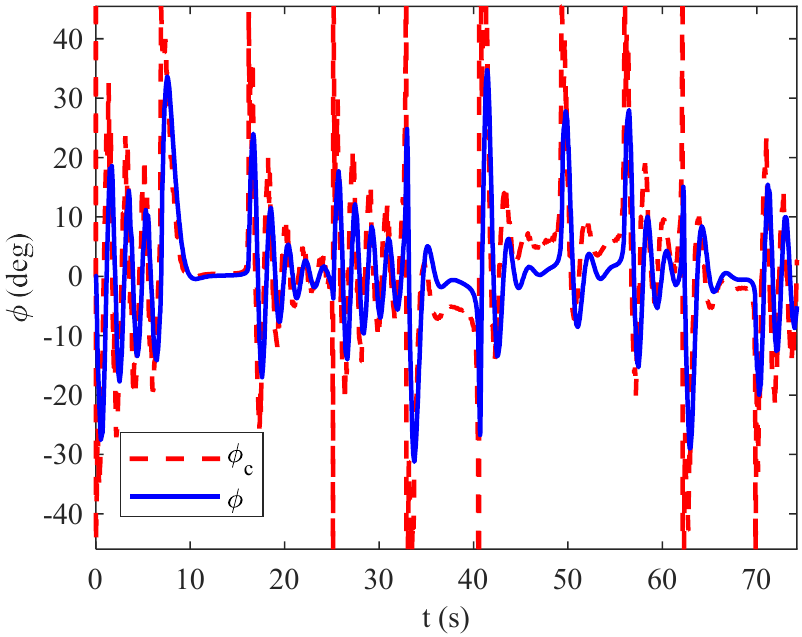}
		&
		\includegraphics[width=0.32\textwidth]{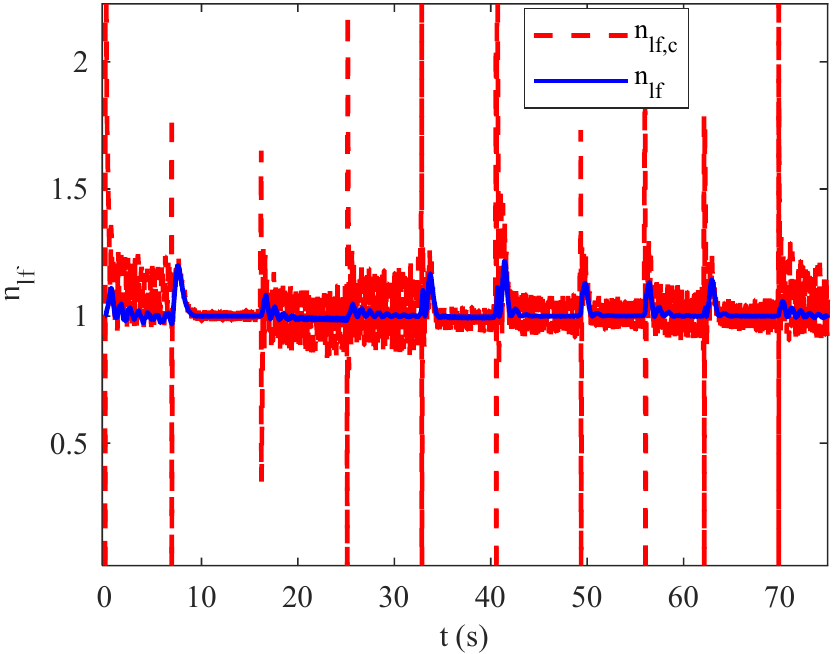}\\
		(a) 
		&
		(b) 
		&
		(c)
	\end{tabular}
	\caption{The tracking curves of the autopilot for the reference signals \(V_{g,c}\), \(\phi_c\), and \(n_{\text{lf},c}\). (a): Tracking curve of \(V_{g,c}\); (b): Tracking curve of \(\phi_c\); (c): Tracking curve of \(n_{\text{lf},c}\).}
	\label{fig:11}
\end{figure*}
The time-series curves of \( \eta^{\text{lat}} \), \( \eta^{\text{lon}} \), and \( \|\eta\|_2 \) for the Optimal-RLLP guidance law are depicted in Figure \ref{fig:12}. Notably, both \( \eta^{\text{lat}} \) and \( \eta^{\text{lon}} \) stabilize within a minimal range around the origin in the lateral and longitudinal planes, with their fluctuation amplitudes consistently constrained below 0.4rad. Under optimal conditions, \( \|\eta\|_2 \) can even exhibit stable fluctuations within a narrow band of less than 0.05rad. This behavior effectively ensures that the UAV's actual trajectory achieves precise tracking of reference waypoints across the north, east, and height dimensions. As shown in Figure \ref{fig:13}, the tracking error for each waypoint is maintained below 1$\text{m}$.
\subsubsection{Comparison for RLLP and classical methods}
\begin{figure*}[htbp]
	\centering
	\begin{tabular}{c}
		\includegraphics[width=0.92\textwidth]{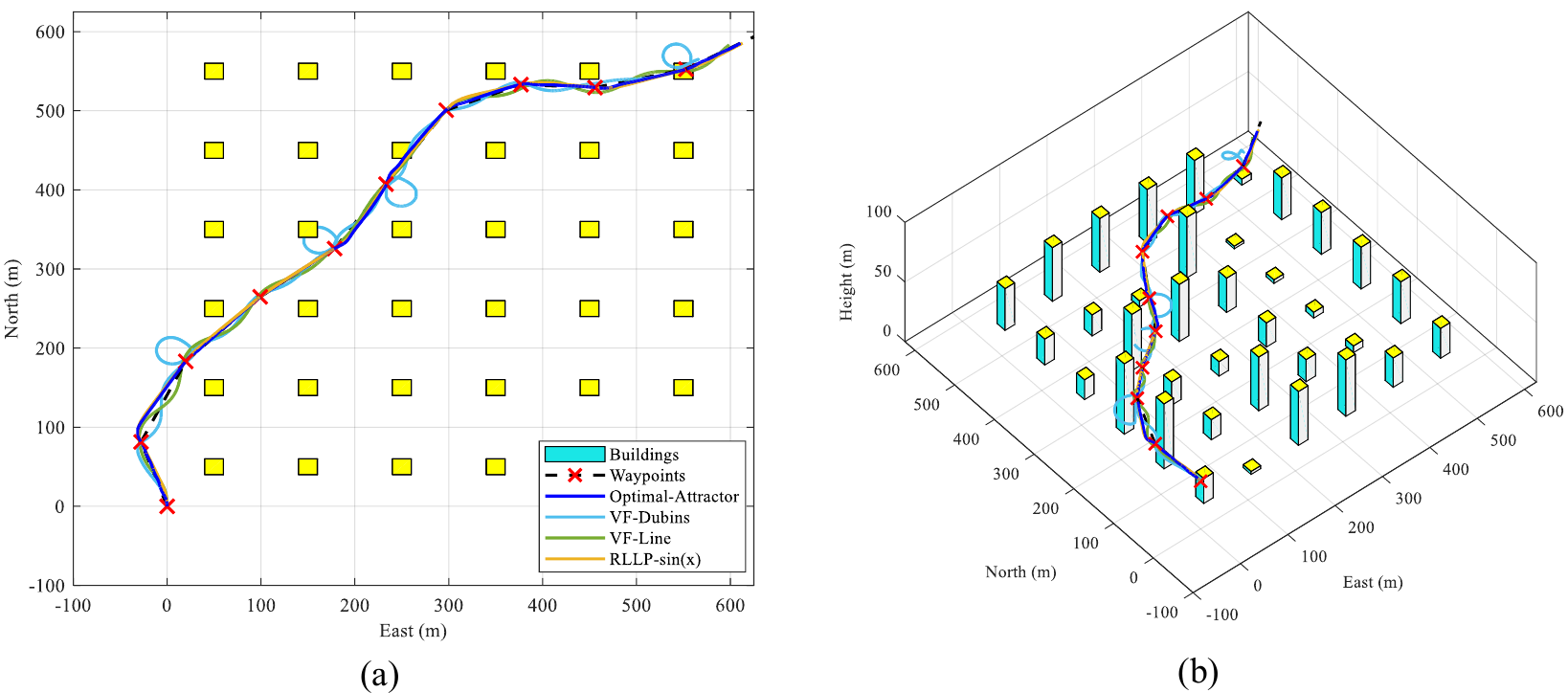} 
	\end{tabular}
	\caption{Simulation comparsion of UAV tracking trajectory for reference waypoints based on Optimal RLLP, VF-Dubins, VF-line, RLLP-$\sin(x)$. (a): Top view; (b): Side view.}
	\label{fig:15}
\end{figure*}

\begin{figure*}[htbp]
	\centering
	\begin{tabular}{cc}
		\includegraphics[width=0.48\textwidth]{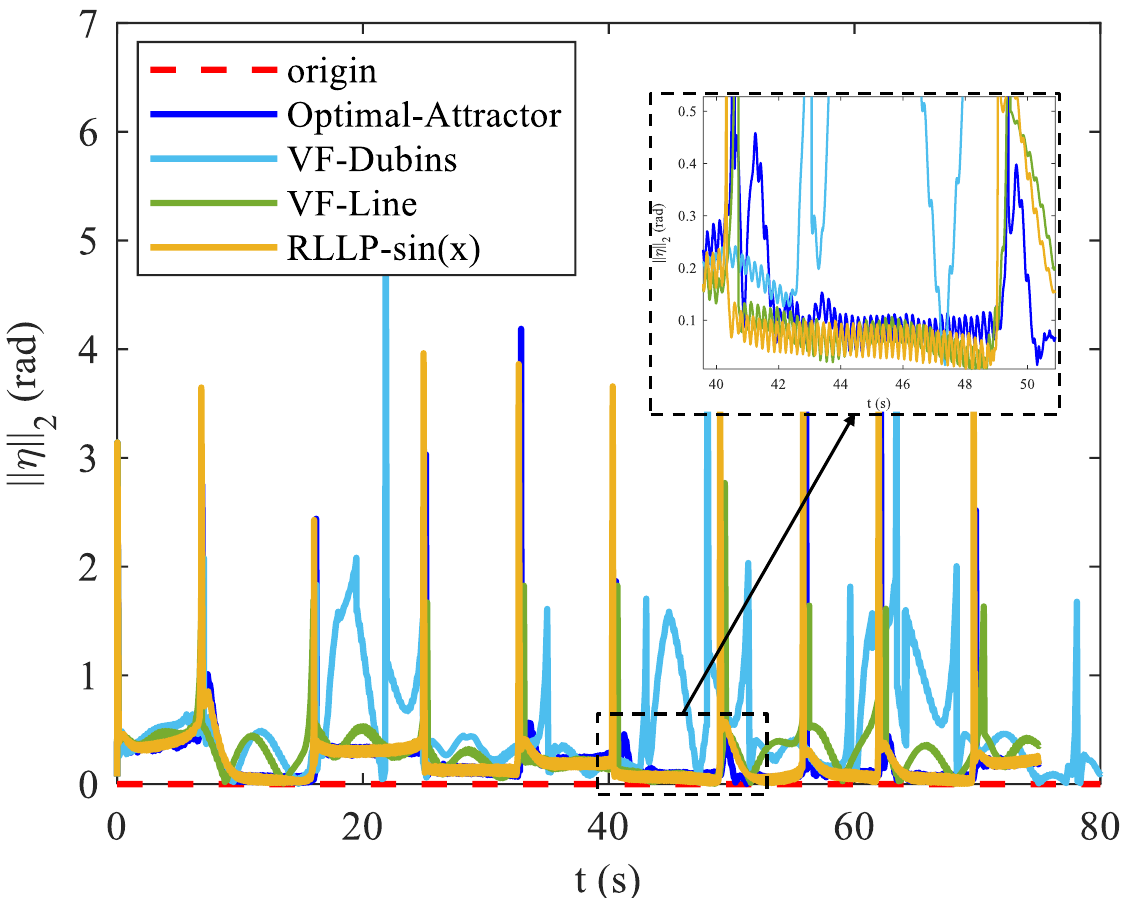} &
		\includegraphics[width=0.48\textwidth]{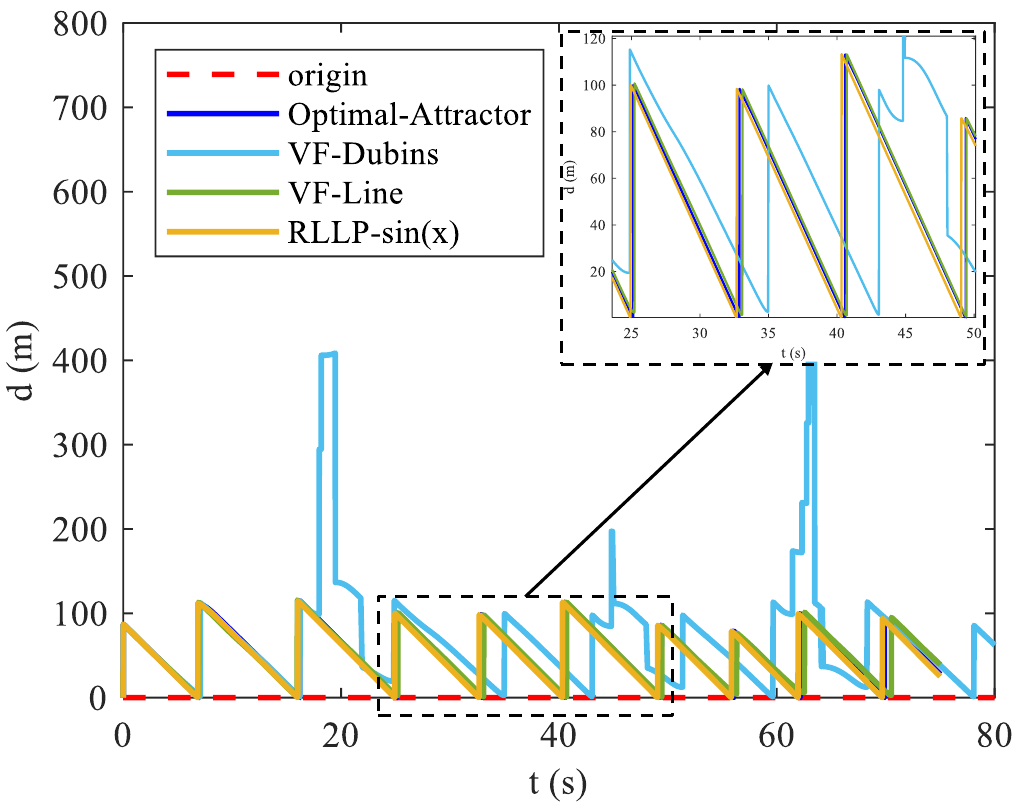} \\
	\end{tabular}
	\caption{Comparison experiments of four algorithms including Optimal RLLP, VF-Dubins, VF-Line, and RLLP-$\sin(x)$ on convergence indicators $||\eta||_2$ and $d$. (a) Comparison experiment on $||\eta||_2$; (b) Comparison experiment on $d$.}
	\label{fig:16}
\end{figure*}

To further demonstrate the superiority of the Optimal-RLLP guidance law, comparative analyses were conducted against the virtual vector field method with Dubins curve connections (VF-Dubins\cite{beard2012small}), the virtual vector field method with straight-line connections (VF-Line\cite{yao2021singularity}), and the RLLP guidance law with the functional form \( f = \sin(x) \) (RLLP-$\sin(x)$). The UAV tracking trajectories for reference waypoints under the four guidance laws are illustrated in Figure \ref{fig:15}. Notably, Optimal-RLLP outperforms the other three methods in straight-line path tracking, a feature indicative of its superior disturbance rejection capability. This advantage is further manifested in smoother waypoint transitions, where the UAV executes turns with smaller curvatures when switching between adjacent waypoints.
The superiority of Optimal-RLLP is attributed to the characteristics visualized in Figure \ref{fig:16}. In the time-series curve of \( \|\eta\|_2 \), only Optimal-RLLP and RLLP-$\sin(x)$ guidance law efficiently stabilize \( \|\eta\|_2 \) within each trajectory segment, while VF-Dubins and VF-Line methods exhibit significant oscillations. Compared with RLLP-$\sin(x)$, Optimal-RLLP demonstrates a higher exponential convergence rate, enabling faster stabilization to a smaller steady-state range under more complex disturbances. This feature is precisely what endows Optimal-RLLP with enhanced robustness.

\subsection{Robustness under atmospheric disturbances}
\begin{figure*}[htbp]
	\centering
	\begin{tabular}{ccc}
		\includegraphics[width=0.32\textwidth]{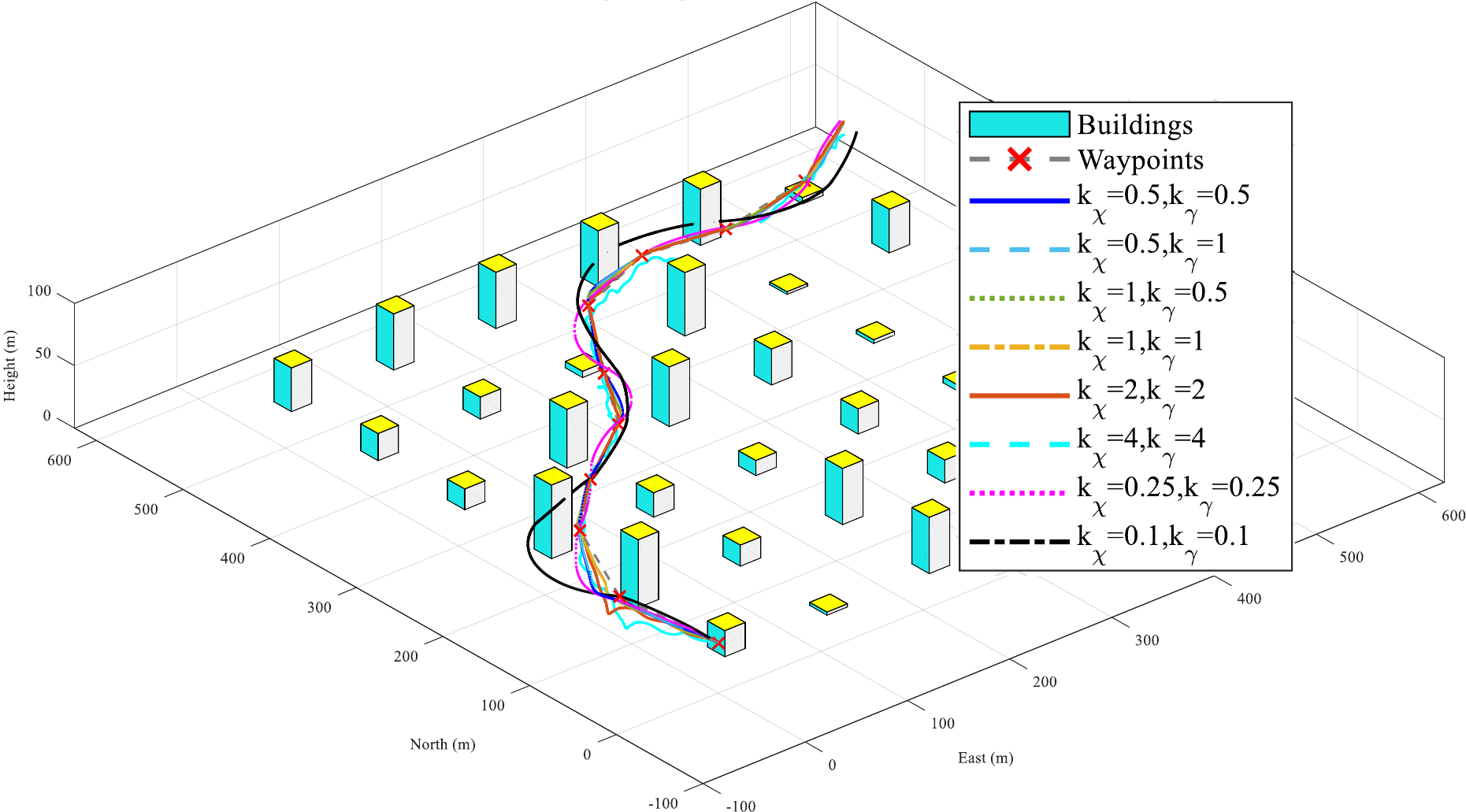} &
		\includegraphics[width=0.32\textwidth]{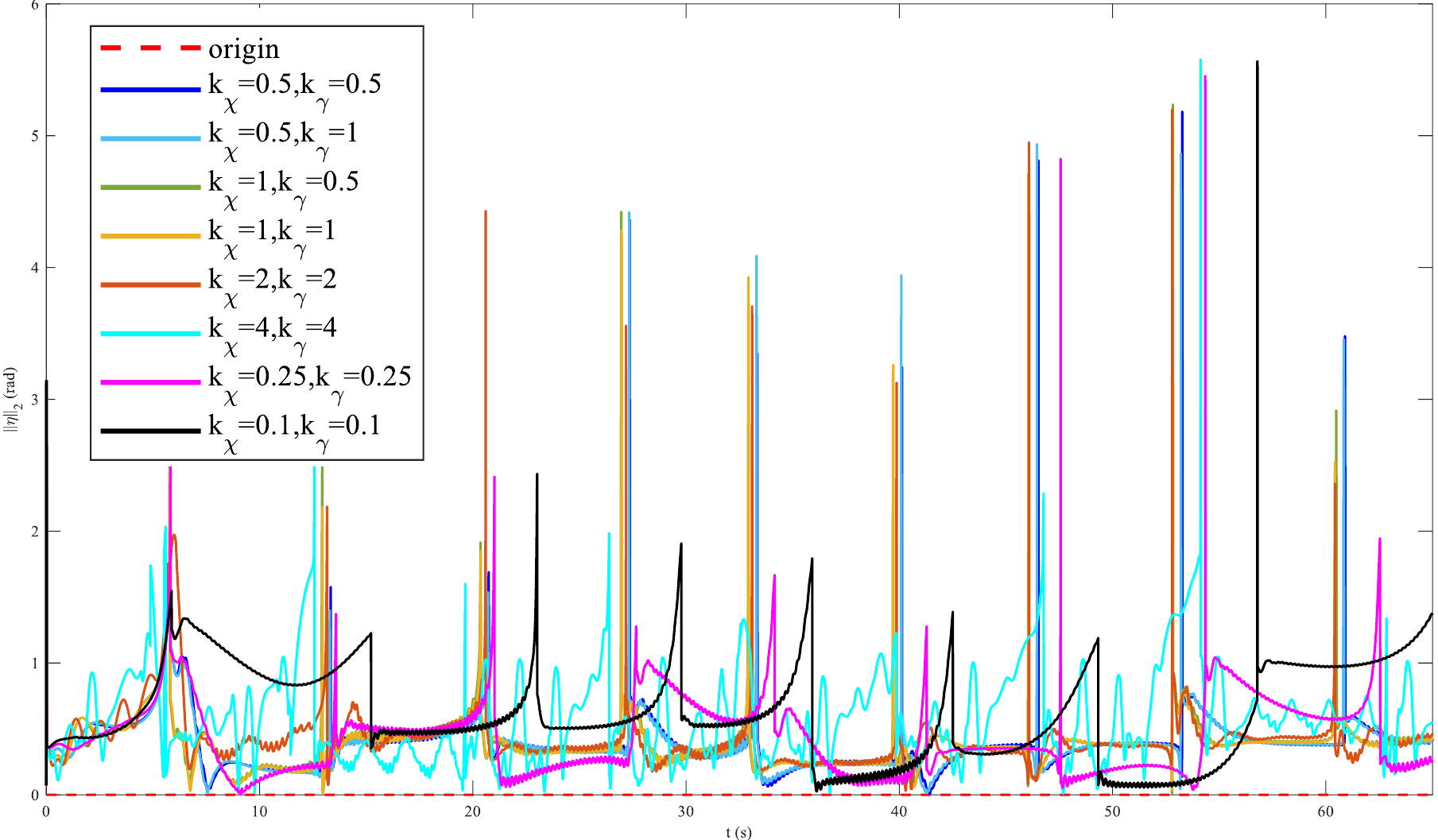} &
		\includegraphics[width=0.32\textwidth]{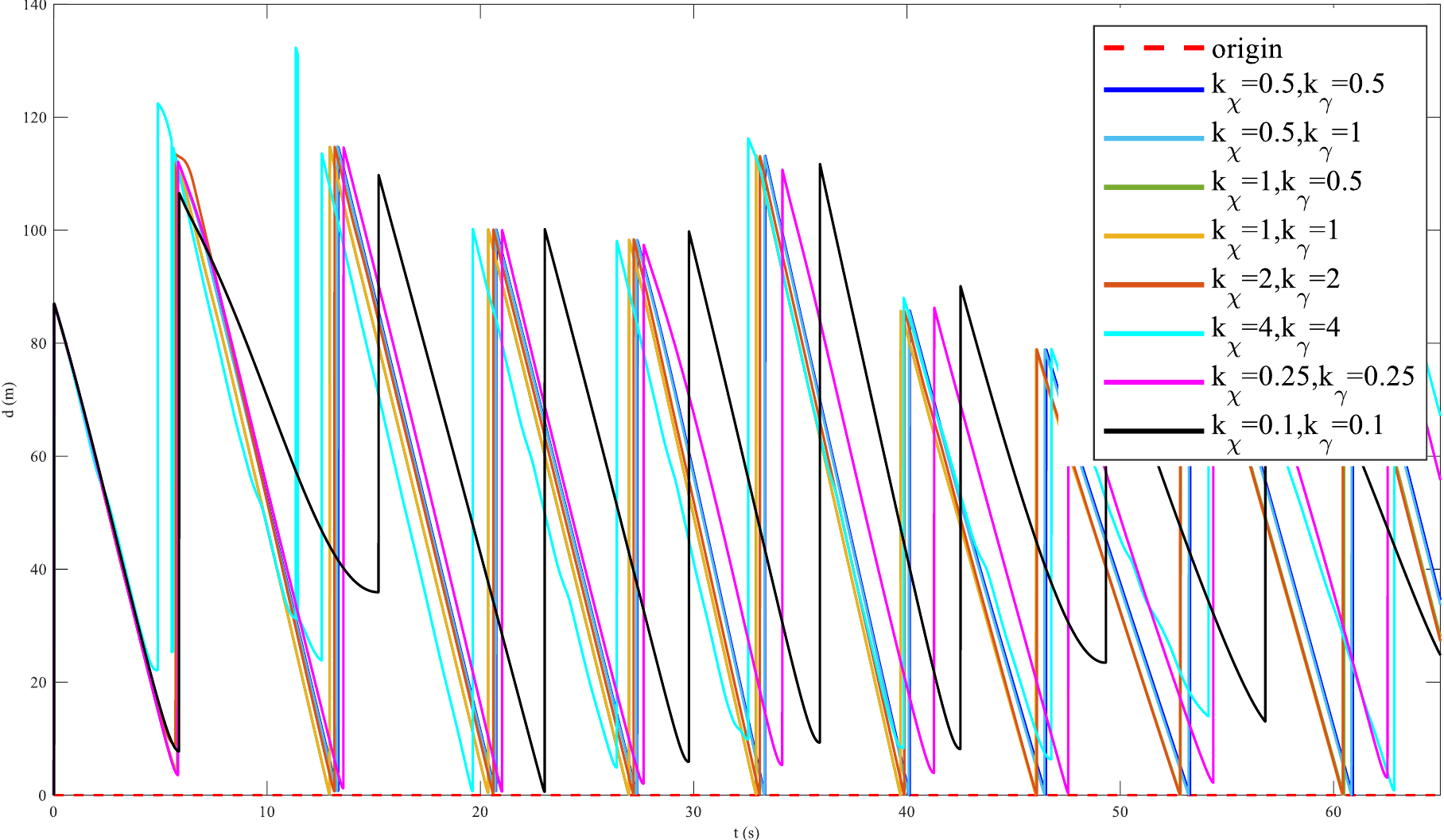} \\
		& (a) & \\
		\includegraphics[width=0.32\textwidth]{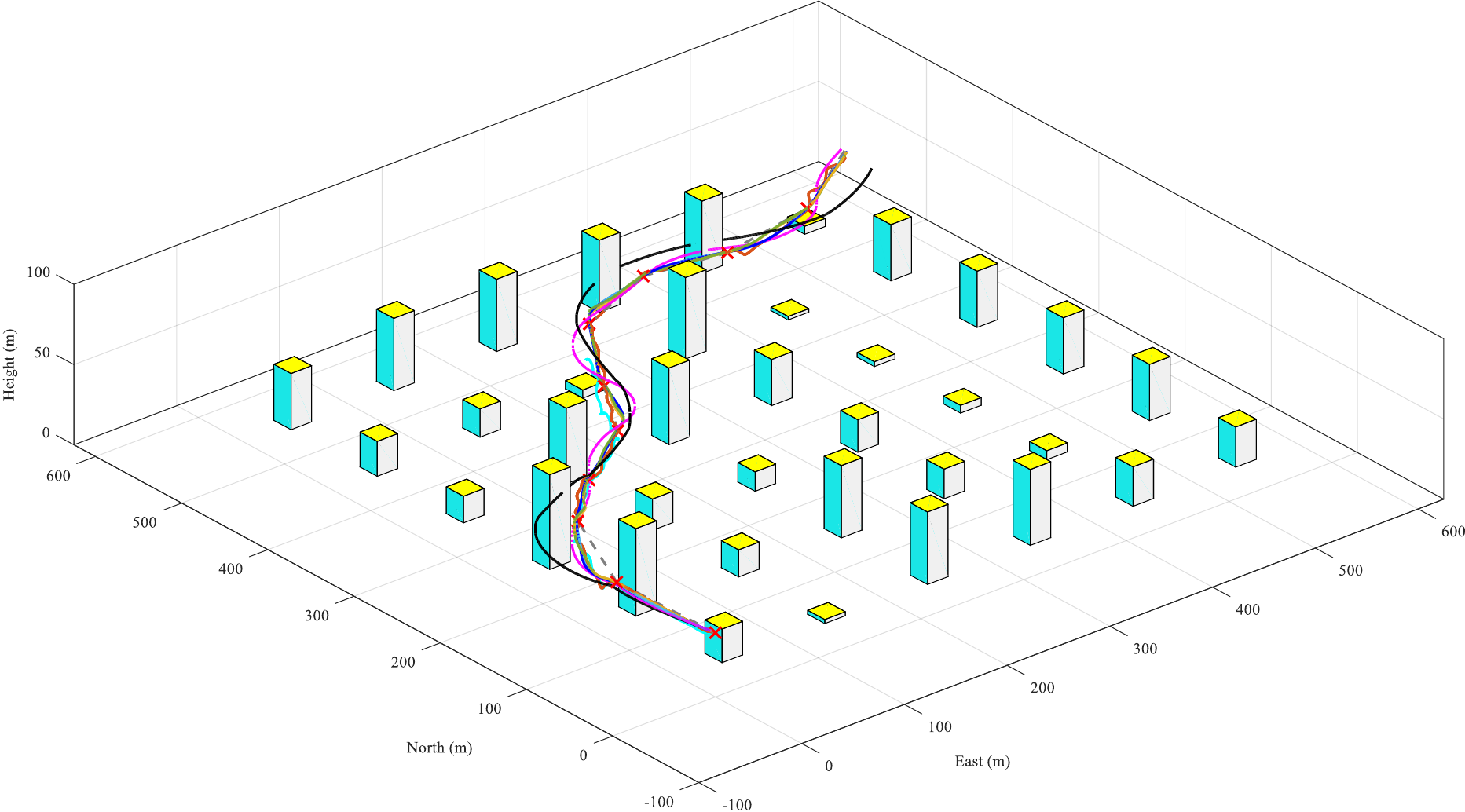} &
		\includegraphics[width=0.32\textwidth]{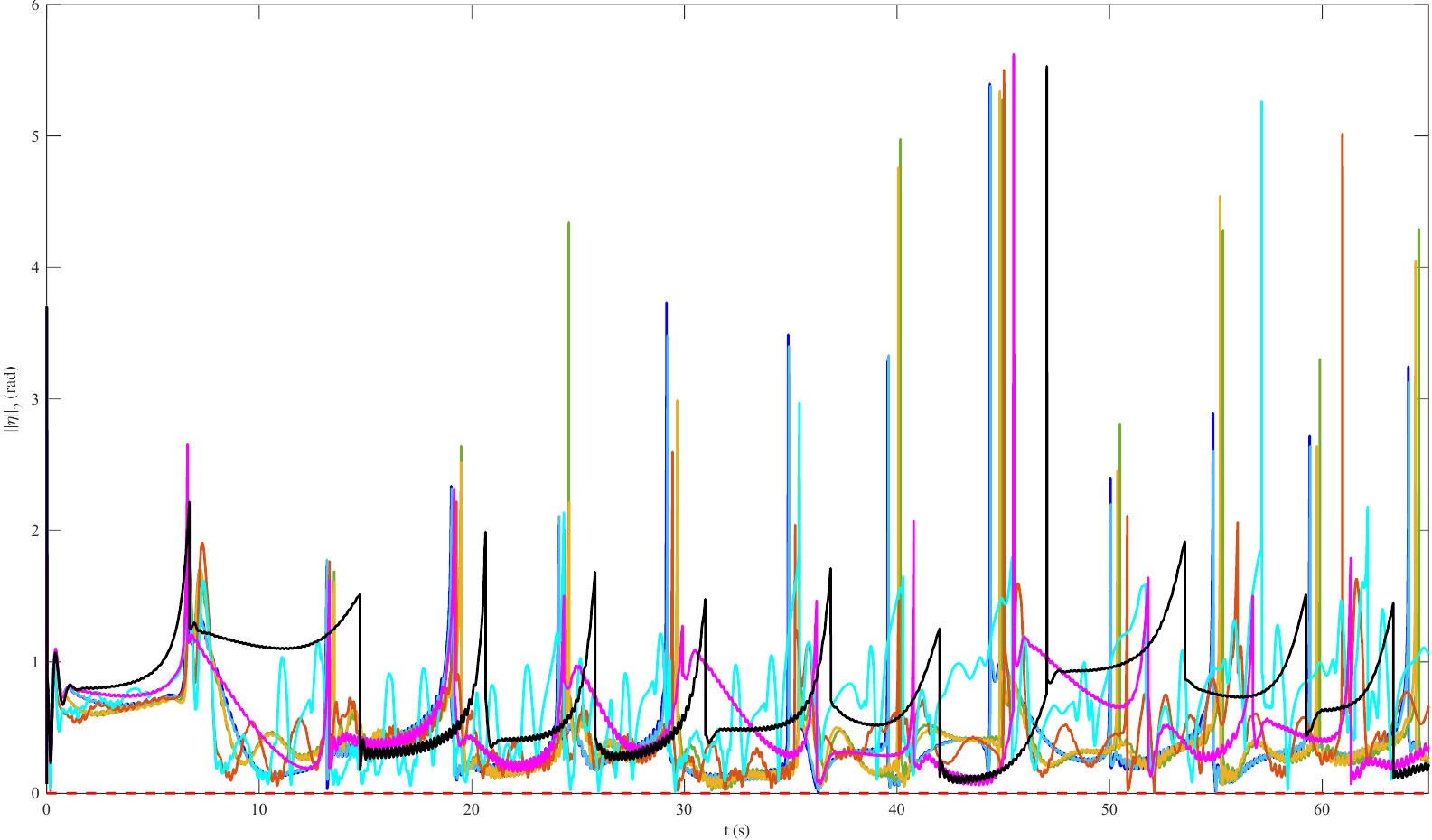} &
		\includegraphics[width=0.32\textwidth]{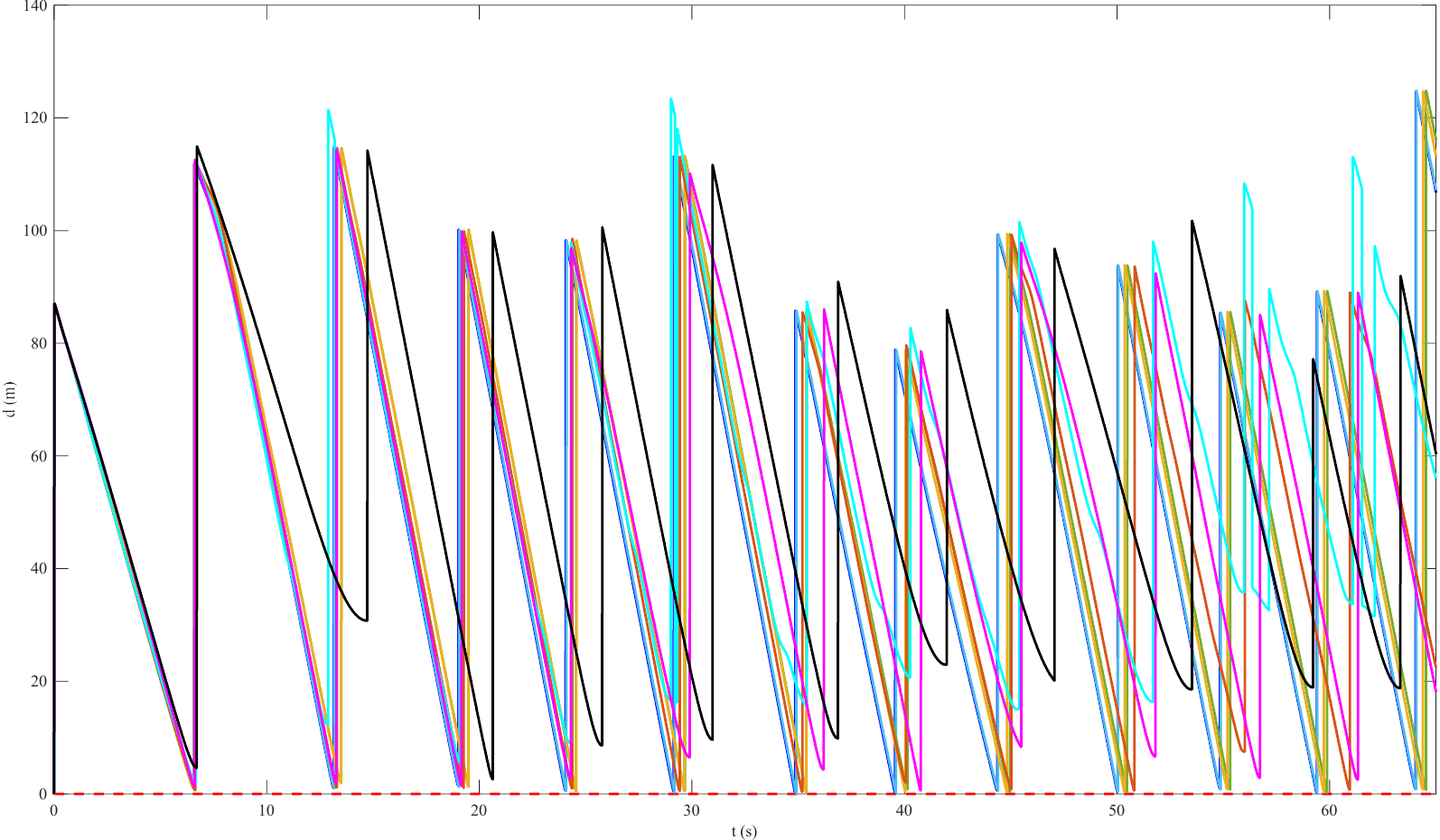} \\
		& (b) & \\
		\includegraphics[width=0.32\textwidth]{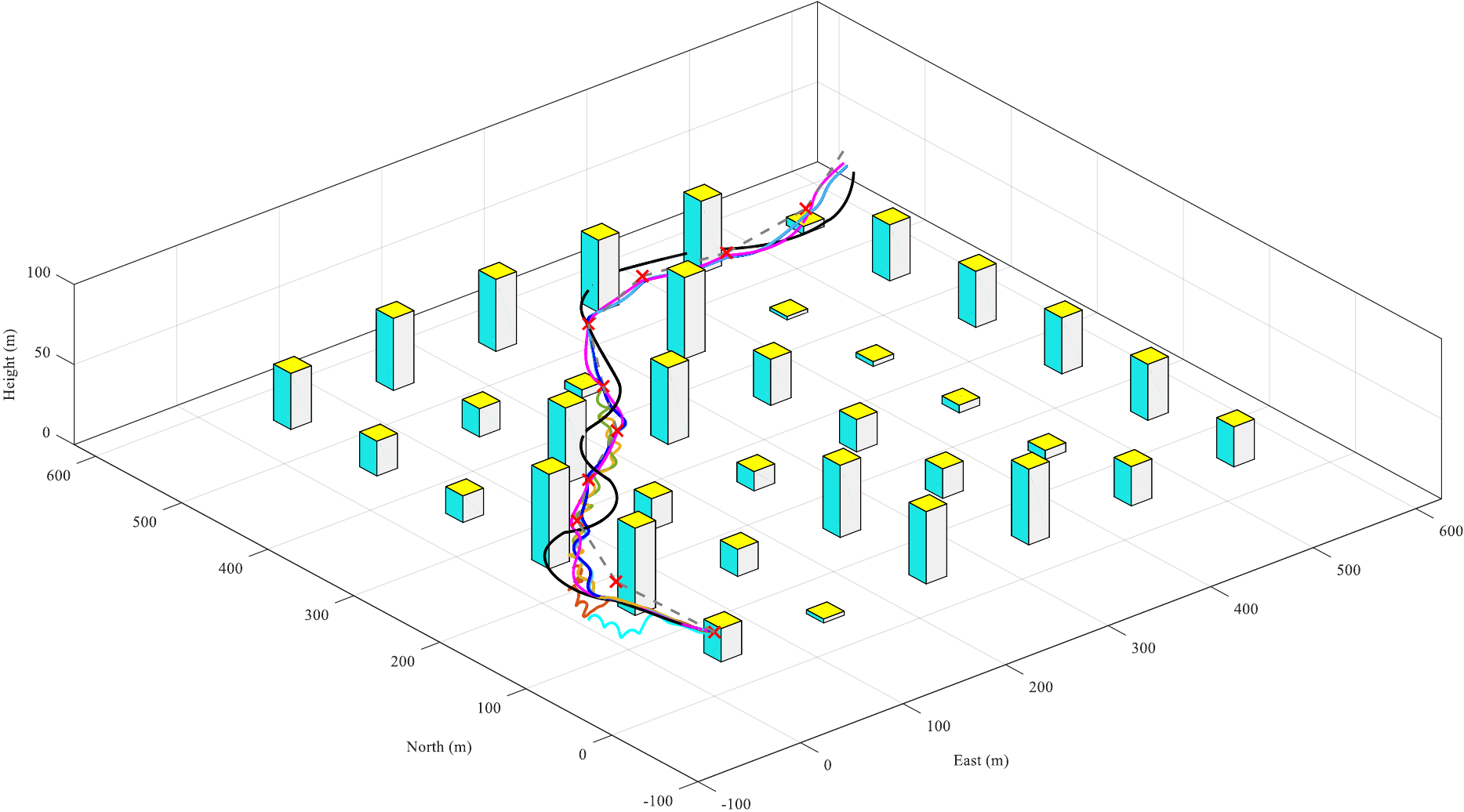} &
		\includegraphics[width=0.32\textwidth]{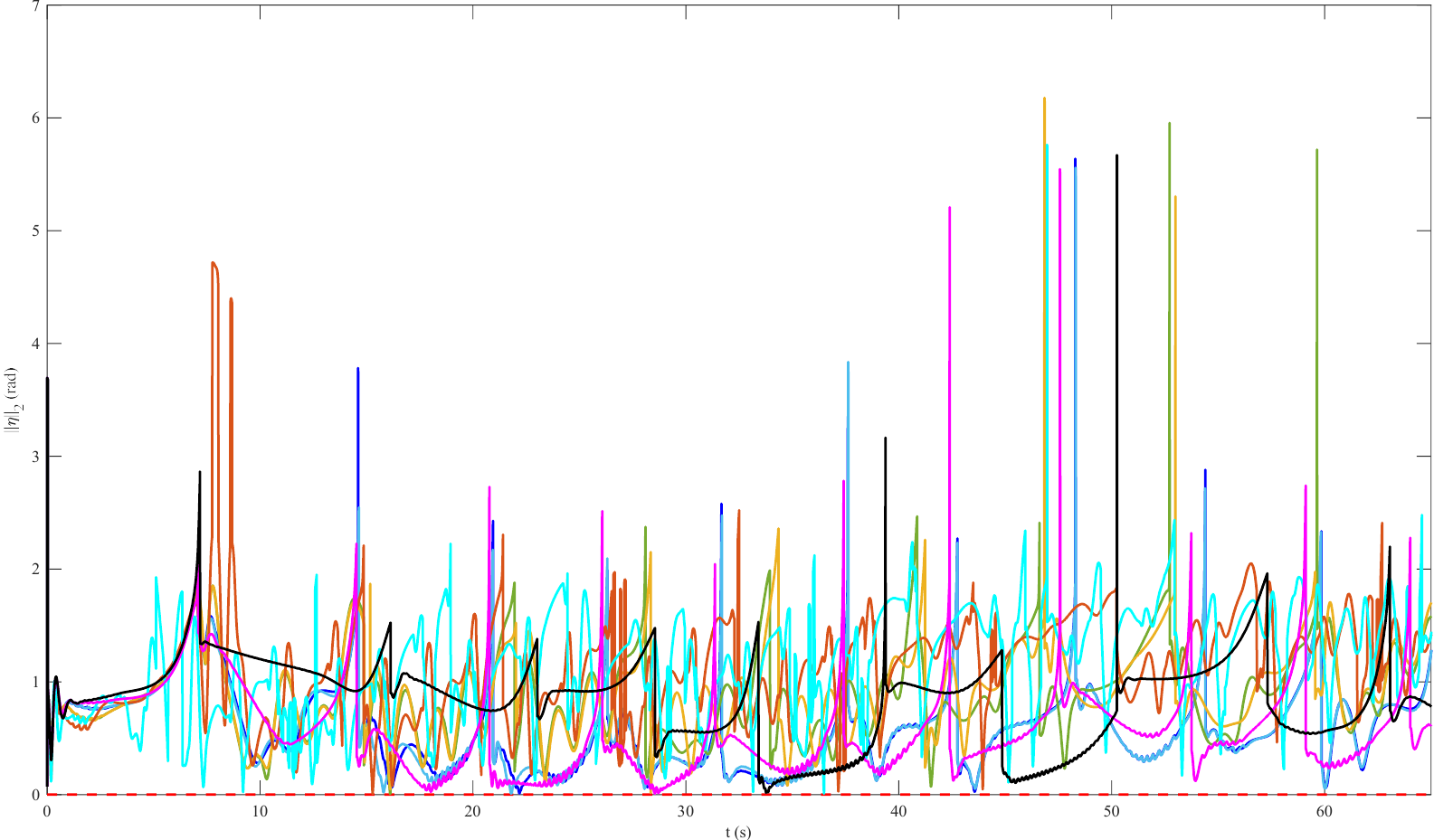} &
		\includegraphics[width=0.32\textwidth]{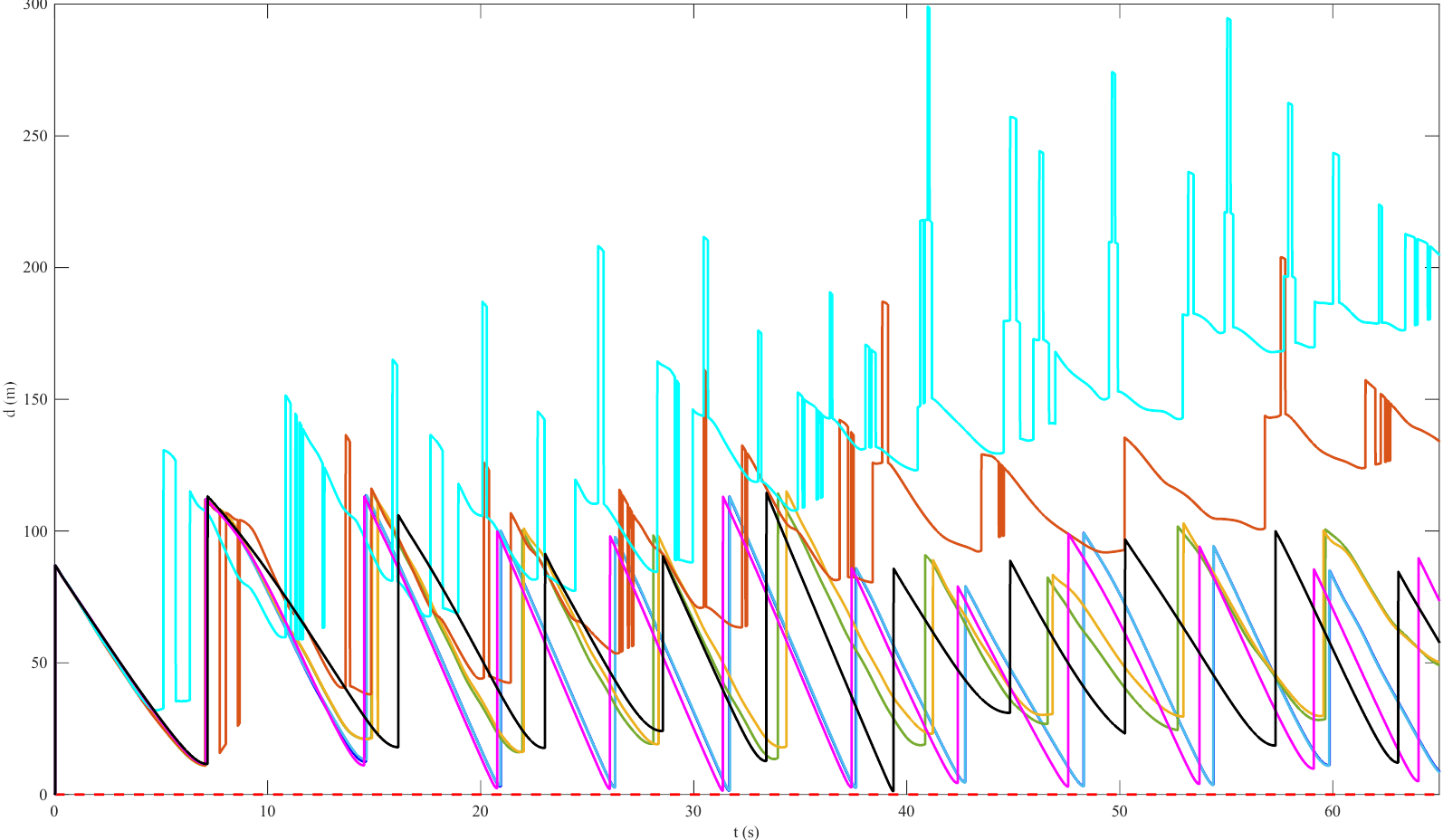} \\
		& (c) & \\
		\includegraphics[width=0.32\textwidth]{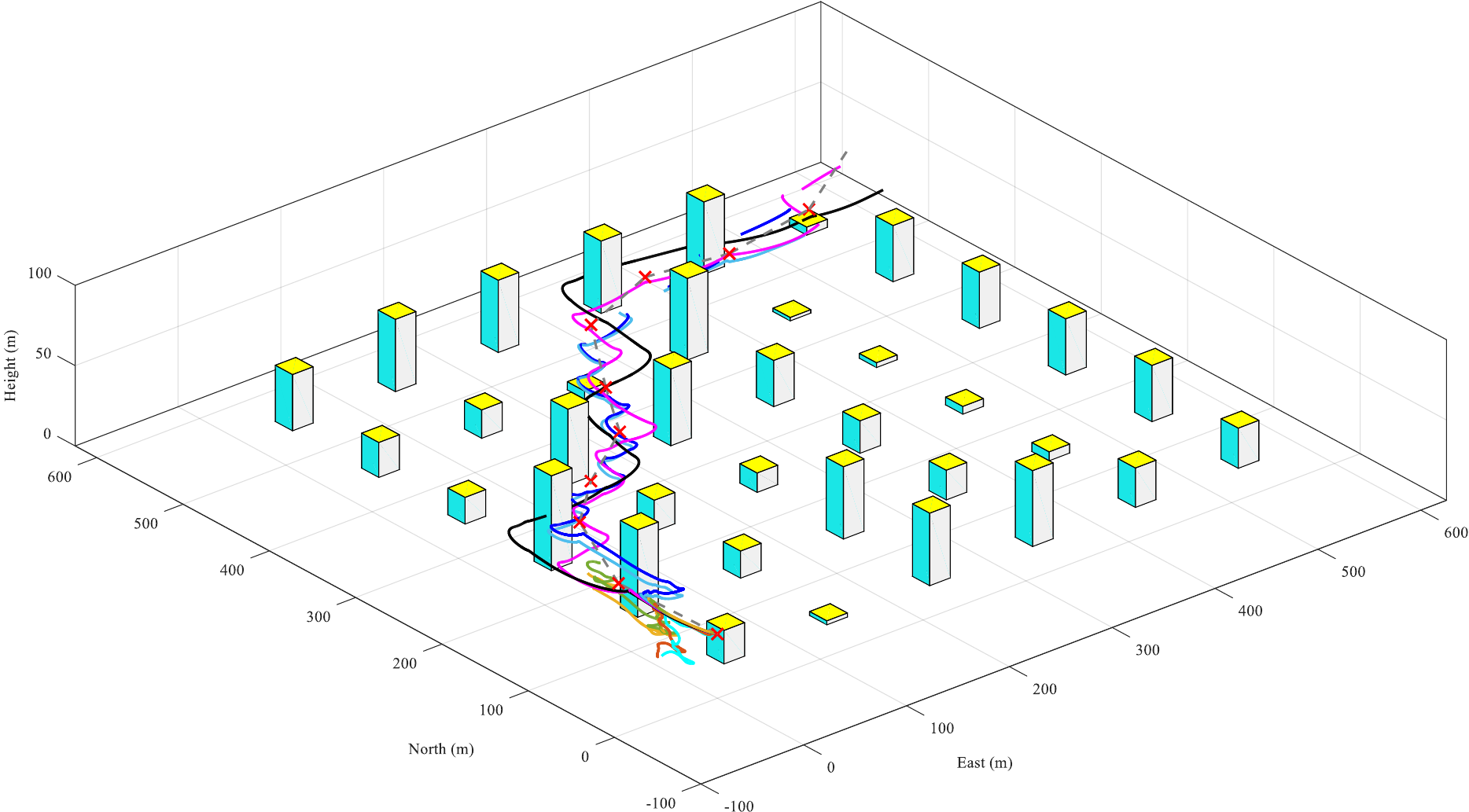} &
		\includegraphics[width=0.32\textwidth]{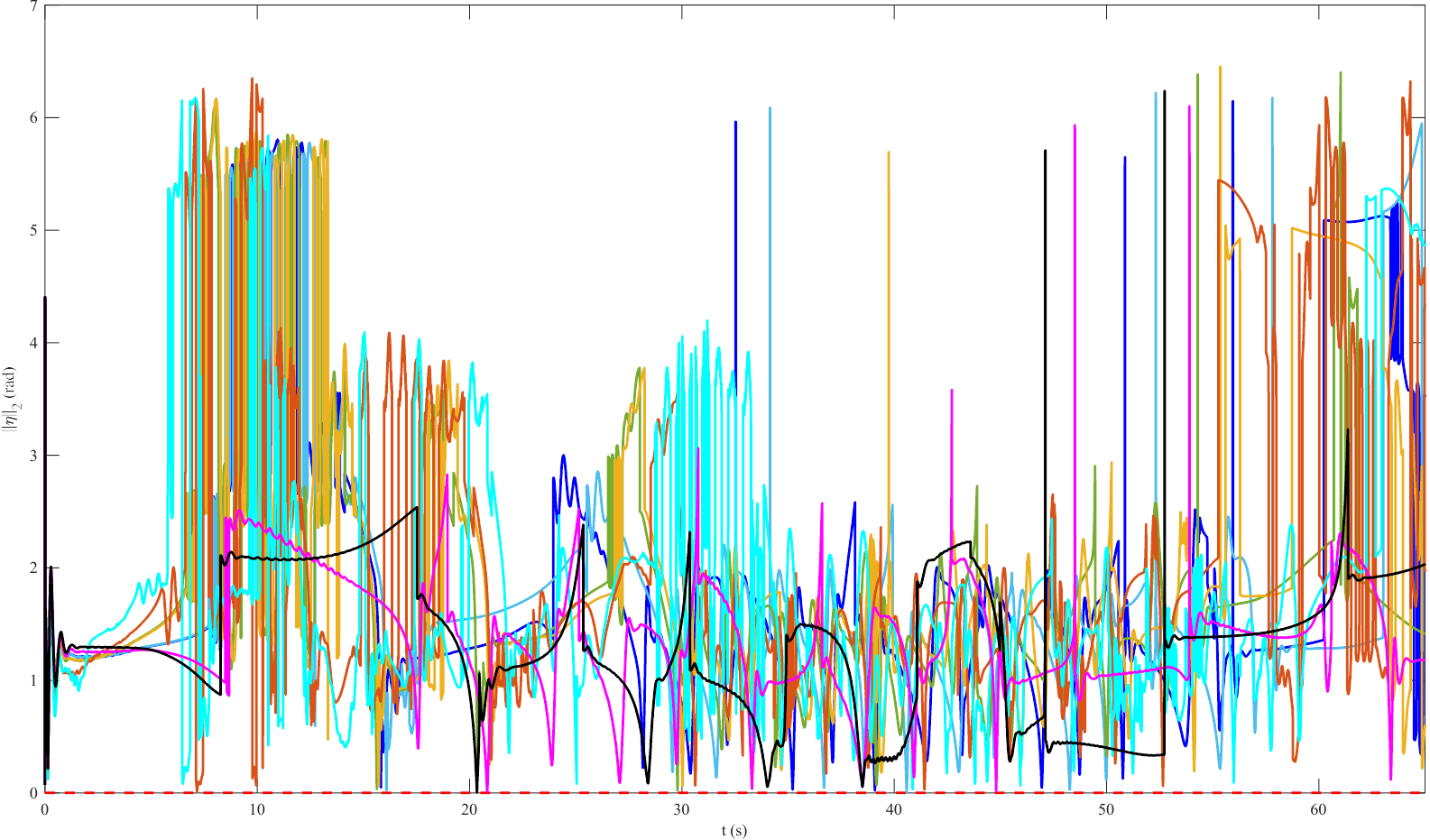} &
		\includegraphics[width=0.32\textwidth]{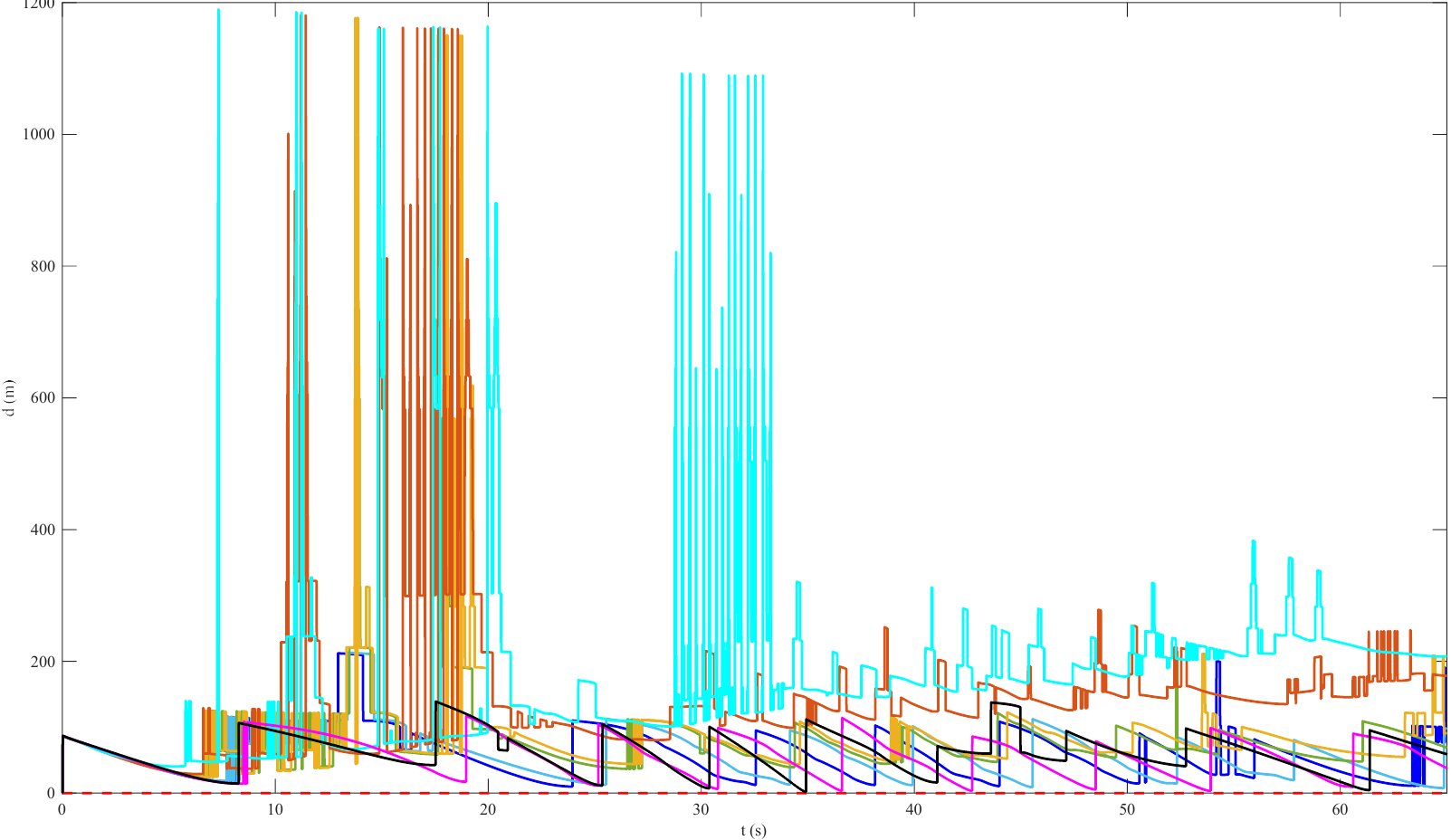} \\
		& (d) & \\
		\includegraphics[width=0.32\textwidth]{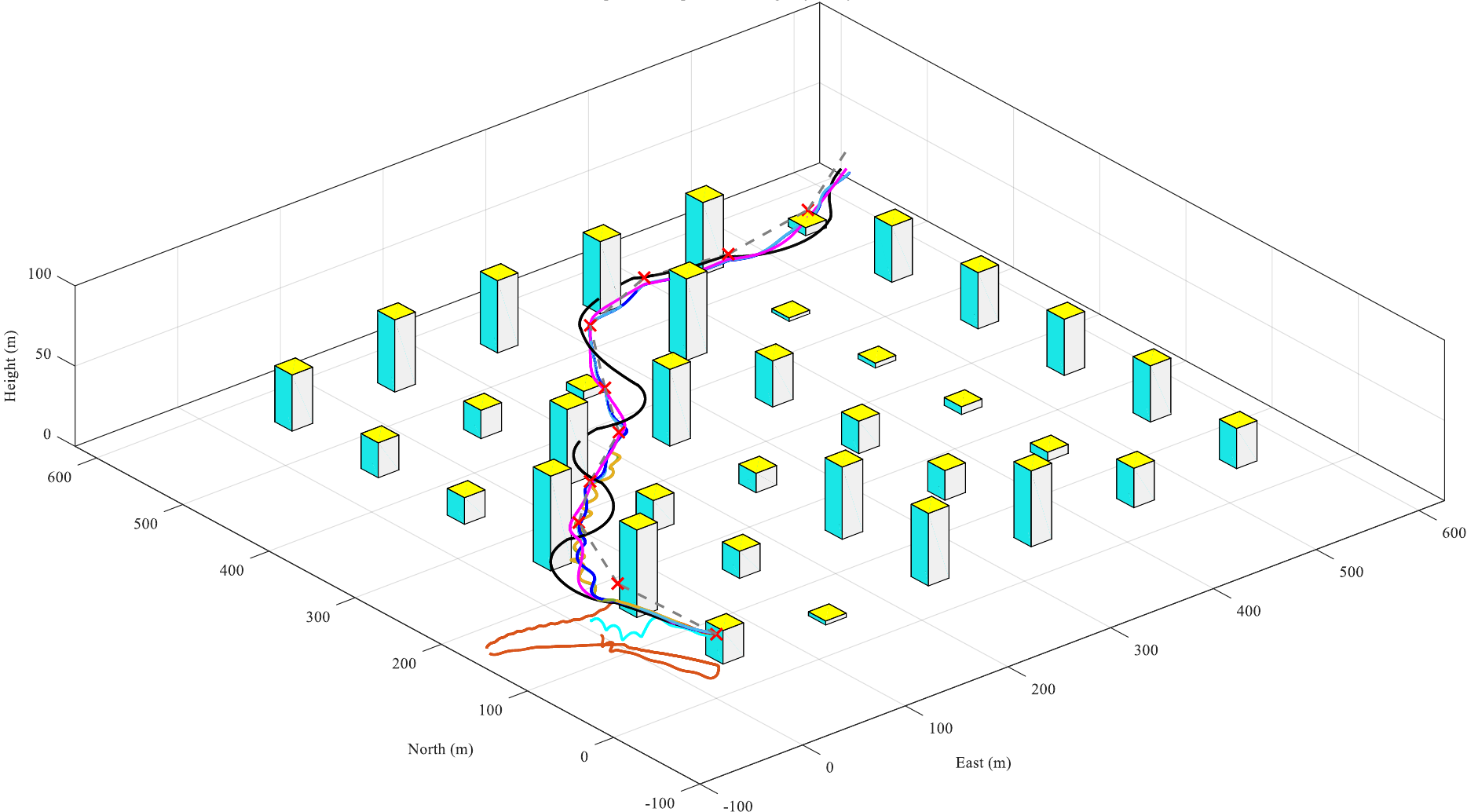} &
		\includegraphics[width=0.32\textwidth]{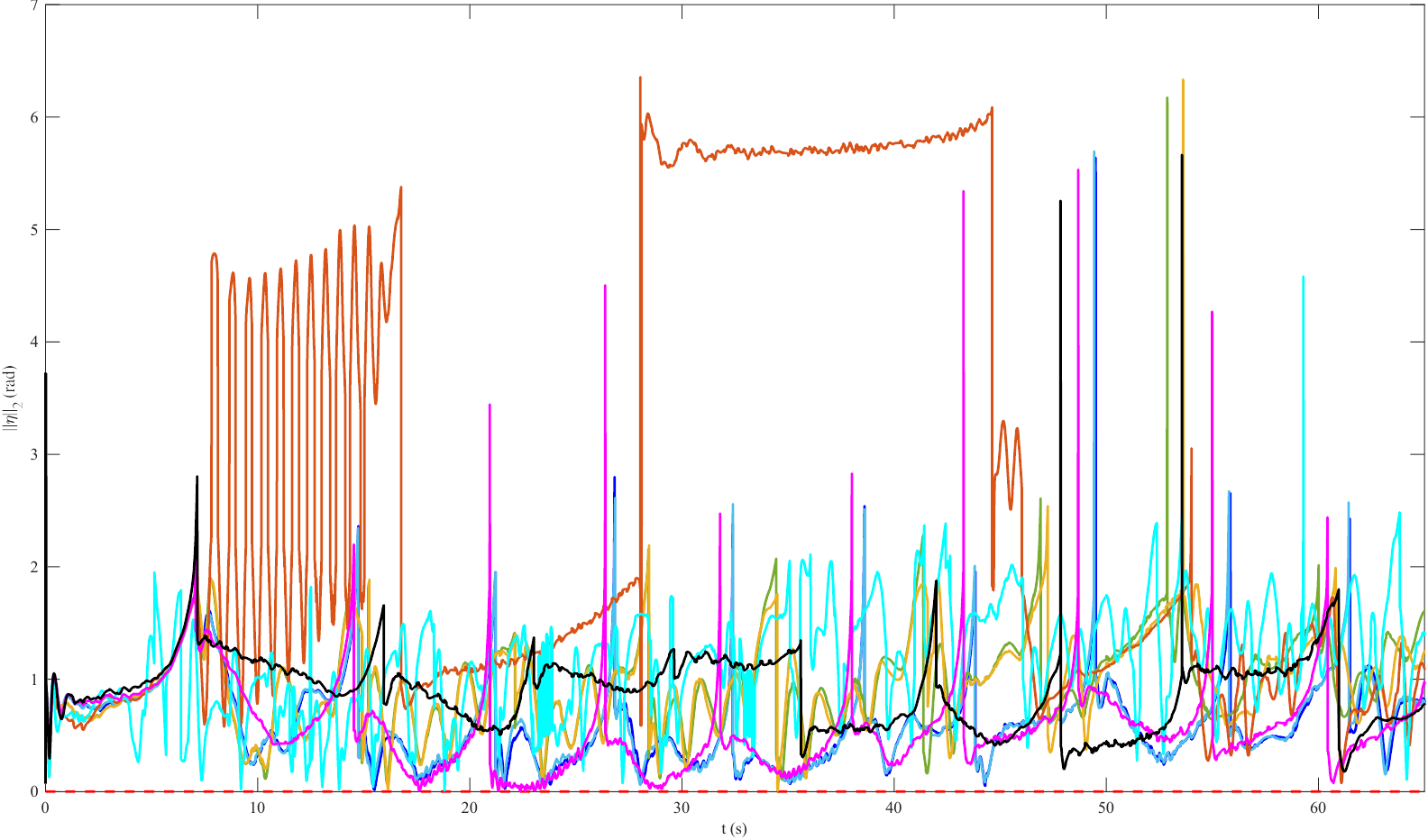} &
		\includegraphics[width=0.32\textwidth]{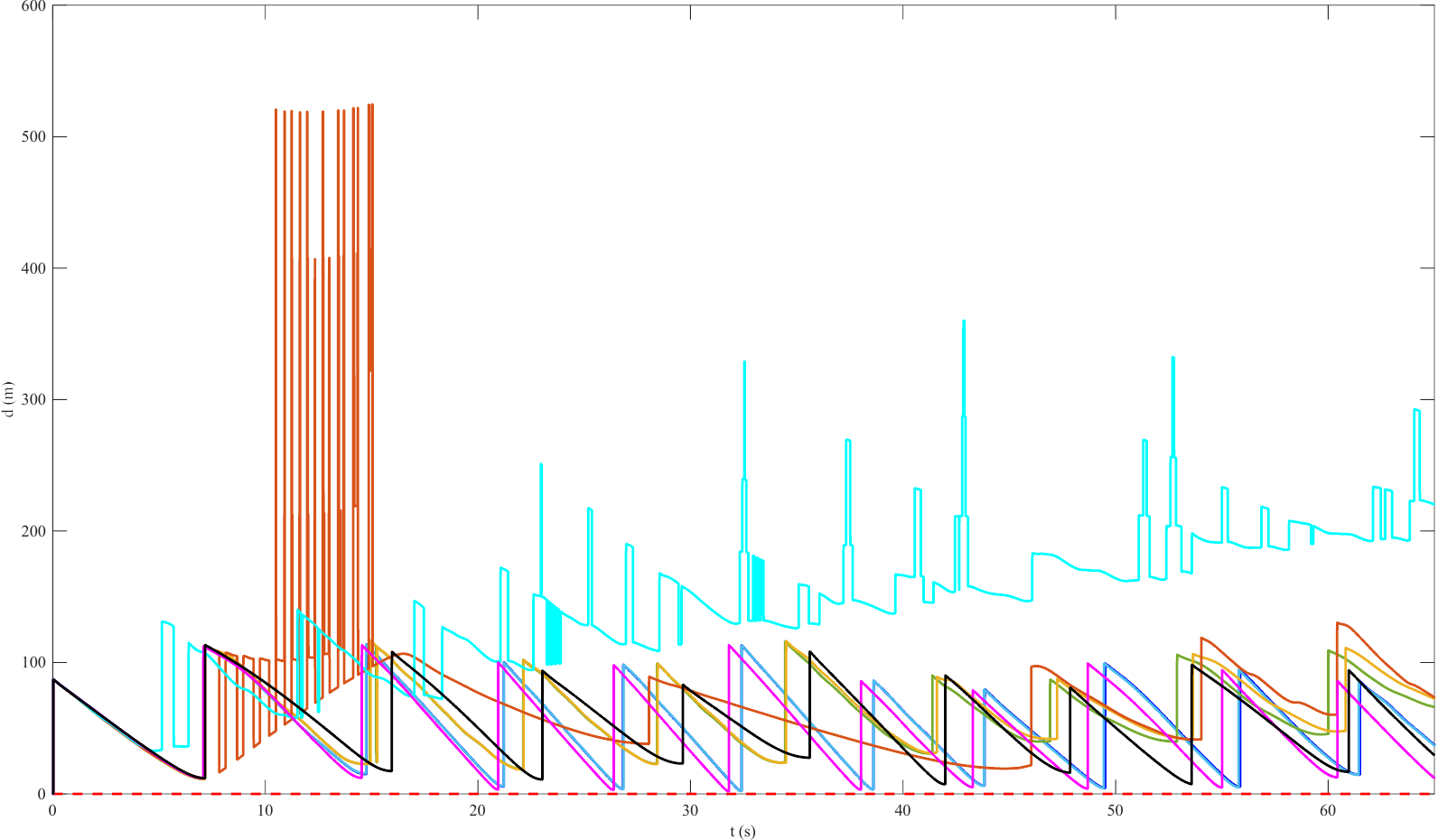} \\
		& (e) & \\
		\includegraphics[width=0.32\textwidth]{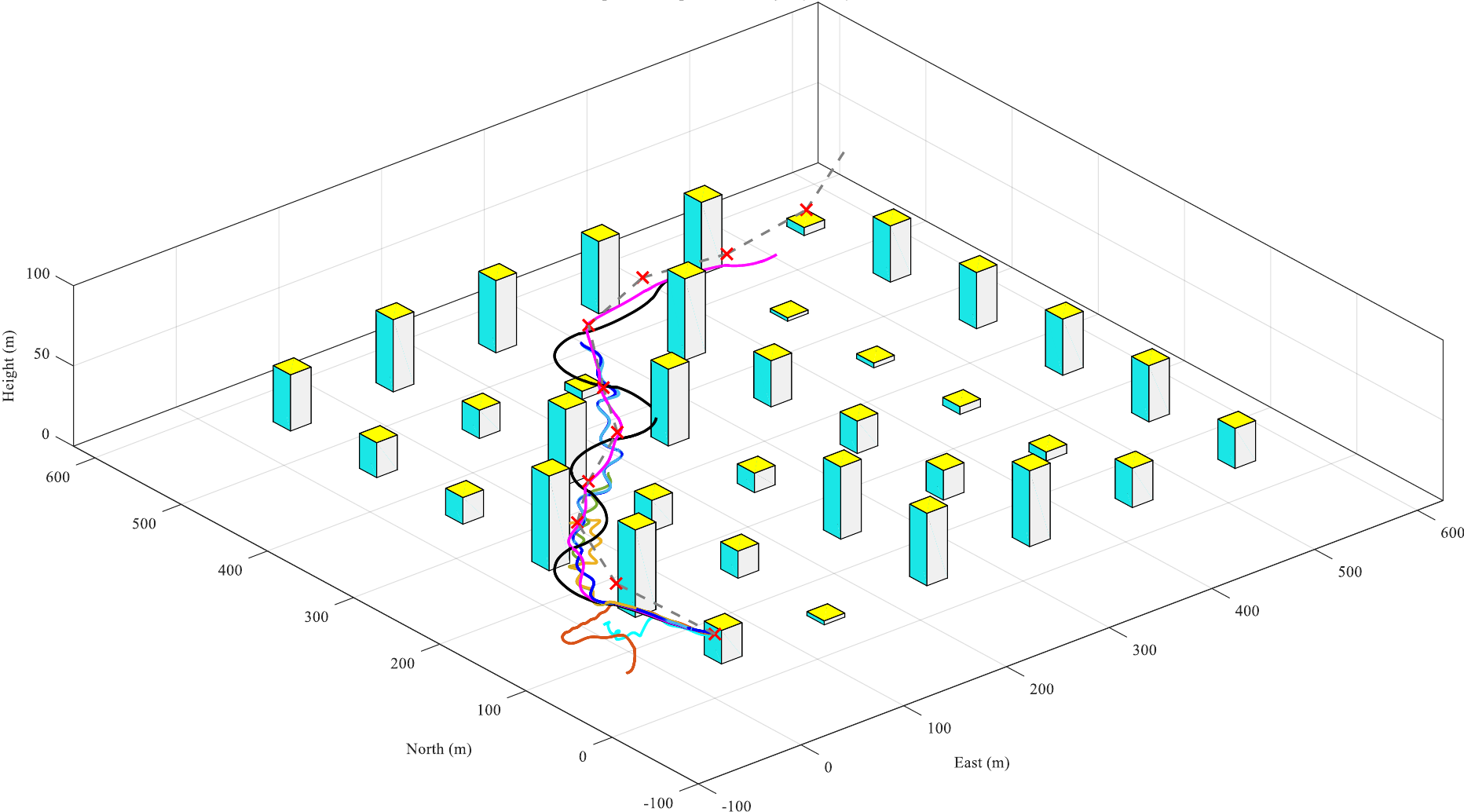} &
		\includegraphics[width=0.32\textwidth]{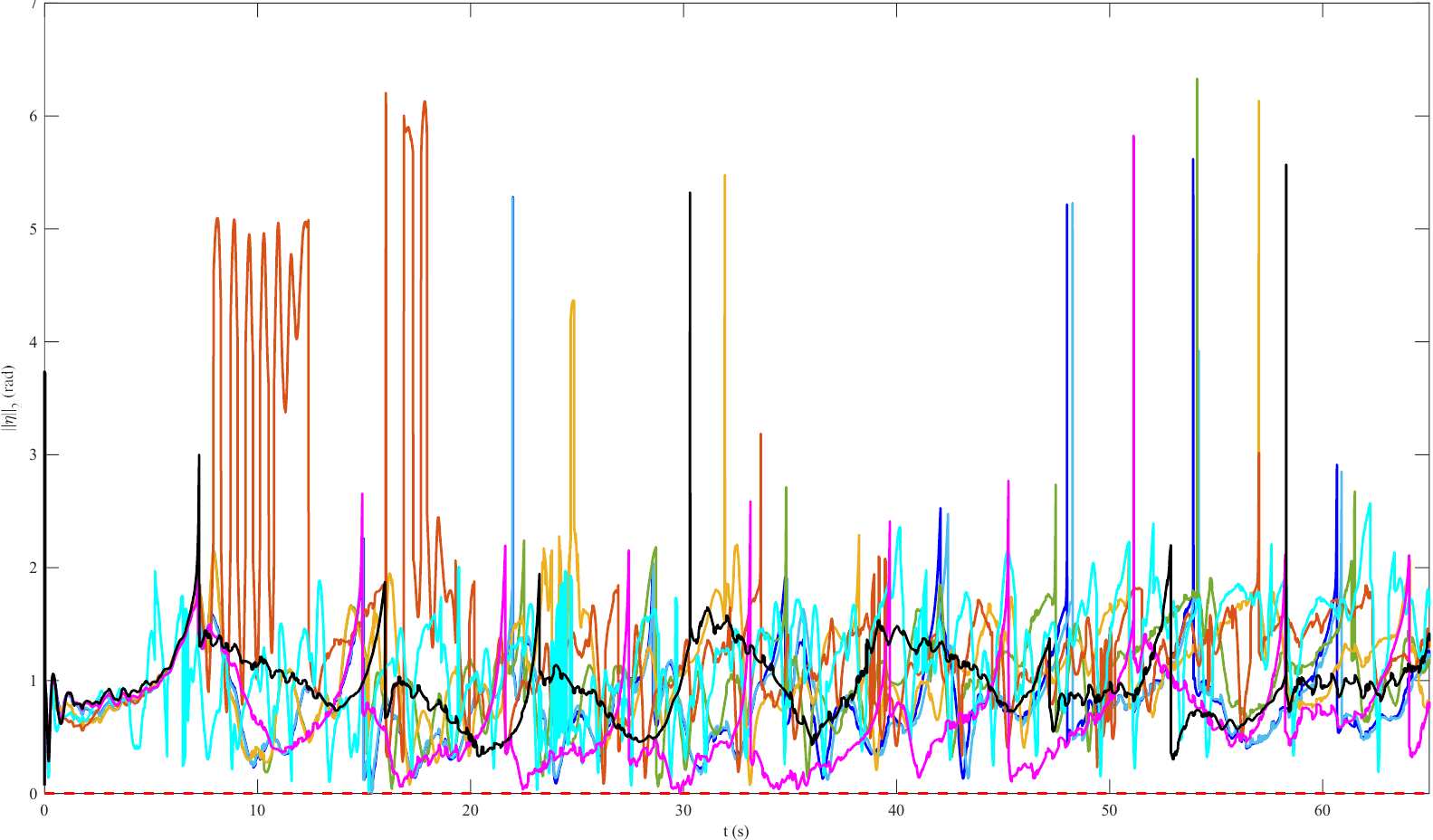} &
		\includegraphics[width=0.32\textwidth]{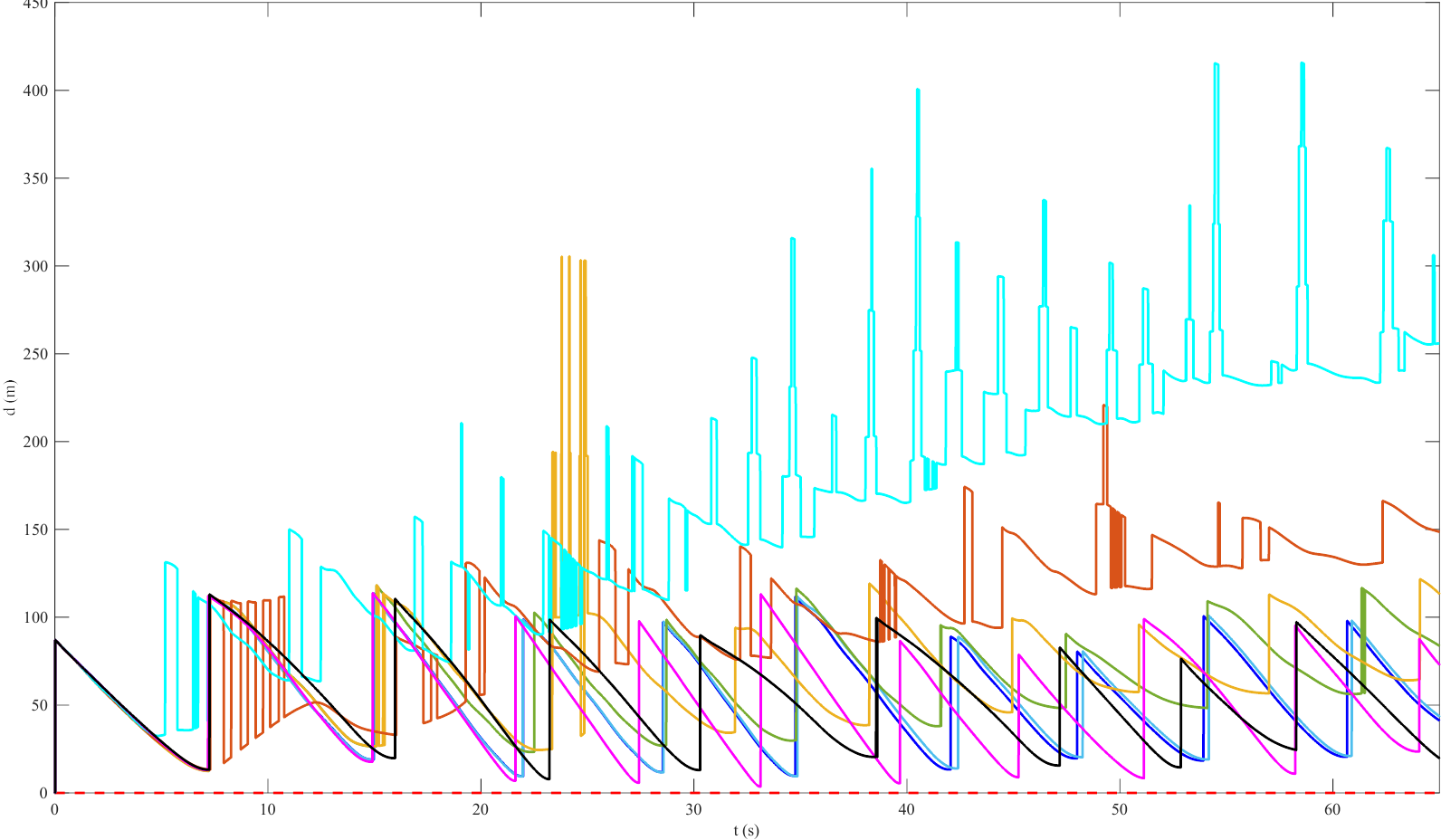} \\
		& (f) & \\
	\end{tabular}
	\caption{When different disturbances \( w_1, w_2, w_3, w_4, w_5, w_6 \) occur, for the conductivity RLLP-sin(x) with different parameters \( k_{\chi} \) and \( k_{\gamma} \), the path-following trajectory of the UAV for waypoints, as well as the corresponding time-series curves of \( d \) and \( \|\eta\|_2 \) are presented. 
	(a): For disturbance \( w_1 \);
	(b): For disturbance \( w_2 \);
	(c): For disturbance \( w_3 \);
	(d): For disturbance \( w_4 \);
	(e): For disturbance \( w_5 \);
	(f): For disturbance \( w_6 \).}
	\label{fig:22}
\end{figure*}

This section discusses wind-induced atmospheric disturbances and their influence on aircraft dynamics, followed by an analysis of RLLP-$\sin(x)$ robustness under varying disturbance intensities and algorithm parameters. Wind and atmospheric disturbances affect the airspeed \( V_a \), angle-of-attack \( \alpha \), and sideslip angle \( \beta \). These parameters mediate the impact of wind on aerodynamic forces and moments, thereby influencing aircraft motion. The airspeed \( V_a \) is defined as the velocity of the airframe relative to the surrounding air mass, while the wind velocity \( V_w \) denotes the velocity of the air mass relative to the ground. For simulation purposes, the total wind vector is decomposed as \( V_w = V_{w_s} + V_{w_g} \), where \( V_{w_s} \) is a constant vector representing steady ambient wind, and \( V_{w_g} \) is a stochastic process modeling wind gusts and other atmospheric disturbances.  
The ambient wind is typically expressed in the inertial frame as \( V_{w_s} = (w_{n_s}, w_{e_s}, w_{d_s})^T \), with \( w_{n_s} \), \( w_{e_s} \), and \( w_{d_s} \) denoting steady wind speeds in the north, east, and down directions, respectively. The stochastic gust component is described in the aircraft body frame as \( V_{w_g} = (u_{w_g}, v_{w_g}, w_{w_g})^T \). Here, the non-steady gust portion is modeled using white noise filtered through a linear time-invariant system based on the von Kármán turbulence spectrum \cite{stengel2005flight} within the Dryden transfer functions \cite{beard2012small}. The Dryden gust model parameters specified in MIL-F-8785C \cite{langelaan2011wind} are: turbulence intensities along body-frame axes \( \sigma_u = \sigma_v = 2.12 \, \text{m/s} \), \( \sigma_w = 1.4 \, \text{m/s} \); spatial wavelengths \( L_u = L_v = 200 \, \text{m} \), \( L_w = 50 \, \text{m} \). The influence mechanisms of specific aerodynamic forces and moments are detailed in Ref.\cite{beard2012small}.
\begin{table}[htbp]
	\centering
	\caption{Some different types of disturbances caused by wind field where $U$ is the uniformly distributed random numbers over the interval [0,1].}
	\footnotesize
	\begin{tabularx}{0.48\textwidth}{lll}
		\toprule
		Types & $V_{w_s} (\text{m}/\text{s})$ & $V_{w_g} (\text{m}/\text{s})$\\
		\midrule
		$w_1$ & (5,0,0) & (0,0,0)\\
		$w_2$ & (5,5,0) & (0,0,0)\\
		$w_3$ & (5,5,2) & (0,0,0)\\
		$w_4$ & (10,10,2) & (0,0,0)\\
		$w_5$ & (5,5,2) & (-0.25+0.5$U$, -0.125+0.25$U$, -0.125+0.25$U$)\\
		$w_6$ & (5,5,2) & (-0.5+$U$, -0.25+0.5$U$, -0.25+0.5$U$)\\
		\bottomrule
	\end{tabularx}
	\label{tab:1}
\end{table}
\begin{table}[htbp]
	\centering
	\caption{Some different types of $k_{\chi}$ and $k_{\gamma}$ for the RLLP-$\sin(x)$.}
	\footnotesize
	\begin{tabularx}{0.45\textwidth}{llll}
		\toprule
		Types & ($k_{\chi},k_{\gamma}$) & $I(f)$ & $R(f)$\\
		\midrule
        $k_1$ & (0.5,0.5) & $0.5\pi$ & 1\\
        $k_2$ & (0.5,1) & $\pi$ & 1\\
        $k_3$ & (1,0.5) & $\pi$ & 1\\
        $k_4$ & (1,1) & $0.25\pi$ & 2\\
        $k_5$ & (2,2) & $0.125\pi$ & 4\\
        $k_6$ & (4,4) & $0.0625\pi$ & 8\\
        $k_7$ & (0.25,0.25) & $\pi$ & 0.5\\
        $k_8$ & (0.1,0.1) & $2.5\pi$ & 0.2\\
		\bottomrule
	\end{tabularx}
	\label{tab:2}
\end{table}
To investigate the path-following performance of RLLP-type algorithms under wind and atmospheric disturbances, as well as the robustness and limitations exhibited by varying parameters of RLLP algorithms (which typically affect their attractor \( I(f) \)), we compare the differences in indices \( d \) and \( \|\eta\|_2 \) of the RLLP-sin(x) algorithm during UAV waypoint path-following. The comparisons are conducted under 6 types of wind and atmospheric disturbances (\( w_1 \)-\( w_6 \)) and 8 parameter categories (\( k_1 \)-\( k_8 \)), with specific parameter settings provided in Table \ref{tab:1} and Table \ref{tab:2}. Experimental results are presented in Figure \ref{fig:22}, from which the following conclusions can be drawn.

Initially, in experiments (a)-(f), as the intensity of applied steady wind speeds increases, the RLLP-sin(x) algorithm under all parameter configurations except \( k_5 \) and \( k_6 \) can maintain a desirable degree of robust stability for \( \|\eta\|_2 \), thereby ensuring that \( d \) achieves fast finite-time stabilization of path-following errors under disturbances. Notably, compared with other parameters, the algorithms corresponding to \( k_5 \) and \( k_6 \) overly prioritize a large exponential convergence rate \( R(f) \) in the time-series curves of \( \|\eta\|_2 \), with the goal of confining path-following errors within the smallest possible attractor \( I(f) \); consequently, they exhibit pronounced oscillatory behavior. In experiments involving stronger wind intensities (d)-(f), such detrimental impact on stabilizing path-following errors is further amplified, rendering it evident that the corresponding index \( d \) fails to achieve strict finite-time error stabilization. This phenomenon primarily stems from the fact that an excessively large exponential convergence rate \( R(f) \) exacerbates oscillations of the angle error \( \eta \) during convergence, thereby impairing the stabilization effect of distance errors.  

Furthermore, observations of the UAV’s path-following trajectories reveal that a larger attractor \( I(f) \) and convergence rate \( R(f) \) typically result in trajectories closer to straight lines with smaller fluctuations, enabling the UAV’s actual trajectory to better evade obstacles in complex terrains. Direct evidence of this is that the trajectory corresponding to parameter \( k_8 \) exhibits a greater curvature relative to others, rendering it more susceptible to collisions with obstacles and subsequent damage. However, it possesses the advantage that the UAV’s trajectory undergoes less significant variation as wind intensity increases, indicating reduced oscillations during the convergence of path-following errors—generally implying stronger adaptability of the algorithm to unknown disturbances.  
\begin{figure*}
	\centering
	\begin{tabular}{cc}
		\includegraphics[width=0.45\textwidth]{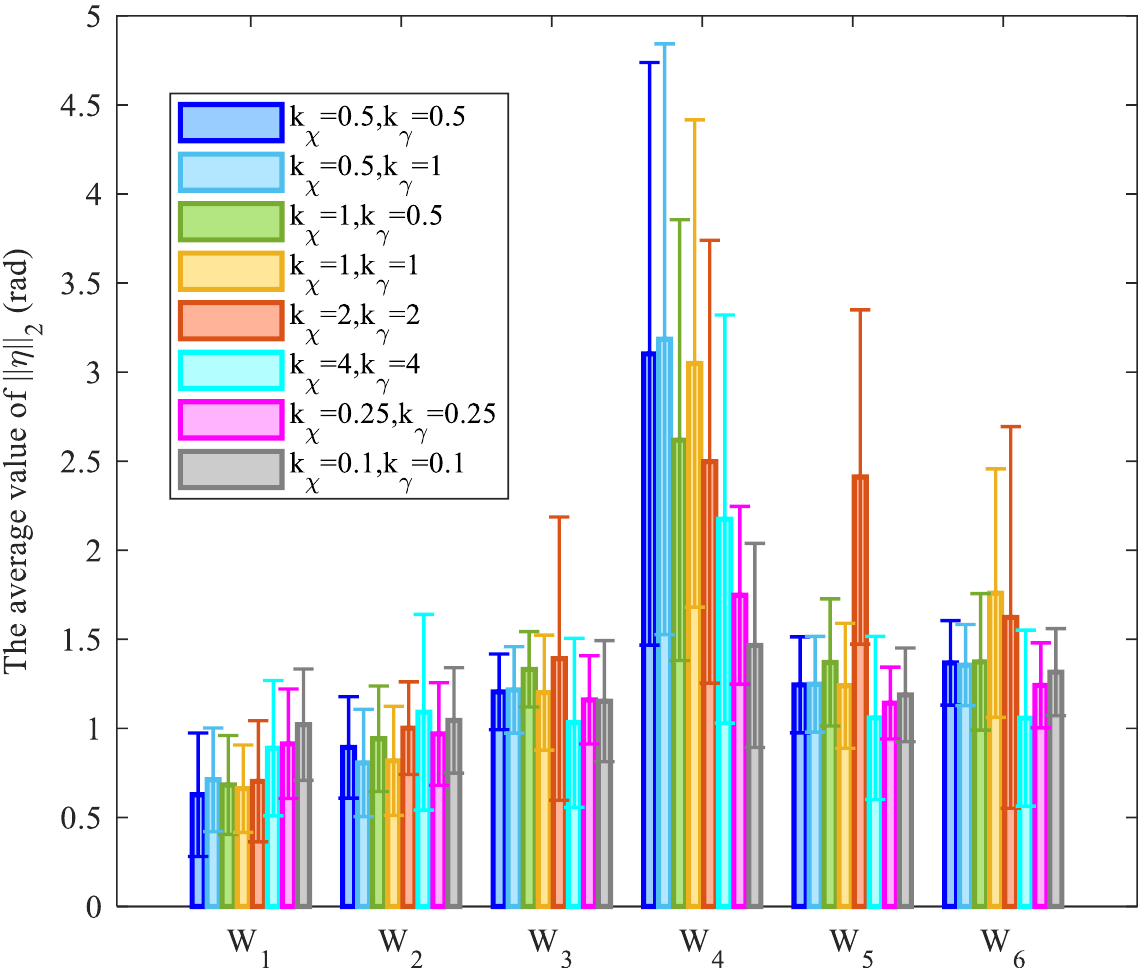} &
		\includegraphics[width=0.45\textwidth]{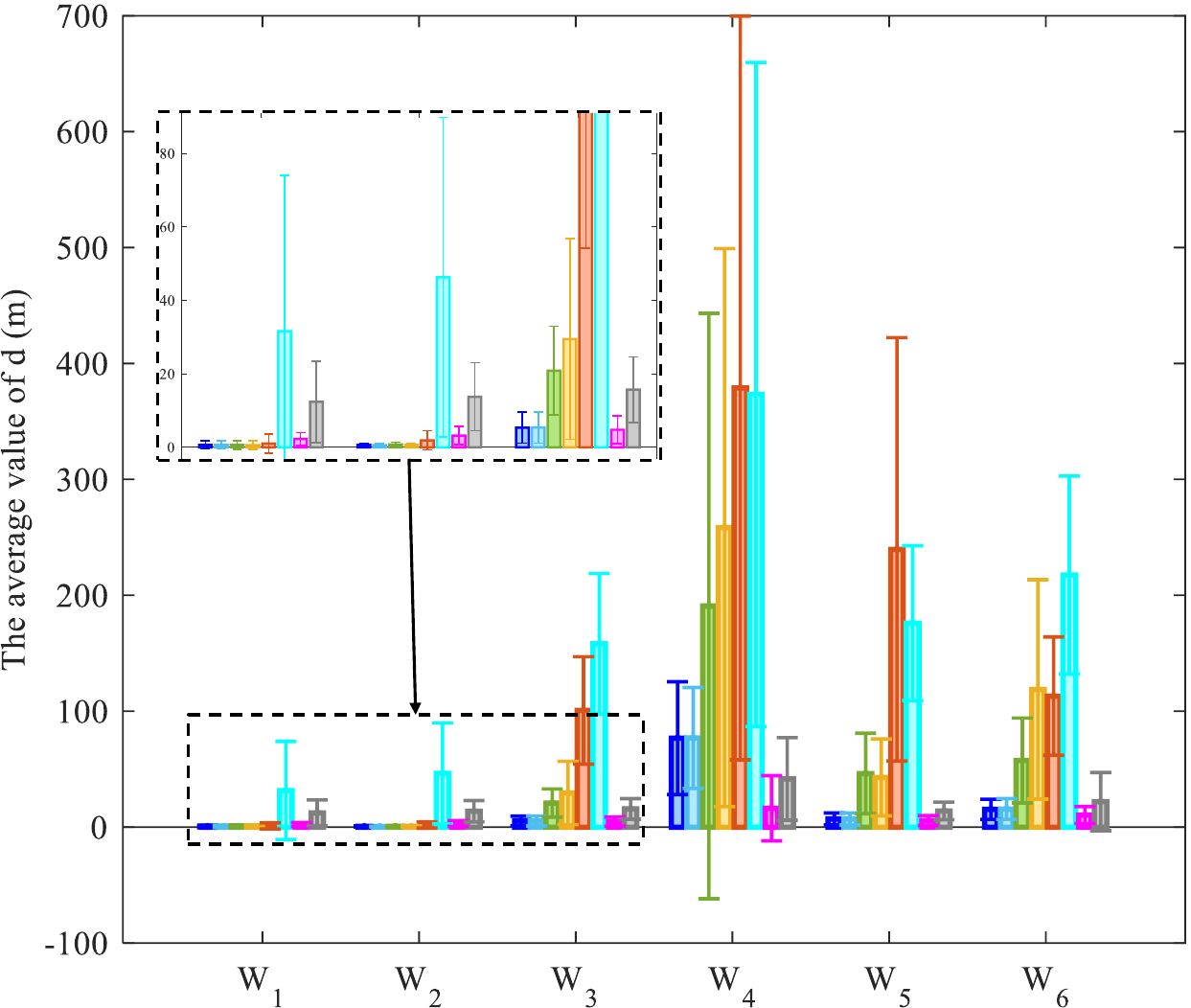} \\
		(a)&(b)
	\end{tabular}
	\caption{Comparison of indices \(\bar{\eta}\) and \(\bar{d}\) during the path-following process under different disturbance conditions \(w_1, w_2\)-\(w_6\) and different parameters \(k_1, k_2\)-\(k_8\). (a): The comparison on \(\bar{\eta}\); (b): The comparison on \(\bar{d}\).}
	\label{fig:23}
\end{figure*}
To quantitatively validate our assertions, we introduce the following metrics \(\bar{\eta}\), $\eta_{\text{std}}$, and \(\bar{d}\), $d_{\text{std}}$ to characterize the error convergence performance of the time-series curves \(\|\eta\|_2\) and \(d\) over the entire path-following trajectory.
\begin{align}
	\bar{\eta} &= \frac{1}{N}\sum_{i=1}^N ||\eta(t_i)||_2 \\
	\eta_{\text{std}} &= \sqrt{\frac{1}{N-1} \sum_{i=1}^N\left(||\eta(t_i)||_2 - \bar{\eta} \right)^2} \\
	\bar{d} & =  \frac{1}{N}\sum_{i=1}^N d(t_i)\\
	d_{\text{std}} & = \sqrt{\frac{1}{N-1} \sum_{i=1}^N
		\left( d(t_i) - \bar{d} \right)^2 }
\end{align}
where \( t_1, t_2, \ldots, t_N \) correspond to the instants when the UAV switches the waypoints being tracked during the path-following process. Smaller magnitudes of \(\bar{\eta}\) and \(\bar{d}\) signify superior error stabilization efficacy and path-following performance. 

As shown in Figure \ref{fig:23}, under disturbance conditions \( w_1 \)-\( w_3 \) without random gusts, both the mean values \( \bar{\eta} \) and \( \bar{d} \) of the indices, as well as their standard deviations \( \eta_{\text{std}} \) and \( d_{\text{std}} \), exhibit a positive correlation with the \( R(f) \) corresponding to parameter \( k \). This trend is particularly evident when \( k_{\chi}=k_{\gamma}=2 \) and \( k_{\chi}=k_{\gamma}=4 \). Under disturbance conditions \( w_4 \)-\( w_6 \) with random gusts applied, parameters corresponding to larger \( R(f) \) (i.e., \( k_{\chi}=k_{\gamma}=4 \)) and those with smaller \( R(f) \) (i.e., \( k_{\chi}=k_{\gamma}=0.25 \)) both ensure a smaller \( \bar{\eta} \). However, the coefficient \( k_{\chi}=k_{\gamma}=0.25 \), which corresponds to a smaller standard deviation, achieves a smaller \( \bar{d} \) and a smaller standard deviation \( d_{\text{std}} \). This indicates that a smaller convergence rate \( R(f) \) tends to guarantee a more stable anti-disturbance effect, albeit at the cost of larger trajectory curvature and partial obstacle avoidance performance.  
\begin{figure*}
	\centering
	\begin{tabular}{cccc}
		\includegraphics[width=0.23\textwidth]{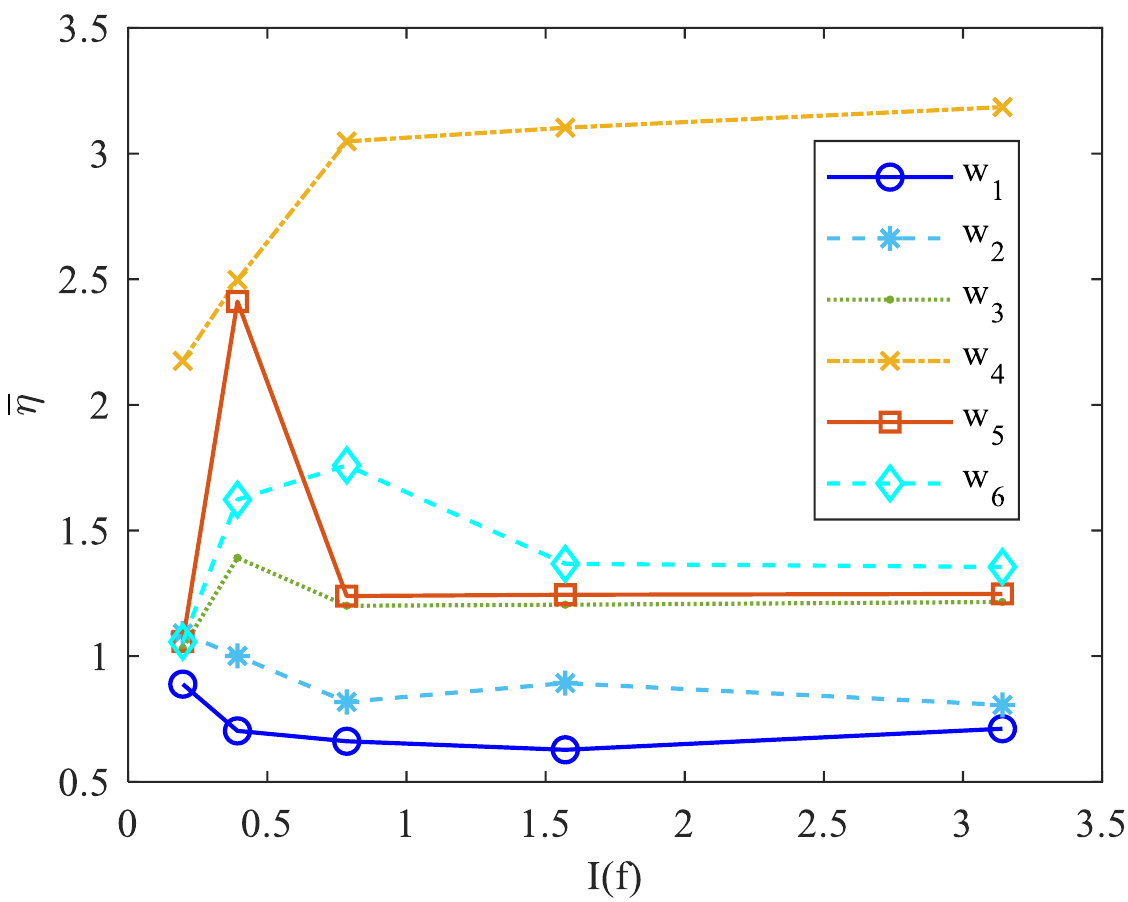} &
		\includegraphics[width=0.23\textwidth]{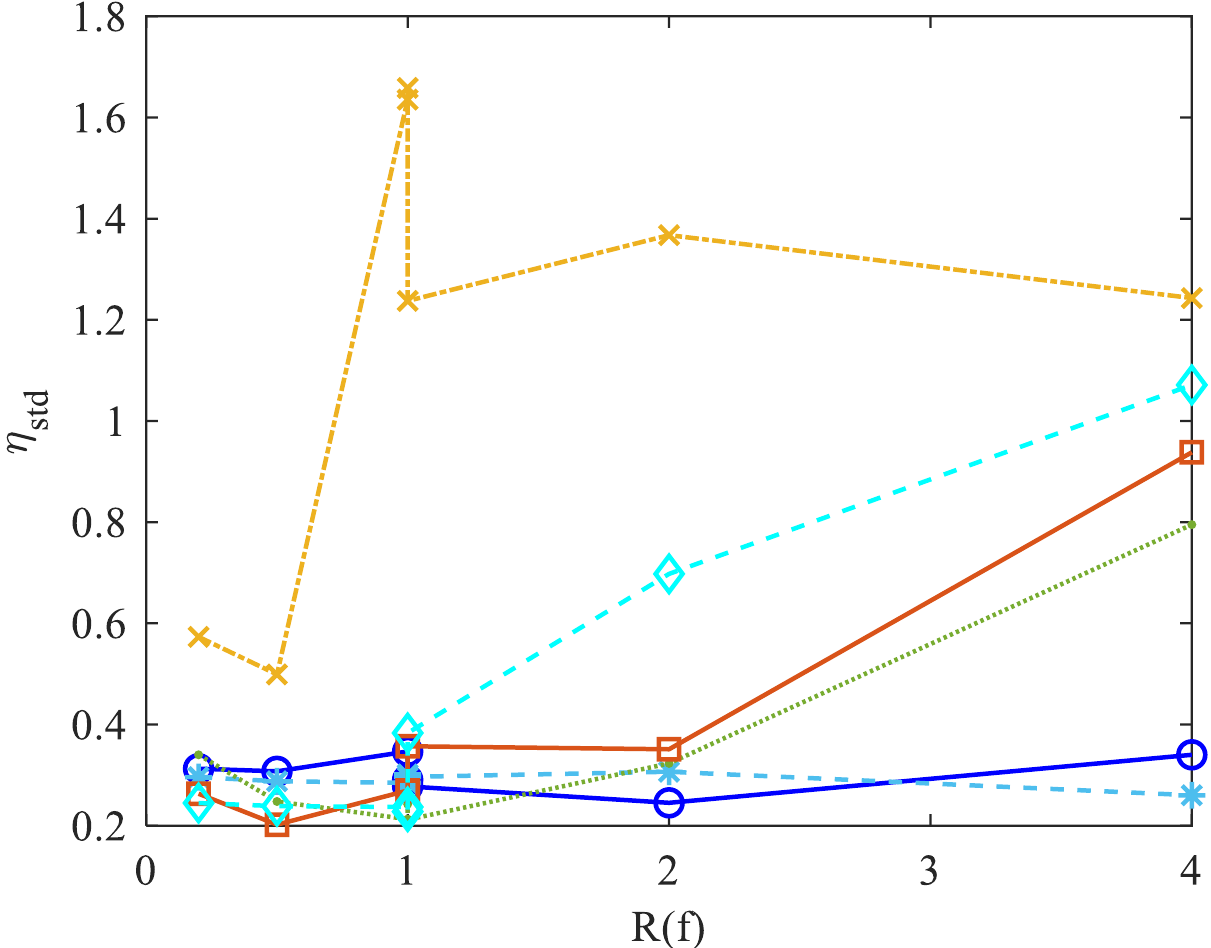} &
		\includegraphics[width=0.23\textwidth]{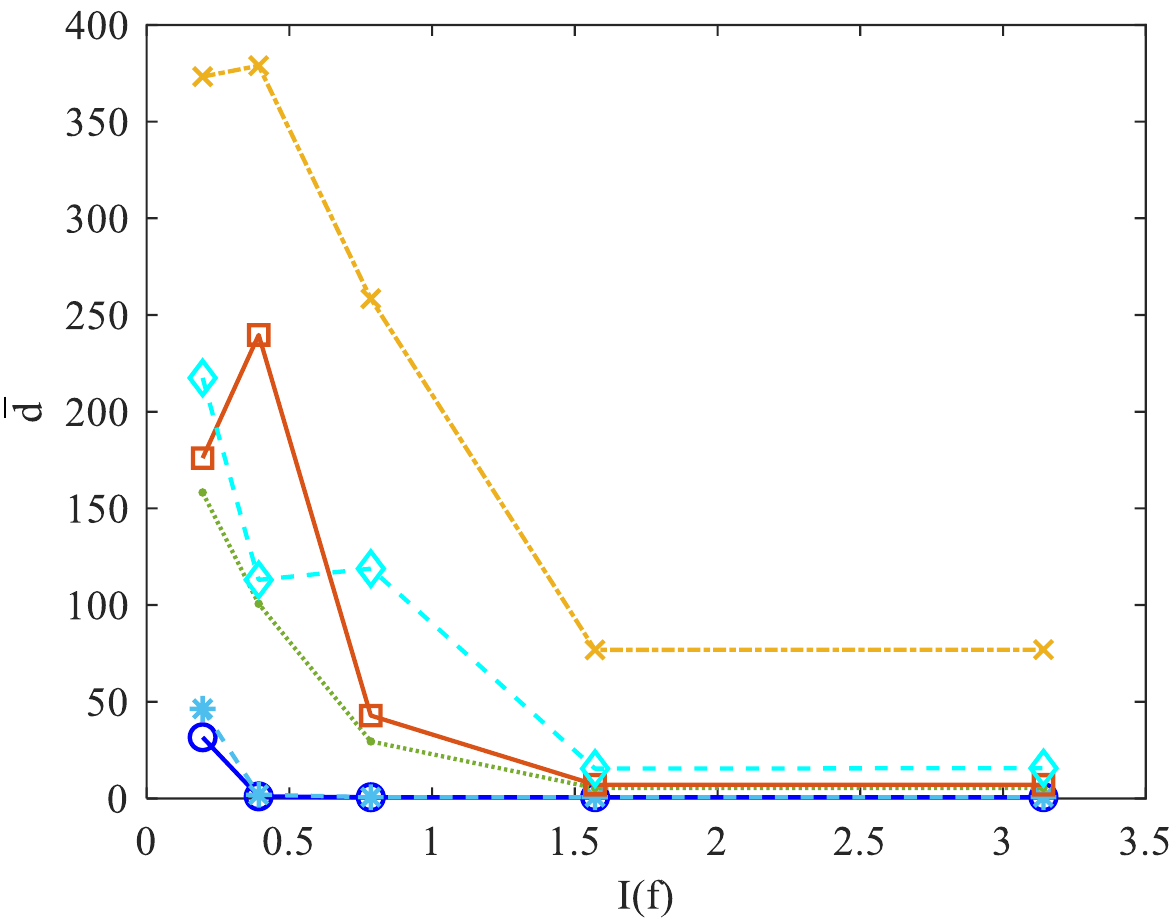} &
		\includegraphics[width=0.23\textwidth]{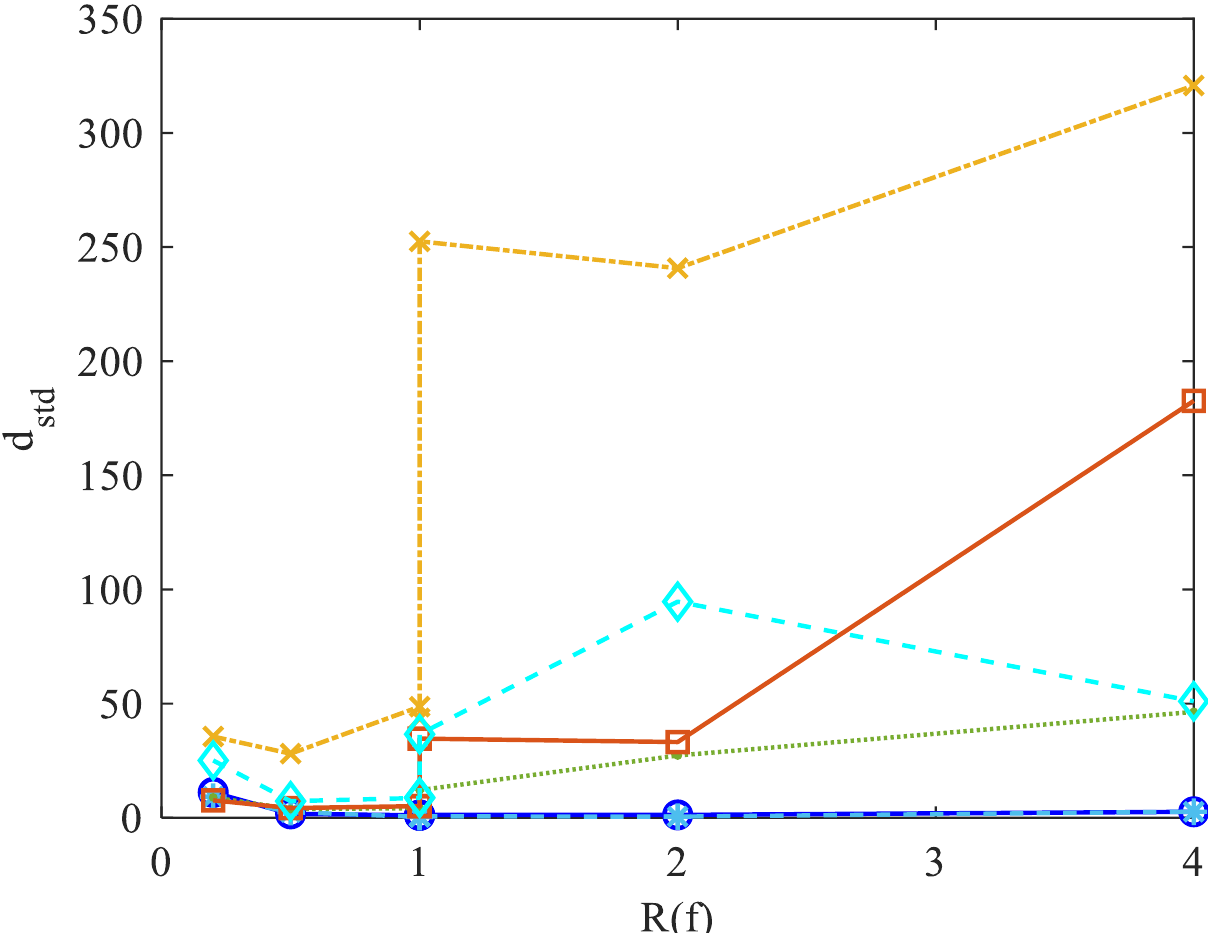} \\
		(a)&(b) & (c)&(d) \\
	\end{tabular}
	\caption{The statistical analysis based on the results in Figure \ref{fig:23} reveals the effects of RLLP-sin(x) under different \( I(f) \) and \( R(f) \) ( corresponding to parameters \( k_1 \)-\( k_8 \) ) on the metrics \(\bar{\eta}\), \(\eta_{\text{std}}\), \(\bar{d}\), and \(d_{\text{std}}\) under various disturbances \( w_1 \)-\( w_6 \).  
		(a): Impact of \( I(f) \) on \(\bar{\eta}\);  
		(b): Impact of \( R(f) \) on \(\eta_{\text{std}}\);  
		(c): Impact of \( I(f) \) on \(\bar{d}\);  
		(d): Impact of \( R(f) \) on \(d_{\text{std}}\).}
	\label{fig:24}
\end{figure*}
As shown in Figure \ref{fig:24}, \(\eta_{\text{std}}\) exhibits a strong positive correlation with \(R(f)\), which directly contributes to the positive correlation between \(R(f)\) and \(d_{\text{std}}\). This indicates that a larger \(R(f)\) increases the oscillatory behavior during the convergence of \(\eta\), thereby amplifying the oscillations in the convergence of distance errors. For \(I(f) > 0.5\), \(I(f)\) has little impact on \(\bar{\eta}\) and \(\bar{d}\). However, when \(I(f) \leq 0.5\), both \(\bar{\eta}\) and \(\bar{d}\) tend to decrease as \(I(f)\) increases. This phenomenon is primarily attributed to the fact that a smaller \(I(f)\) is typically associated with a larger \(R(f)\), which leads to larger oscillation amplitudes in \(\|\eta\|_2\) and \(d\), and consequently results in larger \(\bar{\eta}\) and \(\bar{d}\).

To conclude, under strong wind disturbances, a larger attractor \( I(f) \) often facilitates better finite-time robust stabilization of path-following errors. Nevertheless, a larger \( I(f) \) is in many instances correlated with a larger \( R(f) \). Although a larger \( R(f) \) renders the waypoint-tracking trajectory closer to a straight line (aiding obstacle avoidance), it induces substantial oscillations during the convergence of path-following errors, which frequently leads to failure in stabilizing such errors. Therefore, an optimal parameter configuration principle is to constrain \( I(f) \) within a small range while minimizing \( R(f) \)—even though these two objectives may be mutually conflicting. This principle can, on the one hand, ensure efficient stabilization of path-following errors and avoid excessive error oscillations, and on the other hand, guarantee that the trajectory remains sufficiently close to waypoints to achieve effective obstacle avoidance.

\section*{Conclusions and future work}
\label{sec:Conclusions and future work}
In summary, the innovative aspects of our work are summarized as follows:  

(1) From the perspective of global random attractors, we propose a novel approach to determine the exponential convergence of perturbed nonlinear systems (see Theorem \ref{thm:lemma_A1}), thereby introducing the metrics \( R(f) \) and \( I(f) \) to quantify the exponential convergence rate and the range of the ultimate convergence set, respectively. Additionally, we establish the sufficient conditions for the look-ahead pursuit guidance law to achieve rapid finite-time stability (see Theorem \ref{thm:theorem_1});

(2) Building upon the aforementioned theoretical framework, we develop a robust longitudinal-lateral look-ahead pursuit (RLLP) guidance method based on the traditional longitudinal and lateral look-ahead pursuit approach (see Eq. (\ref{eq:guidance law1})). We further quantify the key metrics characterizing its robustness and propose a method to optimize these robustness metrics (see Algorithm \ref{alg:optimal_pathfolllowing});

(3) Using real obstacle environments, UAV dynamics, and aerodynamic models, we validate the effectiveness, optimality, and robustness of the proposed RLLP guidance algorithm against realistic atmospheric disturbances through simulation experiments. Finally, we discuss the principles for adjusting guidance parameters in more complex environments and under wind field perturbations. 

Future research will focus on refining these principles, specifically involving the rational design of the guidance function \( f \) to minimize \( R(f) \) while confining \( I(f) \) within a specified range, thereby ensuring smooth convergence of path-following errors.

\section*{Acknowledgements}
This work was supported by the National Natural Science Foundation of China (No.51775435). The preprint version is available at: \url{https://arxiv.org/abs/2505.16407}.

\bibliographystyle{unsrt}
\bibliography{refs}

\end{document}